\documentclass[twocolumn,twocolumnappendix]{aastex631}  %
\received{2022 September 1}
\revised{2023 April 5}
\accepted{2023 April 19}
\published{2023 July 19}

\submitjournal{ApJ}

\shorttitle{The Planetary Accretion Shock. III. 2.5D}  %
\shortauthors{Marleau et al.}
\usepackage{etoolbox}

\makeatletter

\patchcmd{\NAT@citex}
  {\@citea\NAT@hyper@{%
     \NAT@nmfmt{\NAT@nm}%
     \hyper@natlinkbreak{\NAT@aysep\NAT@spacechar}{\@citeb\@extra@b@citeb}%
     \NAT@date}}
  {\@citea\NAT@nmfmt{\NAT@nm}%
   \NAT@aysep\NAT@spacechar\NAT@hyper@{\NAT@date}}{}{}

\patchcmd{\NAT@citex}
  {\@citea\NAT@hyper@{%
     \NAT@nmfmt{\NAT@nm}%
     \hyper@natlinkbreak{\NAT@spacechar\NAT@@open\if*#1*\else#1\NAT@spacechar\fi}%
       {\@citeb\@extra@b@citeb}%
     \NAT@date}}
  {\@citea\NAT@nmfmt{\NAT@nm}%
   \NAT@spacechar\NAT@@open\if*#1*\else#1\NAT@spacechar\fi\NAT@hyper@{\NAT@date}}
  {}{}

\makeatother

\def\degr{{\mbox{\textdegree}}\xspace}

\makeatletter
\let\jnl@style=\rm
\def\ref@jnl#1{{\jnl@style#1}}

\def\aj{\ref@jnl{AJ}}                   %
\def\actaa{\ref@jnl{Acta Astron.}}      %
\def\araa{\ref@jnl{ARA\&A}}             %
\def\apj{\ref@jnl{ApJ}}                 %
\def\apjl{\ref@jnl{ApJ}}                %
\def\apjs{\ref@jnl{ApJS}}               %
\def\ao{\ref@jnl{Appl.~Opt.}}           %
\def\apss{\ref@jnl{Ap\&SS}}             %
\def\aap{\ref@jnl{A\&A}}                %
\def\aapr{\ref@jnl{A\&A~Rev.}}          %
\def\aaps{\ref@jnl{A\&AS}}              %
\def\azh{\ref@jnl{AZh}}                 %
\def\baas{\ref@jnl{BAAS}}               %
\def\bac{\ref@jnl{Bull. astr. Inst. Czechosl.}}
\def\caa{\ref@jnl{Chinese Astron. Astrophys.}}
\def\cjaa{\ref@jnl{Chinese J. Astron. Astrophys.}}
\def\icarus{\ref@jnl{Icarus}}           %
\def\jcap{\ref@jnl{J. Cosmology Astropart. Phys.}}
\def\jrasc{\ref@jnl{JRASC}}             %
\def\memras{\ref@jnl{MmRAS}}            %
\def\mnras{\ref@jnl{MNRAS}}             %
\def\na{\ref@jnl{New A}}                %
\def\nar{\ref@jnl{New A Rev.}}          %
\def\pra{\ref@jnl{Phys.~Rev.~A}}        %
\def\prb{\ref@jnl{Phys.~Rev.~B}}        %
\def\prc{\ref@jnl{Phys.~Rev.~C}}        %
\def\prd{\ref@jnl{Phys.~Rev.~D}}        %
\def\pre{\ref@jnl{Phys.~Rev.~E}}        %
\def\prl{\ref@jnl{Phys.~Rev.~Lett.}}    %
\def\pasa{\ref@jnl{PASA}}               %
\def\pasp{\ref@jnl{PASP}}               %
\def\pasj{\ref@jnl{PASJ}}               %
\def\rmxaa{\ref@jnl{Rev. Mexicana Astron. Astrofis.}}%
\def\qjras{\ref@jnl{QJRAS}}             %
\def\skytel{\ref@jnl{S\&T}}             %
\def\solphys{\ref@jnl{Sol.~Phys.}}      %
\def\sovast{\ref@jnl{Soviet~Ast.}}      %
\def\ssr{\ref@jnl{Space~Sci.~Rev.}}     %
\def\zap{\ref@jnl{ZAp}}                 %
\def\nat{\ref@jnl{Nature}}              %
\def\iaucirc{\ref@jnl{IAU~Circ.}}       %
\def\aplett{\ref@jnl{Astrophys.~Lett.}} %
\def\apspr{\ref@jnl{Astrophys.~Space~Phys.~Res.}}
\def\bain{\ref@jnl{Bull.~Astron.~Inst.~Netherlands}} 
\def\fcp{\ref@jnl{Fund.~Cosmic~Phys.}}  %
\def\gca{\ref@jnl{Geochim.~Cosmochim.~Acta}}   %
\def\grl{\ref@jnl{Geophys.~Res.~Lett.}} %
\def\jcp{\ref@jnl{J.~Chem.~Phys.}}      %
\def\jgr{\ref@jnl{J.~Geophys.~Res.}}    %
\def\jqsrt{\ref@jnl{J.~Quant.~Spec.~Radiat.~Transf.}}
\def\memsai{\ref@jnl{Mem.~Soc.~Astron.~Italiana}}
\def\nphysa{\ref@jnl{Nucl.~Phys.~A}}   %
\def\physrep{\ref@jnl{Phys.~Rep.}}   %
\def\physscr{\ref@jnl{Phys.~Scr}}   %
\def\planss{\ref@jnl{Planet.~Space~Sci.}}   %
\def\procspie{\ref@jnl{Proc.~SPIE}}   %

\def\ptp{\ref@jnl{Prog.~Th.~Phys.}}   %
\def\natas{\ref@jnl{NatAs}}           %
\def\amjm{\ref@jnl{AmJM}}             %

\makeatother

\defcitealias{goli04}{G04}
\defcitealias{marl07}{M07}
\defcitealias{scvh}{SCvH}
\defcitealias{bl94}{BL94}
\defcitealias{mc14}{MC14}
\defcitealias{ensman94}{E94}
\defcitealias{lever81}{LP81}

\defcitealias{m16Schock}{Paper~I}
\defcitealias{m18Schock}{Paper~II}

\usepackage{xspace}

\usepackage{amsmath,amssymb,listings,upgreek,nicefrac}
\usepackage{newtxtext,newtxmath}
\usepackage{grffile}
\usepackage{xcolor,xspace}
\usepackage[normalem]{ulem}  %

\usepackage[T1]{fontenc}
\usepackage{textcomp}     %

\def\Hastjern{{\color[wave]{656}{\textbf{\textasteriskcentered}}}\xspace}

\newcounter{numKommG}
\newcommand{\auskommentiert}[1]{}                  %
\newcommand{\neuII}[1]{{\leavevmode{\boldmath\bfseries#1}}}   %
\newcommand{\neuIII}[1]{{\leavevmode\bfseries\textcolor{purple}{\neuII{#1}}}}   %
\newcommand{\neuIV}[1]{{\leavevmode\bfseries\textcolor{orange}{\neuII{#1}}}}   %
\newcommand{\neuV}[1]{{\leavevmode\bfseries\textcolor{red!30!yellow!65!black}{#1}}}   %

\renewcommand{\neuIII}[1]{{\neuII{#1}}}            %
\renewcommand{\neuIV}[1]{{\neuII{#1}}}             %

\renewcommand{\neuII}[1]{{\leavevmode#1}}          %
\renewcommand{\neuIII}[1]{{\leavevmode#1}}         %
\renewcommand{\neuIV}[1]{{\leavevmode#1}}          %

\renewcommand{\neuV}[1]{{\leavevmode#1}}            %
\def\e{\textrm{e}}                                 %
\def\mH{m_\textrm{H}}                              %
\def\sigSB{\ensuremath{\sigma_{\textrm{SB}}}\xspace}        %
\def\aST{\ensuremath{a_{\textrm{r}}}\xspace}       %
\def\MJ{\ensuremath{M_{\textrm{J}}}\xspace}        %
\def\RJ{\ensuremath{R_{\textrm{J}}}\xspace}        %
\def\RJfett{\textbf{\textit{R}}\ensuremath{\mathbf{_J}}\xspace}  %
\def\MSonne{\ensuremath{{M_\odot}}\xspace}         %
\def\LSonne{\ensuremath{{L_\odot}}\xspace}         %

\def\Ha{H\,$\alpha$\xspace}                               %
\def\Hb{H\,$\beta$\xspace}                                %
\def\Pab{Pa\,$\beta$\xspace}                              %
\def\Brg{Br\,$\gamma$\xspace}                             %
\def\PDS{PDS\,70\xspace}                                  %
\def\PDSb{PDS\,70\,b\xspace}                              %
\def\Dlrmb{Delorme\,1\,(AB)b\xspace}                      %

\def\CFL{\ensuremath{\textrm{CFL}}\xspace}                %

\def\Simrmaxkleiner{\texttt{LowAngMom}\xspace}
\def\Simrmaxgroesser{\texttt{HighAngMom}\xspace}
\def\SimMPgroesser{\texttt{HigherMass}\xspace}

\def\MPkt{\ensuremath{\dot{M}}\xspace}                               %
\def\MPktnettoHill{\ensuremath{\dot{M}_{\textrm{Hill,\,net}}}\xspace}       %
\def\MPktPdir{\ensuremath{\dot{M}_{\textrm{p,\,direct}}}\xspace}      %
\def\MPktHa{\ensuremath{\dot{M}_{\textrm{H}\,\alpha}}\xspace}    %

\def\APlObfl{\ensuremath{A_{\textrm{plnt~surf}}}\xspace}              %
\def\AzpSch{\ensuremath{A_{\textrm{CPD~surf}}}\xspace}                %

\def\MP{\ensuremath{M_{\textrm{p}}}\xspace}        %
\def\RP{\ensuremath{R_{\textrm{p}}}\xspace}        %
\def\RPfett{\textbf{\textit{R}}\ensuremath{\mathbf{_p}}\xspace}  %
\def\ffill{\ensuremath{f_{\textrm{fill}}}\xspace}                      %
\newcommand{\Tacc}{\ensuremath{{T_{\textrm{acc}}}}\xspace}  %
\newcommand{\TAkkP}{\ensuremath{{T'_{\textrm{acc}}}}\xspace}  %
\newcommand{\TAkkklass}{\ensuremath{{T_{\textrm{acc,\,class}}}}\xspace}  %

\def\LHa{\ensuremath{L_{\textrm{H}\,\alpha}}\xspace}    %
\def\FHa{\ensuremath{F_{\textrm{H}\,\alpha}}\xspace}             %
\newcommand{\RH}{\ensuremath{{R_{\textrm{Hill}}}}\xspace}       %
\def\RHill{\RH}
\newcommand{\RB}{\ensuremath{{R_{\textrm{Bondi}}}}\xspace}     %
\newcommand{\HP}{\ensuremath{{H_P}}\xspace}        %
\newcommand{\Teff}{\ensuremath{T_{\textrm{eff}}}\xspace}         %
\def\Pram{P_{\textrm{ram}}}                    %
\def\DeltaLuecke{\ensuremath{\Delta a_{\textrm{gap}}}\xspace}
\def\fpg{\ensuremath{f_{\textrm{d/g}}}\xspace}
\newcommand{\TSch}{\ensuremath{T_{\textrm{shock}}}\xspace}

\newcommand{\MStern}{\ensuremath{{M_{\star}}}\xspace}
\newcommand{\kapR}{\ensuremath{{\kappa_{\textrm{R}}}}\xspace}
\newcommand{\Erad}{\ensuremath{{E_{\textrm{rad}}}}\xspace}
\newcommand{\Trad}{\ensuremath{{T_{\textrm{rad}}}}\xspace}

\newcommand{\qth}{\ensuremath{{q_{\textrm{th}}}}\xspace}
\newcommand{\rmin}{\ensuremath{{r_{\textrm{min}}}}\xspace}
\newcommand{\rmax}{\ensuremath{{r_{\textrm{max}}}}\xspace}

\newcommand{\rfrz}{\ensuremath{{r_{\textrm{frz}}}}\xspace}

\newcommand{\Lu}{\ensuremath{{L_{\textrm{u}}}}\xspace}
\newcommand{\Dru}{\ensuremath{{\Delta r_{\textrm{u}}}}\xspace}
\newcommand{\Ls}{\ensuremath{{L_{\textrm{s}}}}\xspace}
\newcommand{\Nth}{\ensuremath{{N_\theta}}\xspace}

\newcommand{\epsPot}{\ensuremath{{\varepsilon_{\textrm{grav}}}}\xspace}

\newcommand{\vFf}{\ensuremath{{v_{\textrm{ff}}}}\xspace}
\newcommand{\vFfinfty}{\ensuremath{{v_{\textrm{ff},\,\infty}}}\xspace}

\newcommand{\tFfglob}{\ensuremath{{t_{\textrm{ff,\,glob}}}}\xspace}
\newcommand{\tFfrfrz}{\ensuremath{{t_{\textrm{ff,\,frz}}}}\xspace}

\newcommand{\vHakrit}{\ensuremath{{v_{\textrm{H}\,\alpha,\,\textrm{crit}}}}\xspace}
\newcommand{\vkrit}{\ensuremath{{v_{\textrm{crit}}}}\xspace}

\newcommand{\thmax}{\ensuremath{{\theta_{\textrm{max}}}}\xspace}

\newcommand{\rXXX}{\ensuremath{{r_{\textrm{30~\kms}}}}\xspace}

\newcommand{\rhoMitt}{\ensuremath{{\rho_{\textrm{mid}}}}\xspace}
\newcommand{\fomg}{\ensuremath{{f_\omega}}\xspace}
\newcommand{\omgk}{\ensuremath{{\omega_{\textrm{crit}}}}\xspace}

\newcommand{\OKep}{\ensuremath{{\Omega_{\textrm{Kep}}}}\xspace}

\newcommand{\vtr}{\ensuremath{{v_{\textrm{pol}}}}\xspace}
\newcommand{\thi}{\ensuremath{\theta_{\textrm{init}}}\xspace}

\newcommand{\DtauR}{\ensuremath{\Delta \tau_{\textrm{Ross}}}\xspace}
\newcommand{\kapStbfpgRoss}{\kappa_{\textrm{dust,\,Ross}}} %

\newcommand{\DDiff}{\ensuremath{D_{\textrm{F}}}\xspace}

\newcommand{\Fradx}{\ensuremath{{F_{\textrm{rad},\,x}}}\xspace}  %
\newcommand{\Fradth}{\ensuremath{{F_{\textrm{rad},\,\theta}}}\xspace}  %

\newcommand{\Fkinx}{\ensuremath{{F_{\textrm{kin},\,x}}}\xspace}  %
\newcommand{\Fkinr}{\ensuremath{{F_{\textrm{kin},\,r}}}\xspace}  %

\newcommand{\fred}{\ensuremath{{f_{\textrm{red}}}}\xspace}       %
\newcommand{\fredr}{\ensuremath{{f_{\textrm{red},\,r}}}\xspace}  %
\newcommand{\fredth}{\ensuremath{{f_{\textrm{red},\,\theta}}}\xspace}  %

\newcommand{\TAkk}{\Tacc}   %

\newcommand{\etakin}{\ensuremath{\eta^{\rm kin}}\xspace}

\newcommand{\SigE}{\ensuremath{\textrm{g}\,\textrm{cm}^{-2}}\xspace}
\newcommand{\rhoE}{\ensuremath{\textrm{g}\,\textrm{cm}^{-3}}\xspace}
\newcommand{\kapE}{\ensuremath{\textrm{cm}^2\,\textrm{g}^{-1}}\xspace}
\def\kms{\ensuremath{\textrm{km}\,\textrm{s}^{-1}}\xspace}    %
\def\MPktEJ{\ensuremath{\MJ\,\textrm{yr}^{-1}}\xspace}        %
\renewcommand{\arraystretch}{1.4}

\usepackage{booktabs}  %

\begin{document}

\title[The Planetary Accretion Shock. III.]{%
The Planetary Accretion Shock. III.\\Smoothing-free 2.5D simulations and calculation of H $\alpha$ emission}

\author[0000-0002-2919-7500]{Gabriel-Dominique Marleau}
\affiliation{%
Institut f\"ur Astronomie und Astrophysik,
Universit\"at T\"ubingen,
Auf der Morgenstelle 10,
D-72076 T\"ubingen, Germany
}
\affiliation{
Fakult\"at f\"ur Physik,
Universit\"at Duisburg-Essen,
Lotharstra\ss{}e 1,
D-47057 Duisburg, Germany
}
\affiliation{%
Physikalisches Institut,
Universit\"{a}t Bern,
Gesellschaftsstr.~6,
CH-3012 Bern, Switzerland}
\affiliation{%
Max-Planck-Institut f\"ur Astronomie,
K\"onigstuhl 17,
D-69117 Heidelberg, Germany
}

\author[0000-0003-2309-8963]{Rolf Kuiper}
\affiliation{
Fakult\"at f\"ur Physik,
Universit\"at Duisburg-Essen,
Lotharstra\ss{}e 1,
D-47057 Duisburg, Germany
}

\author{William B\'ethune}
\affiliation{%
DAAA,
ONERA,
Universit\'e Paris Saclay,
F-92322 Ch\^atillon,
France
}

\author[0000-0002-1013-2811]{Christoph Mordasini}
\affiliation{%
Physikalisches Institut,
Universit\"{a}t Bern,
Gesellschaftsstr.~6,
CH-3012 Bern, Switzerland}

\begin{abstract}
Surveys have looked for \Ha emission from accreting gas giants but found very few objects. Analyses of the detections and non-detections have assumed that the entire gas flow feeding the planet is in radial free-fall. However, hydrodynamical simulations suggest that this is far from reality. We calculate the \Ha emission from multidimensional accretion onto a gas giant, following the gas flow from Hill-sphere scales down to the circumplanetary disc (CPD) and the planetary surface. We perform  azimuthally-symmetric radiation-hydrodynamics simulations around the planet and use modern tabulated gas and dust opacities. Crucially, contrasting with most previous simulations, we do not smooth the gravitational potential and do follow the flow down to the planetary surface, where grid cells are 0.01~Jupiter radii small radially. We find that only roughly one percent of the net gas inflow into the Hill sphere reaches directly the planet. As expected for ballistic infall trajectories, most of the gas falls at too large a distance on the CPD to generate \Ha. Including radiation transport removes the high-velocity sub-surface flow previously seen in hydrodynamics-only simulations, so that only the free planet surface and the inner regions of the CPD emit substantially \Ha. Unless magnetospheric accretion, which we neglect here, additionally produces \Ha, the corresponding \Ha production efficiency is much smaller than usually assumed, which needs to be taken into account when analysing (non-)detection statistics.
\end{abstract}

\keywords{Accretion --- line emission --- gas giant formation --- radiation-hydrodynamics}

\section{Introduction}
 \label{sec:intro}

\refstepcounter{numKommG}

Over the last roughly three percent of a millennium, hundreds of extrasolar super-Jupiters have been discovered \citep{zhudong21}. However, only very few of those objects are young ($\lesssim100$~Myr). These are accessible almost only at large separations from their host star, through direct imaging ($\sim10$--$100$~au; \citealp[e.g.,][]{wagner19,vigan21}). A few more are predicted to be on the verge of being discovered \citep{asensiotorres21} but this would not change the overall abundance qualitatively. Despite dedicated surveys \citep{Cugno+2019,Zurlo+2020,xie20,follette23,hu22}, fewer gas giants yet have been caught accreting.
It is a robust theoretical prediction that gas undergoing a shock at velocity $v_0$ above a critical value, $v_0>\vHakrit\approx30~\kms$, will emit hydrogen lines as the hydrogen ionised by the shock recombines and cools in the geometrically thin postshock region \citep{aoyama18,Aoyama+2020}.
There are only a few planetary-mass companions with observed line emission clearly linked to a shock: \PDSb and~c \citep{wagner18,Haffert+2019} at \Ha, and \Dlrmb at several lines \citep{eriksson20,betti22b,betti22c,ringqvist23}. However, only the \PDS planets are found in a gas disc, and the others are effectively isolated. At newly-discovered AB~Aur b, point-like \Ha emission is observed but scattering of stellar photons cannot be excluded as the source \citep{currie22,zhou22}.

Different factors can explain the scarcity of \Ha-emitting accreting planets detected at large separations. Most planets are possibly forming closer in to their star, as classically expected from core accretion (e.g., \citealp{thommes08}; see also review in \citealp{emsen21a}) and thus inside the inner working angle (IWA) of current-generation detectors \citep{close20}. Planets could also be accreting episodically, with only brief and therefore unlikely-to-be-caught periods of detectably high accretion \citep[e.g.,][]{brittain20}. Another possibility is that the \Ha and the other hydrogen lines are absorbed by the protoplanetary disc (PPD). Since massive planets open a deep gap in \neuV{the gas and dust distributions}, this might not be an important effect. As an example, scaling the simulations of \citet{Sanchis+2020} to the \PDS system, extinction by the PPD is negligible for their specific accretion rate (see discussion in Section~7.6 of \citealt{maea21}). Alternatively, the gas and dust flowing onto the planet could absorb the \Ha, but \citet{maea21} estimated this not to be important for most planet accretion rates and masses.

Two more factors are of particular importance yet underappreciated. One is that line emission in the planetary-mass regime, especially in the absence of magnetospheric accretion, is intrinsically weaker than for stars for a given mass infall rate \citep{AMIM21L}.

The second factor is that most of the infalling and supersonic gas likely does not reach the planet directly but rather shocks onto a circumplanetary disc (CPD) at a significant fraction of the Hill radius \RHill away from the planet. This consequence of angular momentum conservation \neuIII{holds for matter inflowing due only to the action of gravity (i.e., ballistically; \citealp{mendoza09}). In the planet formation context, it} was pointed out by \citet{tanigawa12}, who found in their isothermal hydrodynamical simulations that the gas hitting the CPD around a Jupiter-mass planet was spread uniformly over a region of size $\sim0.1~\RHill\sim100~\RJ$. This shock should heat up the CPD and possibly make it detectable in the near infrared \citep{szul19II}. However, for typical planet masses the velocity of the gas at the shock $v_0\sim1/\sqrt{r}$ (see Equation~(\ref{eq:vFfinfty}) below) is too low at large distances for significant line emission from the shock (i.e., it has $v_0<\vHakrit$). This implies that only the small fraction of the mass inflow that hits the planetary surface and the innermost region of the CPD would be responsible for line emission. These regions were however not resolved in their simulations. Similarly, other recent work addressing \Ha generation is limited in different ways, as we review in Section~\ref{sec:prevwork}.

In this work, we study mass infall onto the planetary surface and the CPD, including its innermost regions.
To enable an appropriately high resolution $\Delta r\ll\RJ$,
we make the compromise of a simplified dimensionality but otherwise use methods matching or improving previous studies. We then predict the \Ha emission using detailed shock emission models designed for the planetary regime.

This paper is structured as follows.
Section~\ref{sec:prevwork} summarises related studies in the literature and Section~\ref{sec:mod} presents our physical model and numerical methods.
In Section~\ref{sec:res} we discuss the properties of the flow, estimate the amount of generated \Ha, and assess the effect of varying
\neuII{some input parameters.}
In Section~\ref{sec:disc}, we compare our results to previous work before presenting a more general discussion.
Finally, in Section~\ref{sec:summconc} we summarise our findings and conclude.

\section{Previous studies of accreting planets}
 \label{sec:prevwork}

\begin{deluxetable*}{c l c r c c r h}  %
\tablecaption{Comparison of some \neuII{(radiation-)}hydrodynamical studies of accreting gas giants.\label{tab:sammenligning/jevnfoering}}
\tablehead{
\colhead{\Ha\tablenotemark{{\footnotesize a}}} &
\multicolumn2c{Study  %
 and Dimensionality}
& \colhead{Domain size\tablenotemark{{\footnotesize b}}} & \colhead{
\neuIII{Thermodynamics}\tablenotemark{{\footnotesize c}}
} & \colhead{Smoothing\tablenotemark{{\footnotesize d}} \epsPot}  %
& \colhead{Resolution\tablenotemark{{\footnotesize d}} $\Delta r$}  %
& \nocolhead{Thermal mass \qth}
}
\startdata
 & \citet{machida08}       & \textbf{3D} 
   & $(30,~120,~8)~\HP$  &  isothermal   &  0.0060~\HP  &  0.0070~\HP  &  3  \\
 & \citet{tanigawa12}       & \textbf{3D} 
   & $(24,~24,~6)~\HP$  &  isothermal   &  0.0007~\HP  &  0.0004~\HP  &  3  \\
 & \citet{b19b}             & \textbf{3D} 
   & 128~\RP             &  isothermal   & \textbf{none}        &
       \textbf{0.1~\RJfett}  &  4 \\
 & \citet{fung19}           & \textbf{3D} 
   & \textbf{global}             &  isothermal\tablenotemark{{\footnotesize $\dagger$}} 
   & 0.05~\HP        &  0.02~\RH  &  4 \\  %
\Hastjern & \citet{szul20} & \textbf{3D}   %
   & \textbf{global}              &  \textbf{FLD}          & 20~\RJ       &  0.9~\RJ  &  \textbf{8?} \\
 & \citet{dong21} & 2.5D 
   & 10~\RP              &  isothermal    & \textbf{none}        &  \textbf{0.001~\RPfett}  &  \textbf{?} \\
\Hastjern & \citet{takasao21} & 2.5D 
   & $0.03~\RHill\approx100~\RP$       & $\gamma=1.01,\,1.05$\tablenotemark{{\footnotesize $\ddagger$}} & \textbf{none}        &  \textbf{0.005~\RPfett}  &  2~\MJ\tablenotemark{a} \\
  &  \citet{maeda22} &  \textbf{3D}
   &  $(24,~24,~6)~\HP$    &  isothermal      & 0.0002~\HP           &  0.0004~\HP  &  7.5\\  %
\Hastjern & This work & 2.5D 
   & $1~\RHill\approx4100~\RJ$ & \textbf{FLD}         & \textbf{none}        &
      \textbf{0.001~\RJfett}  &  7 \\
\enddata
\tablecomments{%
\neuV{Particularly commendable settings are highlighted in bold.}
All simulations reported here are for gas giants and/or have $\qth\gtrsim2$ \neuII{(see
Section~\ref{sec:freepar} for this parameter and others mentioned here)}. %
\citet{dong21} is included for comparison even though they do not allow infall from the outer edge.
We take the case 
$\qth=4$ of \citet{b19b} and \citet{fung19},
$\MP=3~\MJ$ of \citet{szul20},
\neuII{and $\RHill/\HP=1.36$ ($\qth=7.5$) of \citet{maeda22}}
because they are closest to our fiducial values (Table~\ref{tab:par}).
\tablenotetext{a}{
An \Ha-coloured asterisk (\Hastjern) marks studies predicting \Ha emission (through very different approaches).
}
\tablenotetext{b}{
\neuII{Three values: size in $(x,~y,~z)$ in Cartesian coordinates; one value: radial extent in polar/spherical coordinates.}
``Global'' \neuII{refers to work} simulating at least a significant radial ring of the PPD.
}
\tablenotetext{c}{
\neuIII{Simulations with radiation transfer use flux-limited diffusion (FLD) and have $\gamma=1.43$.}
}
\tablenotetext{d}{
Smallest smoothing length of the gravitational potential and smallest radial cell size if they are not constant or, in nested-grid simulations, if they depend on the grid level.
}
\tablenotetext{\dagger}{They also perform adiabatic simulations for comparison \neuII{but this is likely far from reality}.}
\tablenotetext{\ddagger}{\neuIV{``Nearly isothermal'' would not be an appropriate term because they obtain an extremely hot postshock region (Section~\ref{sec:vglHa}).}}
}
\end{deluxetable*}

Previous studies have looked at the accretion flow toward a forming planet. Pioneering work was presented by \citet{ab09a,ab09b,ayliffe12}. However, most investigations \citep[e.g., those and][]{cimerman17,kurokawa18,b19b,schulik19,schulik20,mai20,bailey21,mold21,krapp21,mold22} were concerned with low planet masses (super-Earths to at most $1~\MJ$), for which no \Ha from an accretion shock can be expected because the infall velocity is too low (see Equation~(\ref{eq:vFfinfty}); \citealp{aoyama18}). The few works exploring accretion onto gas giants have limited spatial resolution ($\Delta r\sim\RJ$) and, most problematically, a sizeable gravitational potential smoothing length  $\epsPot\sim10~\RJ$ or larger, which even modern computational resources impose \citep[e.g.,][]{machida08,tanigawa12,szul17b,lambrechts19,fung19,szul20}.
Table~\ref{tab:sammenligning/jevnfoering} provides an overview.
This significant $\epsPot$ affects the flow already on scales of a few times $\epsPot$ (i.e, out to roughly $r\sim30~\RJ$ to~100~\RJ or more) by weakening the effective mass of the planet. This could thus change qualitatively the flow pattern 
at the length scales that set the shock velocity,
which is a sensitive factor for the \Ha emission.
In Table~\ref{tab:sammenligning/jevnfoering}, studies predicting \Ha emission are highlighted by \neuV{an asterisk} (\Hastjern).

A notable exception to the issue of a large smoothing of the gravitational potential is the work of \citet{takasao21}, who consider the full potential ($\epsPot=0$) of their 12-\MJ planet.
Their cell size $\Delta r\sim0.01~\RJ$ close to the planet is also adequate to resolve the flow details.  %
However, they \neuIV{do not include radiation transfer and} \neuIII{adopt}  %
a heat capacity ratio $\gamma=1.01$ or $\gamma=1.05$.
\neuII{We will compare with their work in Section~\ref{sec:vglHa} and find crucial differences \neuIV{in the post-shock structures and thus} in the emission of \Ha.}
Therefore, including radiation transfer while setting $\epsPot=0$ seems desirable for more realistic simulations.  %
In this work, we complement the studies in the literature by considering the full, non-smoothed potential with a high spatial resolution while including radiation transport.
As a compromise,
instead of resolving the 3D structure of the flow, we assume axisymmetry around the planet.
We describe our model in detail in the next section.

\section{Physical model and numerical methods}
 \label{sec:mod}

\subsection{Approach: Local simulations in 2.5D}
 \label{sec:phil}

We consider a super-Jupiter forming by runaway accretion in a PPD and study the flow around the planet from Hill-sphere scales down to length scales much smaller than a Jupiter radius.
The dynamical timescales near the planet are much shorter than the dynamical or even viscous timescales of the PPD. Therefore, we do not evolve the background PPD and take it as a fixed boundary condition, assuming a circular orbit for the planet.
We
expect the flow around the planet to reach a quasi-steady state over %
a few free-fall times from the Hill sphere down to the planetary surface. While this is not a true steady state because of accretion, this state may represent an instantaneous snapshot in the accretion history of the planet.
We also assume that a quasi-steady state is reached much faster than the mass or accretion rate of the planet change while it forms. %

Initially, the domain contains a negligible amount of mass and thermal energy (see Section~\ref{sec:timeevol}). However, there is no well-defined final amount of mass and thermal energy because the quasi-steady state is an essentially constant accretion flow ($d\MPkt/dt\sim0$), not the absence of accretion ($\MPkt\sim0$). Consequently, we do not attempt to predict quantities such as the CPD thickness, size, or temperature, or the interior luminosity of the accreting planet. They will likely depend on the simulation history, which is not guaranteed to be equivalent to a global calculation of planet formation in an evolving PPD.  %
Instead, quantities such as the CPD thickness or planet luminosity should be seen as independent parameters %
that can be measured in the simulation results. They change slowly in the quasi-steady state.

The physical domain of our simulations extends to the Hill sphere. This is a compromise between, on the one hand, simulating the whole PPD structure, which would be computationally expensive but also sensitive to several poorly-constrained modelling choices (e.g., strength and spatial dependence of viscosity, presence of disc winds, dust grain size evolution and feedback on the gas), and, on the other hand, considering a region only several planetary radii large,
which would let the chosen boundary conditions at that location determine too strongly the gas flow.
Our assumption of axisymmetry around the planet will break down significantly further away than the Hill sphere, where the star's gravity dominates. Therefore, simulating out to \RHill is a natural choice.
This paper extends to a more realistic geometry our previous work in this series (\citealp{m16Schock,m18Schock}; hereafter \citetalias{m16Schock} and \citetalias{m18Schock}), in which we simulated and analysed the properties of the accretion flow and the accretion shock with highly-resolved 1D models for purely radial infall. The emphasis in the present work is
on the flow geometry when including angular momentum conservation.

\subsection{Numerical methods}
 \label{sec:nummeth}

To solve the radiation-hydrodynamics, we use the open-source (magneto)hydrodynamics code \texttt{Pluto} \citep{mignone07,mignone12} 
in combination with the
radiation transport package \texttt{Makemake} \citep{kuiper20}.
The Courant--Friedrichs--Lewy number is \neuIII{kept at %
$\CFL=0.4$, except for the simulation of Section~\ref{sec:varmass}, which uses $\CFL=0.33$.}
We use the non-equilibrium (two-temperature) flux-limited diffusion (FLD)  module \citep{kuiper10,kuiper20} as in \citetalias{m16Schock} and \citetalias{m18Schock}.
As argued in \citet{tanigawa12}, we do not include viscosity since we study the supersonic flow, not the CPD structure.
Also, self-gravity is negligible.
\texttt{Makemake} has been extensively tested and used in a variety of contexts (see \citealt{kuiper20} and references thereto). %

We do not set floor values on quantities such as the density or pressure. The only exception is a minimum on the radiation temperature $\Trad=(\Erad/\aST)^{1/4}\sim10^{-10}$~K (where \aST is the radiation constant)
during the numerical iterations to solve the FLD equation system
to prevent $\Erad\leqslant0$~K from ever being reached.
The smallest temperatures in the converged profiles are however much higher ($T\gtrsim10$~K), as expected.

\subsubsection{Coordinate system, domain size, forces, and resolution}

We consider a simulation region in the $r$--$\theta$ (poloidal, or vertical) plane centred on the planet and averaged over $\phi$, where $r$, $\theta$, and $\phi$ are the usual radial, spherical polar (co-latitudinal), and azimuthal coordinates. We simulate the upper hemisphere %
and assume symmetry at the midplane ($\theta=90$\degr).
The radial grid ranges from $\rmin=1.9~\RJ$ to \rmax, with \rmax a near-unity factor of \RH. Our fiducial simulation has $\rmax=\RH$.
With $\rmin=1.9~\RJ$, the shock that defines the planet surface is \neuII{usually} at $\RP\approx2~\RJ$, changing only slowly \neuII{during the simulation}. This commonly-used size is \neuII{thought to be appropriate}  %
for forming or young planets \citep{marl07,morda12_II,zhu15}.

We include the gravity of the planet by adding a radial acceleration $g=-G\MP/r^2$ everywhere, where \MP is the constant mass of the planet. \neuIII{Thus we avoid any smoothing and set $\epsPot=0$.} Since the mass in the whole simulation domain
is always negligible, no self-gravity is needed.
We do not include the vertical component of the star's gravity directly in the simulation. Above the planet, it would dominate over the planet's gravity for $z\gtrsim3^{(1/3)}\RHill$, making the star's gravity only a small correction to the dynamics in the region we simulate. However, we do impose the appropriate density stratification at the outer edge of the domain, \neuIII{which is sufficient} (see Equation~(\ref{eq:rho-Schichtung}) below). 

We follow the azimuthal component of the velocity $v_\phi$ despite the azimuthal symmetry, making the simulation 2.5D.
The simulations include the Coriolis and centrifugal terms due to the planet's Keplerian orbit around the star.
This is done in a linearised, conservative form \citep{kley98} by enabling the \texttt{ROTATING\_FRAME} option of \texttt{Pluto}.

We use a \neuII{very} fine radial gridding close to the inner edge with $\Delta r=10^{-3}~\RJ$, where the atmosphere is in rotation-modified hydrostatic equilibrium. The cell size increases smoothly outwards. Near the shock that terminates the atmosphere and defines the radius of the planet, cells have $\Delta r\sim10^{-2}~\RJ$. Beyond $r\approx2.5~\RJ$, the cell size increases logarithmically with 76~cells per decade, reaching $\Delta r\approx100~\RJ$ at \RHill. Appendix~\ref{sec:Gitter} gives further details.

The standard polar grid is uniform with $\Nth=181$~cells from pole to equator ($\Delta \theta\approx0.5$\degr). 
Simulations with $\Nth=91$ and even $\Nth=51$ ($\Delta \theta\approx1.8$\degr) yielded the same results overall. The only difference is that only in the middle- and high-resolution simulations is a thin supersonic flow beneath the surface of the CPD visible.
We discuss this in 
Appendix~\ref{sec:thres}.

\subsubsection{Boundary conditions}
 \label{sec:bc}

The radial boundary conditions are described in the next two sections.
In the polar ($\theta$) direction, we use reflective and equatorially symmetric boundary conditions at the pole and midplane, respectively.

\paragraph{Boundary conditions at the outer edge}

We assume the surrounding PPD
to be vertically isothermal and in hydrostatic equilibrium.
\neuIII{We do not include the stellar potential, but we fix} the density at the outer edge to  %
\begin{equation}
 \label{eq:rho-Schichtung}
  \rho(\rmax,\theta)=\rhoMitt \exp\left[-0.5\left(\frac{z}{\HP}\right)^2\right],
\end{equation}
where $z=r\cos\theta$ is the height above the midplane, and $\rhoMitt=\Sigma/(\sqrt{2\pi}\HP)$ is the midplane volume density in the gap with constant surface density $\Sigma$ (see below Expression~(\ref{eq:par})). Equation~(\ref{eq:rho-Schichtung}) thus correctly mimics the influence of the central star.

The poloidal components of the velocity are set as follows at \rmax.
We set $v_\theta=0$.
The radial velocity has $dv_r/dr=0$ to allow both inflow ($v_r<0$) and outflow ($v_r>0$) but it is limited in magnitude on the negative side to the freefall velocity from infinity,
\begin{equation}
 \label{eq:vFfinfty}
  \vFfinfty(r) = \sqrt{\frac{2G\MP}{r}} 
  = 60~\kms\sqrt{\frac{\MP/2~\MJ}{r/2~\RJ}}.
\end{equation}

Locally in the rotating frame, the flow of the PPD reduces to a simple linear shear \citep{hill78,gold65}, which we average along $\phi$ to obtain the azimuthal component of the velocity. %
Taking $x$ to point away from the star along the star--planet direction and $y$ along the orbit of the planet, the shear is $v_y = -q\Omega_0 x$, where the Keplerian orbital angular frequency of the planet is $\Omega_0=\sqrt{G\MStern/a^3}$ for a semi-major axis $a$, and $q=3/2$ for a Keplerian potential. Since $x=R\sin\phi$, with $R$ the cylindrical radius, we have $v_\phi=-q\Omega_0 R\sin^2\phi$, with a $\phi$ average $\langle v_\phi\rangle=-\frac{1}{2}q\Omega_0R$.
We therefore set
\begin{equation}
 \label{eq:vphirmax}
 v_\phi(\rmax)=\langle v_\phi\rangle = -\frac{3}{4}\Omega_0\rmax\sin\theta.
\end{equation}
This is an approximation since a shear flow is an exact description of the motion of the gas only close to the planet while neglecting its presence at the same time.

Averaging the same way the radial component gives $\langle v_r\rangle=0$, which we obviously do not use for $v_r$ because it would prevent accretion into the domain. In reality, due to the planet's gravity there is no pure shear flow but rather complex horseshoe orbits with ``U-turns'' and other features that can be captured only in 3D \citep[e.g.,][]{tanigawa12,schulik20}. Thus, setting $dv_r/dr=0$ as detailed above is a simple attempt to circumvent the limitation of a formally-averaged 2.5D approach.

Finally, we take a zero-gradient boundary condition for the gas pressure ($dP/dr=0$), which however is unimportant because the gas is supersonic for $\rmax<\RB$, which will hold for our cases of interest.
We set $d(r^2\Erad)/dr=0$ for the radiation energy density \Erad; if the radiation is free-streaming at \rmax (as it does turn out to be), this corresponds to a zero-gradient condition on the luminosity \citepalias{m16Schock}.

\paragraph{Boundary conditions at the inner edge}

Young planets have been observed to spin at \neuIII{5\,\%\ to 20\,\%\ of} their break-up frequency \citep[e.g.,][]{bryan18,bryan20}. Therefore we let the planet rotate at \rmin as a solid body
by setting
$v_\phi(\theta) = \fomg\omgk \rmin\sin\theta$, where the critical or break-up frequency is given by (see e.g.\ Section~4 of \citealt{paxton19})
\begin{equation}
 \label{eq:omgk}
    \omgk \approx \sqrt{\frac{G\MP}{\RP^3}}, %
\end{equation}
with the normalised spin \fomg set to~0.1.
\neuII{Planets spinning at near-break-up rates ($\fomg\approx0.8$) might shed mass more than accrete \citep{dong21,fu23} but smaller values should not influence significantly the transport of either mass or momentum in the CPD, and certainly not in the supersonic part of the flow.} \neuIII{Therefore, we do not vary \fomg.}

As in \citetalias{m16Schock} and \citetalias{m18Schock}, the inner edge is closed, without flow of matter.
Therefore, we set $d\rho/dr=0$ at \rmin and \neuII{use a reflecting condition on the radial velocity: $v_r(\rmin^-)=-v_r(\rmin^+)$, where $\rmin^{+\,(-)}$ is above (below) \rmin}. This lets an atmosphere in equilibrium build up beneath the settling zone.
We use a no-slip boundary condition: $v_\theta=0$.
The pressure is determined by hydrostatic equilibrium in the presence of rotation:
\begin{equation}
 \label{eq:dPdrj}
 \frac{dP}{dr} = -\rho\left( g - \frac{v_\phi^2}r\right),
\end{equation}
where $g=-G\MP/r^2$ (see above). 
At the inner edge, \neuIII{we choose a small luminosity $L(\rmin)=10^{-7}~\LSonne$ and 
impose accordingly across the interface 
$d\Erad/dr = L/(4\pi r^2)/ \DDiff$,
where $\DDiff=\lambda c/(\kapR \rho)$ is the local diffusion coefficient, with $\lambda$ the flux limiter (see details in \citetalias{m16Schock}) and \kapR the Rosseland mean opacity.
The luminosity}
increases outwards due to the compression of the accreting gas \citepalias{m18Schock}.

\subsubsection{Microphysics}

As in our previous work,
Rosseland- and Planck-mean opacities are taken from \citet{malygin14} for the gas, and from \citet[][model \texttt{nrm\_h\_s}]{semenov03} for the dust, with the dust sublimation as in \citet{isella05}.
The maximal dust-to-gas mass ratio is $\fpg=10^{-4}$. This reduction with respect to the ISM value reflects the fact that the ``pressure bump'' induced by the planet in the PPD could keep out opacity-carrying grains from the gap (\citealp{dr19,chachan21,karlin23}; but see \citealp{szul22}).
This is however uncertain
and \fpg could be varied in future work.

For the \neuIV{equation of state} (EOS), we use a perfect gas with a constant mean molecular weight $\mu=2.3$ and adiabatic index $\gamma=1.4$, appropriate for a solar mixture of H$_2$ and He with a hydrogen mass fraction $X=0.75$. In \citetalias{m16Schock} and \citetalias{m18Schock} we have shown that the choice of $\mu$ and $\gamma$ does not affect the hydrodynamic structure of the accretion flow (as seen also in \citealt{chenbai22}). The CPD properties, in particular its thickness, might be affected especially by $\mu$ but the accretion history is likely as important a factor. We recall that we do not wish to predict quantitative properties of the CPD here. Thus the choice of the EOS will not bear qualitatively on the results.
\subsubsection{Reaching quasi-steady state}
 \label{sec:timeevol}

Gas is free to flow into the simulation domain from the outer edge. For a steady state to be reached, at least a few global free-fall times
need to elapse. Evaluating at $r=\rmax$ the free-fall time from any radius $r$ to a much smaller position (e.g., \citealp{mungan09}) yields the global free-fall time
\begin{equation}
 \label{eq:tFfglob}
  \tFfglob = %
  \pi \sqrt{\frac{\rmax^3}{8 G\MP}}.
\end{equation}
By definition of free-fall, Equation~(\ref{eq:tFfglob}) 
ignores angular momentum. In reality, the latter will reduce the radial velocity of the gas and thus increase its fall time. For reference, if \rmax is always set to \RHill, the free-fall time becomes
$\tFfglob=(2\pi/\Omega_0)/\sqrt{96}$ %
or
one tenth of an orbital period.

We initialise the simulation with small density and temperature values \neuIII{that decrease outward independently of angle}, and set $\Trad=T$ (where $T$ is the gas temperature), $v_r=v_\theta=0$, and $v_\phi=-0.75\Omega_0R$ throughout. This state is quickly ``forgotten'' over a timescale comparable to \tFfglob as the gas begins to fall due to gravity of the planet.
The accreting gas accumulates in a CPD whose outer edge is defined by a radial shock and grows slowly over time. 

To speed up the computation, we use two phases. In Phase~I, while 
we let the large-scale and supersonic flow reach a quasi-steady state that erases the initial conditions,
we do not compute the hydrodynamics of the innermost region, 
whose early-time properties will be unimportant once the gas from \rmax reaches it.
We use \texttt{Pluto}'s \texttt{FLAG\_INTERNAL\_BOUNDARY} to make
inactive the cells between \rmin and 
a ``freeze radius'' 
set to $\rfrz=10~\RJ$.
Importantly, the inactive cells
are ignored when determining the hydrodynamics timestep\footnote{By default
  in \texttt{Pluto}, all cells were considered; we changed this.%
} $\Delta t$.
\neuIII{From the Courant condition, $\Delta t$ increases with cell size and decreases with temperature. Therefore, not having to take the innermost cells into account, which are the smallest (Figure~\ref{fig:Gitter}) and the hottest (Figure~\ref{fig:T2D}),
speeds up computations considerably.}
The radiation transport is always solved over the full domain, both in Phase~I and the subsequent Phase~II (described next).

\neuIII{After a few global free-fall times in Phase~I,}
we restart the simulation but now evolve the density and velocity everywhere as usual. This is Phase~II. After a brief transition period, no features remain at \rfrz and all quantities $(\rho,v,P,\Trad)$ are smooth.
We run Phase~II for %
thousands of free-fall times from \rfrz\xspace \neuIII{(numerical details are given in Section~\ref{sec:parval})}. By Equation~(\ref{eq:tFfglob}), the free-fall time at \rfrz\xspace \neuIII{down to ``$r=0$''} is roughly $\tFfrfrz=(\rfrz/\RHill)^{1.5}\tFfglob\approx\tFfglob/8000$.
Thus, over the course of Phase~II the inner regions are in quasi-steady state given the large-scale flow, while the large-scale flow cannot change appreciably since Phase~II lasts for $\lesssim\tFfglob$.

A feature of our set-up is that the CPD has to build up from the infalling gas since we do not put in any structure initially. %
The accreting gas naturally accumulates in a CPD whose outer edge is defined by a radial shock and increases over timescales of hundreds of \tFfglob. The thickness (the aspect ratio) of the CPD does not vary much while it grows.
\neuIII{The formation of the CPD causes a spherical shock to propagate outward through the infalling material, with part of shock at the position of the expanding outer edge of the CPD.}
Consequently, the flow pattern close to the planet is not quite in steady state.
We estimate in Appendix~\ref{sec:sigh-pattern} how much this affects our analysis and find 
\neuIII{that it should not change our conclusions}.
Therefore, for simplicity we will call the Phase~II state with a qualitatively constant flow pattern
a quasi-steady state, and analyse this, keeping in mind that over much longer timescales there are likely quantitative fluctuations.

\subsection{Free parameters}
 \label{sec:freepar}

\neuIII{One can parametrise}
the degrees of freedom of the problem
\neuIII{in different ways.}
To help bridge simulations and observations, we choose as independent parameters %
\begin{equation}
 \label{eq:par}
  \left(\MStern, a, \Sigma, h, \rmax, \MP, \RP \right),
\end{equation}
where, \neuIII{repeating some definitions,} \MStern is the stellar mass,
$a$ the semi-major axis of the planet, 
$\Sigma$ the PPD surface density at $a$ as reduced by gap opening,  %
$h$ the PPD aspect ratio at $a$,
\MP the planet mass, %
and \RP the physical radius of the planet.
Instead of setting $\Sigma$ directly, one could choose a value for the viscosity parameter $\alpha$ \citep{shak73} and an unperturbed surface density of the PPD $\Sigma_0$,
and, following \citet{kanagawa18a}, let
$\Sigma = \Sigma_0/(1+K/25)$, where $K=q^2/(\alpha h^5)$.

\neuIII{To be consistent with the approximation of symmetry around the planet, one should choose \rmax to be smaller than the width of the gap, which we do not model. This choice also predicts the reduced surface density to be constant across the gap \citep{kanagawa17}, which we assume when setting $\rho(\rmax)$ (Equation~(\ref{eq:rho-Schichtung})).}
\neuIII{Also, we introduced the parameter \rmax because we do not simulate the whole PPD. However, }%
\neuII{%
\rmax should be seen 
\neuIII{primarily} not as a numerical parameter but rather as a 
(simple) 
\neuIII{way of controlling}
the incoming angular momentum of the gas.
We vary \rmax in Section~\ref{sec:varrmax}.}

The other characteristic quantities follow from Expression~(\ref{eq:par}):  %
the planet--star mass ratio $q=\MP/\MStern$;
the Hill radius $\RH=a\left(\MP/[3\MStern]\right)^{1/3}$;
the Bondi radius $\RB=aq/h^2$;
the pressure scale height $\HP=a h$;
the Keplerian orbital angular frequency of the planet $\Omega_0$ (defined \neuII{above Equation~(\ref{eq:vphirmax})});  %
and in particular \qth,
the ratio between the planet mass and the ``disc thermal mass'' (for short, ``the thermal mass''; e.g., \citealp{korycansky96,machida08,fung19}):
\begin{equation}
 \label{eq:qth}
 \qth=q/h^3=\RB/\HP=3(\RH/\HP)^3.
\end{equation}
When $\RP\ll\min(\RB,\RH)$,
in the isothermal and inviscid limit, \neuIII{one may expect \qth to be the only parameter controlling} the flow in a local region around a planet on a Keplerian orbit (\citealp{korycansky96}; \neuIII{see however \citealp{b19a,b19b}}).
The radiation transfer %
introduces a physical scale through the temperature- and density-dependent opacities, but qualitatively \qth should be key in determining the flow.

One characteristic quantity emerges from our set-up: the net mass inflow rate into the Hill sphere \MPktnettoHill, that is, what flows in minus what flows out. \neuV{This in turn is set by more global PPD physics \citep[e.g.,][]{nelson23,choksi23}.} The growth rate of the planet cannot be controlled directly but is at most \MPktnettoHill; it is less if some of the large-scale flow feeds instead the CPD. We need to measure \MPktnettoHill from the simulation output because we only set the gradient of the radial velocity at \rmax (Section~\ref{sec:bc}), so that we do not know a priori how much mass will flow in or out as a function of angle.
However, the gas at \rmax will turn out to be in (inward) freefall at all angles.
Then, \MPktnettoHill is maximal and given by $\MPktnettoHill=\int 4\pi r^2\rho |\vFfinfty|\sin\theta\,d\theta$.

The flow patterns that we will obtain should not depend sensitively on our choice of $\Sigma$ and hence \MPktnettoHill.
This would hold exactly in pure-hydrodynamics simulations,
but here the optical depth introduces a length scale. However, in practice this is not an important effect since the radiative transfer and hence the thermodynamics in the free-fall flow do not depend strongly on the density, and even large variations in the Rosseland optical depth do not modify the flow, at least in 1D \citepalias{m16Schock}.

\subsection[Parameter values guided by PDS 70 b]{Parameter values guided by \PDSb}
 \label{sec:parval}

\begin{deluxetable}{llh}  %
\tablecaption{Chosen and derived fiducial parameters.\label{tab:par}}
\tablehead{
\colhead{Quantity} & \colhead{Symbol and Value}
}
\startdata
\multicolumn{2}{l}{\textit{Chosen parameters (Expression~(\ref{eq:par}))}} & \\  %
Stellar mass                & $\MStern = 0.9~\MSonne$ &   \\
Semi-major axis             & $a = 22$~au       &   \\
PPD surface density in gap  & $\Sigma = 0.021$~\SigE   & \\
PPD aspect ratio at $a$     & $h= 0.067$  &  \\
Outer radius of domain      & $\rmax  = 1\times\RH$  & \\
Planet mass                 & $\MP = 2~\MJ$  &  \\
Planet radius               & $\RP =2~\RJ$ & \\
\hline
\multicolumn{2}{l}{\textit{Derived parameters}}\\
Disc thermal mass    &   $\qth= 3(\RHill/\HP)^3 = 7.1$ &  \\  %
Hill radius     &   $\RH=4100~\RJ = 1.33\HP$ &  \\
Bondi radius    &   $\RB=  22$~k$\RJ = 10.4$~au  & \\
Orbital period  &   $2\pi/\Omega_0 = 3.4\times10^9$~s  & \\
Free-fall time from $\rmax$ & $\tFfglob = 3.5\times10^8$~s & \\  %
Mass flux into Hill sphere  & $\MPktnettoHill = 6.9\times10^{-6}~\MPktEJ$ &  \\  %
Midplane density in gap     & $\rhoMitt = 3.8\times10^{-16}~\rhoE$    &  \\
Gap 50\,\%\ full width      & $\DeltaLuecke=7.1$~au  & \\
Free-fall velocity on planet & $\vFf(\RP)=59.5$~\kms & \\
\hline
\hline
\multicolumn{2}{l}{\textit{Other chosen parameters}} & \\
Planet normalised spin        & $\fomg = \omega/\omgk = 0.1$ & \\  %
Maximal dust fraction         & $\fpg =0.0001=0.01\times\textrm{ISM}$  &\\
\hline
\enddata
\tablecomments{%
To set $\Sigma$, we chose $\Sigma_0$ and $\alpha$ from \citet{bae19},
\neuIII{and the gap width follows from \citet{kanagawa17}} (see text).
}
\end{deluxetable}

We consider parameters that could be appropriate for \PDSb \citep{keppler18,bae19,toci20,wang21vlti}, without however attempting to match observational properties exactly.
We take $\MP=2~\MJ$, guided by the posterior distribution of \citet{wang21vlti} and other tentative indications of a low, few-\MJ mass \citep{bae19,Stolker+20b,uyama21}. A higher value is also possible and is considered in Section~\ref{sec:varmass}.
The surface density is set to $\Sigma=0.021~\SigE$ for the gas in the gap of the background PPD at $a=22$~au,
coming from $\Sigma_0=2.7~\SigE$ with $\alpha=10^{-3}$ following\footnote{\neuIII{How this fits with the proposed age of 8--10~Myr for the star \citep{rZerjal23} instead of 5~Myr \citep{mueller18}, should be re-assessed.}} \citet{bae19}.
\citet{bae19}, or \citet{toci20} with their $\alpha\approx0.005$, obtain surface densities closer to $\Sigma\approx10^{-3}~\SigE$, due also to their higher \MP.  %
We return to $\Sigma$ in Section~\ref{sec:discPDS}.
\neuII{Using the expressions of} \citet{kanagawa17}, the bottom of the gap, with a constant surface density, is $\DeltaLuecke=7.1$~au wide, centered on $a=22$~au.
Our assumption of a constant $\Sigma$ over the outer boundary of the simulation domain is thus justified, since $\RHill=4100~\RJ=2.0$~au is smaller than $\DeltaLuecke/2$.

In Table~\ref{tab:par}, we summarise our choices and the resulting relevant quantities including the PPD pressure scale height, free-fall time, and disc thermal mass of the planet. We are in the high-mass regime with $\qth=7.1\gg1$. \neuII{This high value of \qth is the same as in one of the simulations of \citet{maeda22}, who however use a very different set-up (Table~\ref{tab:sammenligning/jevnfoering}) and do not study the accretion close to and at the surface of the planet.}

\neuIII{For the fiducial case, we let Phase~I run for $7.0\times10^8$~s (2.0~\tFfglob) before switching to Phase~II. The snapshot used for the analysis was taken at $2.4\times10^8$~s after the beginning of Phase~II, which represents 0.7~\tFfglob but thousands of free-fall times from $\rfrz=10~\RJ=5~\RP$ to \RP.
As a check, we kept running separately the simulation of Phase~I up to $t\sim30\tFfglob$. As espected, the overall flow remained the same while the CPD grew in size and slightly in thickness. Therefore, the structure of the flow and of the CPD in Phase~II are representative of a possible steady state. A similar description---several \tFfglob for Phase~I, and snapshots taken at more than hundreds of free-fall times from \rfrz in Phase~II---applies qualitatively also to simulations varying \MP and \rmax (Sections~\ref{sec:varrmax} and~\ref{sec:varmass}).}

\section{Results}
 \label{sec:res}

Here, we present the flow of the gas from the Hill sphere down to the planet and CPD. In Section~\ref{sec:flow}, we analyse what fraction of the gas entering the Hill sphere reaches the planet directly and what fraction has sufficient velocity to generate \Ha. In Section~\ref{sec:Ha} we estimate the resulting \Ha luminosity and compare it to the assumption that the preshock velocity is \vFf.
In Section~\ref{sec:varrmax} we assess the effect of the 2.5D approximation by varying the angular momentum of the incoming gas, and in Section~\ref{sec:varmass} we consider a higher planet mass.

\subsection{Gas flow from the Hill sphere to the planet}
 \label{sec:flow}
\begin{figure*} %
 \centering
 \includegraphics[width=0.77\textwidth]{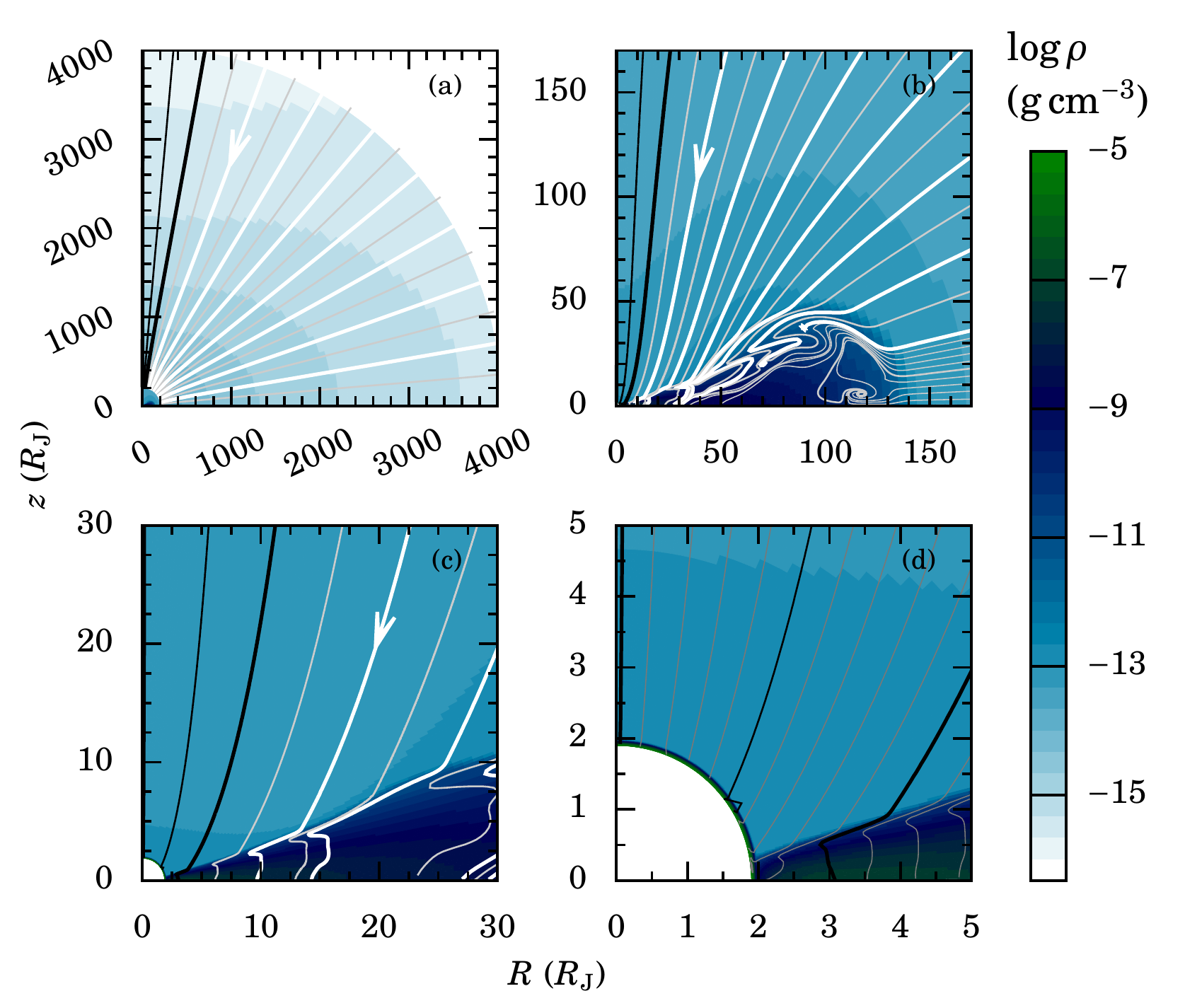}
\caption{%
Density structure (colour) and flow pattern (lines)
from \RHill to \RP scales. %
The streamlines start at \RHill at $\thi\equiv\theta(\rmax)=0.5$\degr and from 5\degr up to 85\degr in steps of 5\degr (panels~(a)--(d); thick: multiples of 10\degr) or also 1\degr (panels~(b) and~(d)).
Due to angular momentum conservation, the arrow-bearing streamline with $\thi=20$\degr, for example, \neuV{hits} the CPD with $\theta\gg20$\degr and not on the planet surface.
Only gas within $\thi\approx10$\degr shocks with sufficient velocity to generate \Ha ($\thi=0.5$, 5, 10\degr: black).
}
\label{fig:rho-Ueberblick}
\end{figure*}

Figure~\ref{fig:rho-Ueberblick} shows the large- and small-scale structure of the flow and the gas density. At the Hill sphere, the gas is in radial freefall at all latitudes, so that all the gas entering the Hill sphere will accrete onto \neuII{(that is, become part of)} the planet or CPD. \neuII{For comparison, in 3D a fraction of the flow would flow back out on perturbed horseshoe orbits \citep{machida08,lambrechts17,maeda22}. Therefore, our inflow rate corresponds to the \textit{net} inflow in 3D (see also Section~\ref{sec:freepar}).}  %
\neuIII{Everywhere outside of the planet and CPD (i.e., where the gas is infalling)},
\neuII{%
the radiative flux\footnote{\neuII{Since
   we perform radiation-hydrodynamical simulations, this automatically includes the interior fluxes from the planet and the CPD as well as the accretion luminosities (the kinetic energy transformed into radiation at the respective shocks).}}}
\neuIII{is almost purely radial (see Appendix~\ref{sec:T+L-Strukt}). Accordingly,}
the temperature %
is approximately constant along $\theta$ at a given radius.
\neuII{Then,}
the density stratification
leads to a \neuII{positive} pressure gradient: $dP/d\theta\propto d\rho/d\theta\propto -d\rho/dz>0$. In turn, the pressure gradient pushes the gas outside of $r\approx100~\RJ$ poleward. It does so by at most 5\degr\xspace \neuIII{compared to a radial trajectory},  %
and inside of $r\approx100~\RJ$, angular momentum conservation deflects the gas outwards.
This deviation from a radial trajectory becomes more important closer in to the planet. For example, the streamline that started at $\thi\equiv\theta(\rmax)=20$\degr at \RHill
joins the CPD at $\theta=73\degr\gg\thi$ (Figure~\ref{fig:rho-Ueberblick}c).
The gravitational potential energy of the gas serves to increase all three components of the velocity.

The key result seen in Figure~\ref{fig:rho-Ueberblick} is that
most streamlines reach the CPD at a large distance (hundreds of Jupiter radii) from the planet. Only a small fraction of the total mass influx from \RHill falls in close to the planet. 
This consequence of angular momentum conservation was seen by \citet{tanigawa12}
with a very different set-up \neuII{(Table~\ref{tab:sammenligning/jevnfoering})}.
\neuV{Independent work by Chen \&\ Bai (in prep.) also finds this.}
It has been derived analytically for ballistic (pressure-free) trajectories starting from an outer edge in solid-body rotation, in the context of star formation \citep{ulrich76,mendoza09}.
We have now obtained that most gas reaches the CPD far from the planet also when radiation transfer and thermal effects are included. This conclusion will be seen to hold for other parameter combinations.

Two partial \neuII{mass influx} rates are of interest,
\neuII{in particular for simulations that cannot resolve down to these scales.}
One is the %
gas falling directly onto the planetary surface:
\begin{equation}
\label{eq:MPktPdir}
 \MPktPdir = 4\pi \RP^2 \int \rho(\theta) v_r(\theta) \sin\theta\,d\theta,
\end{equation}
integrated at $r=\RP$ from the pole ($\theta=0$) down to the surface of the CPD where it connects to the planetary surface (in a so-called ``boundary layer''; e.g., \citealp{kley89a,hertfelder17}). %
This gives $\MPktPdir=3.0\times10^{-8}~\MPktEJ$ at $\RP=2~\RJ$, where the CPD is roughly 10\degr thick.
Since $\MPktnettoHill=6.9\times10^{-6}~\MPktEJ$ (Table~\ref{tab:par}), only $\MPktPdir/\MPktnettoHill=0.4$~percent
of the gas entering the Hill sphere reaches the planetary surface directly.
\neuIII{The expressions of \citet{ab22} predict a qualitatively similar result, with differences because they assumed at the Hill sphere a uniform density and solid-body rotation, and neglected pressure forces (cf.\ our Section~\ref{sec:flow}), as in \citet{mendoza09}.}  %

\begin{figure*}[t]
 \centering
 \includegraphics[width=0.49\textwidth]{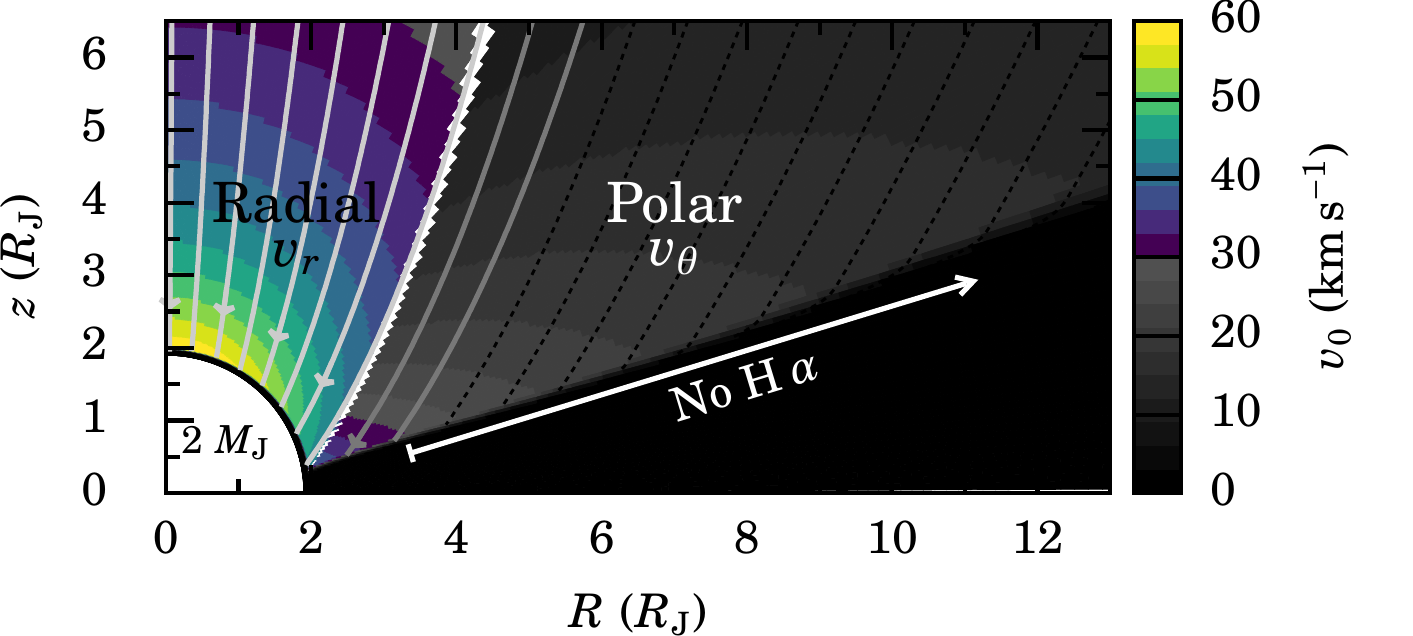}
 \includegraphics[width=0.49\textwidth]{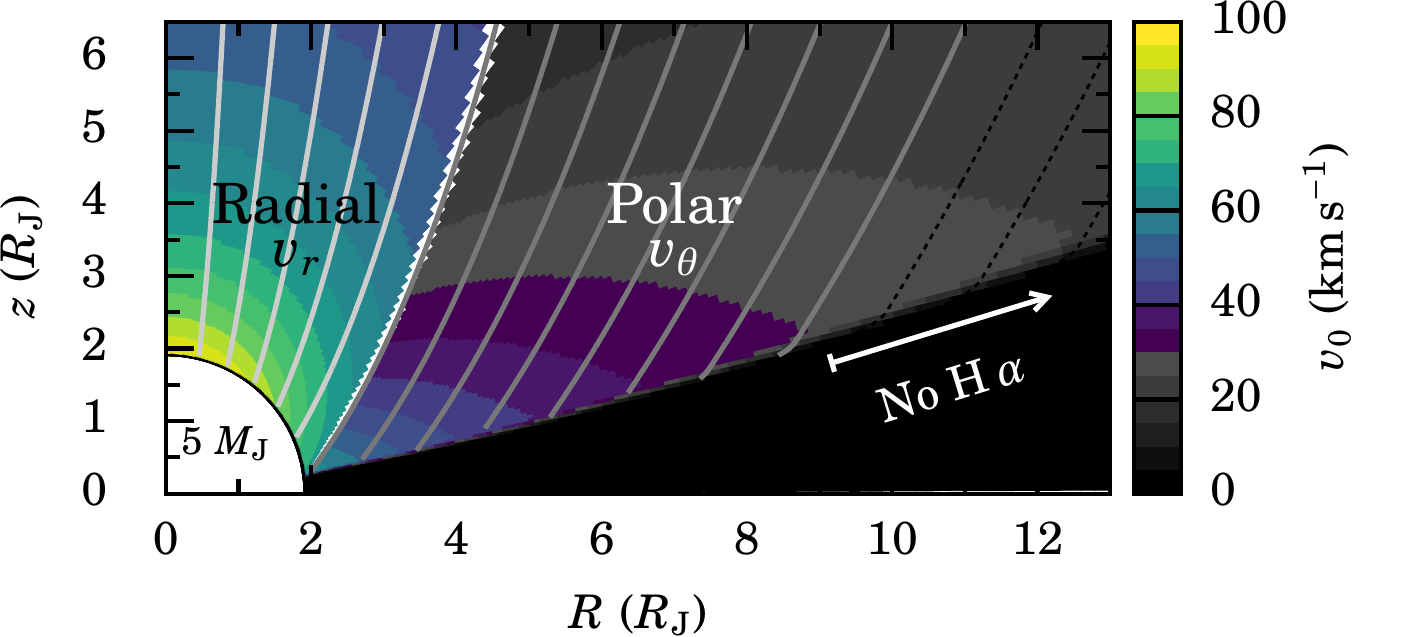}
\caption{%
Components of the velocity \neuII{(in \kms) perpendicular to each shock $v_0$}:
radial $v_r$ around streamlines that will hit the planet surface, and polar $v_\theta$ for the ones that will hit the CPD.
\Ha emission requires $v_0>\vHakrit\approx30~\kms$ (solid grey streamlines; \neuII{black dotted otherwise}).
Streamlines start at \RHill in steps of $\Delta\theta=1\degr$. %
\neuII{\textit{Left:}~Fiducial simulation ($\MP=2~\MJ$); \textit{right:}~\SimMPgroesser ($\MP=5~\MJ$). The two velocity scales differ but both have colours only above \vHakrit. In both cases, the preshock velocity over the whole free planetary surface is high enough for \Ha emission, but only a part of the CPD emits.}
}  %
\label{fig:vnormSchock}
\end{figure*}

The second partial \neuII{mass flow} rate, \MPktHa, measures the mass inflow able to generate \Ha, that is, the gas whose preshock velocity $v_0$ exceeds %
$\vHakrit$.
To determine \MPktHa, we look at the component of the velocity that is perpendicular to the planetary surface and to the CPD surface,
shown in Figure~\ref{fig:vnormSchock}.
For the shock at the surface of the planet, the normal component is the radial velocity $v_r$ because the planet surface is nearly spherical.
\neuII{Even for this small mass of $\MP=2~\MJ$, over the whole free surface of the planet, the gas is fast enough to generate \Ha (Equation~(\ref{eq:vFfinfty})).}
Out to at least $\approx10~\RJ$ for the CPD in our simulation, the CPD (shock) surface is \neuIII{nearly} radial and is flat,
such that the normal component is nearly equal to the polar velocity $v_\theta$ evaluated above the shock surface. For simplicity, we take $v_0=v_\theta$.
We identify the largest radius \rXXX on the CPD out to which $v_0\geqslant\vHakrit$
and integrate \neuII{the mass flux} at that radius from the pole down to the surface of the CPD:
\begin{equation}
\label{eq:MPktHa}
 \MPktHa = 4\pi \rXXX^2 \int \rho(\theta) v_r(\theta) \sin\theta\,d\theta.
\end{equation}
Due to 
\neuIII{time-independence}, this is equivalent to an integral over the shock surfaces.
We find a maximum radius $\rXXX=3.3~\RJ$
and a CPD height of 14\degr there.
This yields
$\MPktHa=4.7\times10^{-8}~\MPktEJ$,
or $\MPktHa/\MPktnettoHill=0.7$~percent.

Both \MPktPdir and \MPktHa are small fractions of \MPktnettoHill. Correspondingly, they originate from a narrow polar region, in which the specific angular momentum of the gas $j_z = Rv_\phi\propto r^2\sin^2\theta$ (Equation~(\ref{eq:vphirmax})) is low. Indeed, tracing the streamlines that define \MPktPdir and \MPktHa back to \rmax,
we find starting angles of $\thi\approx7$\degr and $\thi\approx9$\degr, as seen in Figure~\ref{fig:vnormSchock}.

Comparing \MPktHa and \MPktPdir, we see that
\neuII{$(1-\MPktPdir/\MPktHa)\approx60$~percent}
of the total \Ha-generating gas falls on the CPD surface and not on the planet.
\neuII{This assumes that all of \MPktPdir produces \Ha, which holds.}
However, the local \Ha flux \FHa depends strongly on $v_0$ (crudely, $\FHa\sim{v_0}^3$; \citealp{aoyama18}), so that it is not clear a priori whether the CPD or the planetary surface dominates the total emission. We look at this in more detail in the next section.

\subsection[Approximate H alpha emission]{Approximate \Ha emission}
  \label{sec:Ha}
\begin{figure}
 \centering
 \includegraphics[width=0.47\textwidth]{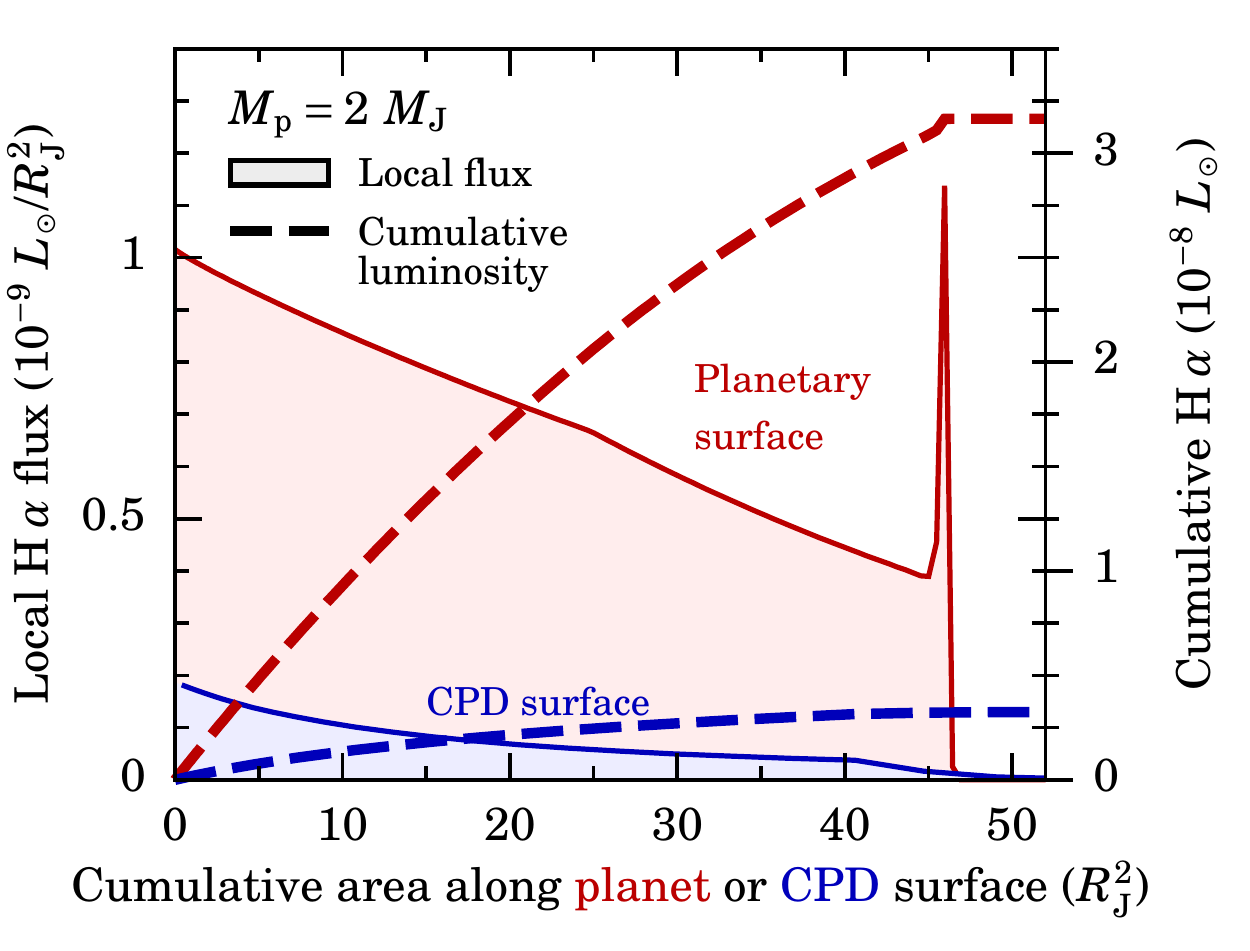}
\caption{%
Flux from the accretion shocks on the planet surface (solid red curve with a filled area) and on the CPD surface (blue) %
as a function of the cumulative area along the planetary surface, starting at the pole, or along the CPD surface, starting at the planet at $\RP=2~\RJ$ (Equation~(\ref{eq:cumulsurf})).
\neuII{With this $x$ axis, the area under each curve is proportional to its contribution to the luminosity.}
The spike comes from the high-velocity surface flow but barely contributes to the total flux,
\neuII{contrary to what \citet{takasao21} find (see Section~\ref{sec:vglHa})}.
Dashed curves show the cumulative integral of each contribution (right axis).
}
\label{fig:LHaPzpSch}
\end{figure}

We estimate the observable \Ha luminosity from the planet-surface and CPD-surface shocks.
Detailed 2D radiation transport of the generated \Ha is beyond the scope of this paper.
However, based on \citet{maea21} and for $\fpg=10^{-4}$, we expect the incoming gas and dust to be very optically thin to \Ha photons\footnote{\neuIII{However, the Planck mean opacity is high enough for the gas and radiation temperatures to be equal, except in the Zel'dovich spikes (Appendix~\ref{sec:T+L-Strukt}).}} for a \neuII{mass influx} rate $\MPktnettoHill\sim10^{-5}~\MPktEJ$ or even much higher.
Also, given that $\qth\gtrsim1$ planets open gaps, extinction by the PPD is \neuII{possibly} negligible.
Therefore, summing the local \Ha production along the planetary and CPD surfaces (the radiative source terms) gives a reasonable estimate of the luminosity leaving the system. 

We display in Figure~\ref{fig:LHaPzpSch} the \Ha flux per emitting area $\FHa$, which depends only on the local preshock density $\rho_0$ and preshock velocity $v_0$ above either shock.
We use the data of \citet{aoyama18} for $\FHa(\rho_0,v_0)$.
For the surface shock, the flux is plotted against the distance from the pole, and for the CPD shock, outwards from the surface of the planet, where the CPD begins. Instead of the linear distance, we use the respective cumulative areas (including both hemispheres):
\begin{subequations}
\label{eq:cumulsurf}
\begin{align}
  \APlObfl(\theta) &= \int_0^\theta 4\pi\RP^2 \sin\theta'\,d\theta',\\  %
  \AzpSch(r) &= \int_{\RP}^r 4\pi r'\sin\theta\,dr',\label{eq:AzpSch}
\end{align}
\end{subequations}
where we have assumed a $\theta=\mbox{constant}$ CPD surface in Equation~(\ref{eq:AzpSch}), which holds approximately for the region whose \Ha emission dominates (see Figure~\ref{fig:vnormSchock}).
For the actual analysis, we look for the temperature peak (the Zel'dovich spike) for each $r=\mbox{constant}$ ring in the $r$--$\theta$ plane, and use the cell above it to define the surface.
With Equation~(\ref{eq:cumulsurf}) to measure distance, the ratio of the areas under the curves gives the relative contribution of each shock.
Emission comes from the exposed planetary surface from the pole down to the surface of the CPD at \thmax, which corresponds to a filling factor $\ffill=1-\cos\thmax=0.83$,
and %
from the CPD from the planetary radius out to
$r\approx\rXXX\approx3.3~\RJ$ (as found in Section~\ref{sec:flow}). %
\neuII{The region without emission is labelled in Figure~\ref{fig:vnormSchock}.}

The integrated luminosities (dashed lines \neuII{in Figure~\ref{fig:LHaPzpSch}}) are
$\LHa=3.2\times10^{-8}~\LSonne$ for the planetary surface shock and
$\LHa=0.45\times10^{-8}~\LSonne$ for the contribution by the CPD.
Approximately, \neuIII{ignoring the angular dependence of the radiation by} summing the two terms, an observer looking at the system would see an \Ha luminosity $\LHa\approx3.7\times10^{-8}~\LSonne$.
We compare this to the observations of \PDSb in Section~\ref{sec:discPDS}.
The supersonic surface flow on the CPD, discussed in Appendix~\ref{sec:thres}, causes a local spike in the emission (filled region in Figure~\ref{fig:LHaPzpSch}). However, this thin layer
contributes negligibly to the integrated emission.
This contrasts strongly with the results of \citet{takasao21},
to which we return in Section~\ref{sec:vglHa}.
\refstepcounter{numKommG}

Figure~\ref{fig:LHaPzpSch} shows
that the planetary surface \neuII{shock} largely dominates the \Ha emission.
The preshock densities, $\rho_0\sim10^{-13}~\rhoE$, 
which depend only weakly on position, are similar to within 0.1~dex between both shocks,
and the emitting areas are similar \neuII{(near $45~\RJ^2$)}.
However, the difference in the preshock velocities
is much more \neuIV{consequential}.
\neuIV{The velocities are different in part because}
the emitting region of the CPD \neuIV{is at} slightly \neuIV{greater $r$ (Figure~\ref{fig:vnormSchock}: $r\approx3~\RJ$ and $r\approx(3$--10)~\RJ for the two simulations)} than \neuIV{the planet surface ($\RP=2~\RJ$).}
The other, and more important, factor is that for the CPD, the shock velocity \neuII{(i.e., the component orthogonal to the shock surface)} is \neuII{the polar velocity} $v_\theta$, and this is even smaller than \neuII{the local radial velocity} $v_r$.

We can compare the \Ha luminosity \neuII{from the free planetary surface to the luminosity} %
expected from purely radial accretion at \vFf for the same \MPktPdir and \ffill.
Explicitly, the combination $(\MPktPdir, \MP, \RP, \ffill)$ that we have here implies an average preshock number density
$n_0=X\MPktPdir/(4\pi\RP^2\ffill\vFf\mH) = 6.3\times10^{10}~\textrm{cm}^{-3}$ (Equation~(A9) of \citealt{Aoyama+2020})
and thus
$\LHa'=4\pi\RP^2\ffill\times\FHa(n_0,\vFf) = 4.7\times10^{-8}~\LSonne$, which is higher by 50\,\% than what we found.
The difference is due mainly to the strong dependence of $\FHa$ on $v_0$ with, roughly, $\FHa\propto v_0^3$ \citep{aoyama18}.
Indeed, in our simulation, the radial velocity at the pole is equal to the freefall value $\vFf=59.5$~\kms, but at the equator it is lower by about 30\,\%.
This is because the gas gains more velocity in $\theta$ and in $\phi$ further away from the pole, with the three components summing up to
$v_r^2+v_\theta^2+v_\phi^2 = \vFf^2$
everywhere in the free-falling region by conservation of energy. %
In other words, centrifugal forces due to angular-momentum conservation are slowing down the infalling gas at low latitudes.
This reduction of the preshock velocity leads to less emission ($\sim \rho_0v_0^3$) for the same total \neuII{mass flow} rate ($\sim \rho_0v_0$) compared to the simple assumption of radial freefall. %

\subsection{Varying the incoming angular momentum}
 \label{sec:varrmax}
\begin{figure*}
 \centering
 \includegraphics[width=0.77\textwidth]{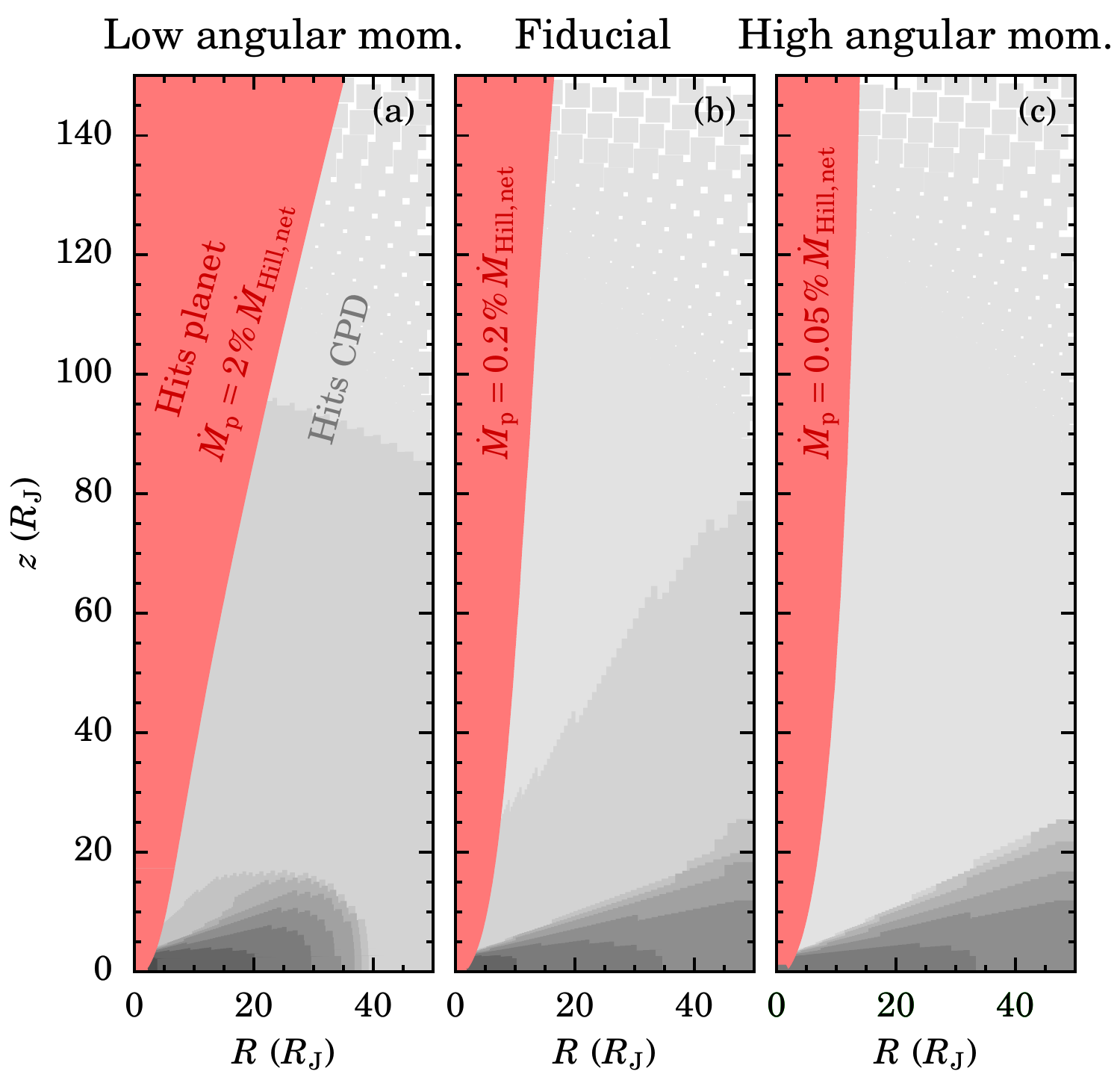}
\caption{%
Fraction of the total \neuII{inflow rate} reaching the planet surface for different amounts of angular momentum at the Hill sphere, probed by varying the position of the outer radius of the simulation domain: $\rmax/\RHill=0.7$, 1.0 (fiducial run), 1.3 \neuII{(left, middle, and right panel, respectively)}. In all cases, the gas flows inwards for all angles from \rmax down to the planet or CPD. The greyscale shows the density (logarithmic). \neuII{Each simulation extends to $\rmax=4100~\RJ$.}
}
\label{fig:varrmax}
\end{figure*}

The specific angular momentum of the gas entering the domain depends on the choice of the outer radius \rmax (Equation~(\ref{eq:vphirmax})).
As argued in Section~\ref{sec:phil}, setting $\rmax=\RHill$ is a natural choice for simulations assuming axisymmetry around the planet, but it remains approximate. The incoming angular momentum sets what fraction of the gas can reach the planet, and \neuIII{in general} it might determine whether an outflow near the midplane occurs or not. \neuIII{Here, we have infall at all angles, but} \neuII{an outflow does occur}
in the azimuthal average of 3D simulations \citep[e.g.][]{schulik20}.
Therefore,
we performed two additional simulations with the same parameters except for $\rmax=0.7\RHill$ and $\rmax=1.3\RHill$, named \Simrmaxkleiner and \Simrmaxgroesser, to give the accreting gas less or more angular momentum, respectively.
\neuII{We have chosen}
planet radii\neuII{near $\RP=2~\RJ$ by keeping $\rmin=1.9~\RJ$, with the exact values set by how much mass is accreted and how it cools until a quasi-steady-state is established as described in Section~\ref{sec:timeevol}. The radii turn out to be respectively} $\RP=2.15$ and 1.95~$\RJ$, which is similar enough for our purposes, \neuII{especially since the radius does not affect directly the infall of matter (see also Section~\ref{sec:verschAsp})}.
Since $\RHill=1.33\HP$ (Table~\ref{tab:par}), \Simrmaxkleiner has $\rmax\approx\HP$.  %
We took $\Nth=51$ instead of $\Nth=181$ since it does not influence the accretion flow.
We find that also for these simulations the radial velocity of the gas at \rmax quickly reaches and remains at the free-fall velocity for all angles.
Because we keep the density at \rmax the same, the \neuII{mass influxes} at the Hill sphere are somewhat lower and higher \neuII{respectively},
with
$\MPktnettoHill=5\times10^{-6}~\MPktEJ$ for \Simrmaxkleiner and
$\MPktnettoHill=9\times10^{-6}~\MPktEJ$ for \Simrmaxgroesser,
instead of $\MPktnettoHill=7\times10^{-6}~\MPktEJ$ for the fiducial run.
\neuII{As summarised in Section~\ref{sec:freepar}, judging from \citet{m16Schock} this difference in density will not influence the flow pattern.}

Figure~\ref{fig:varrmax} compares the flow of the gas on scales of $r\sim50$--$150~\RJ\sim0.01$--$0.03\RHill$ for $\rmax/\RHill=0.7$, 1.0, 1.3.
As expected, with larger \rmax, only gas from a smaller cone around the pole reaches the planetary surface directly (highlighted in pink).
The starting angle of the last streamline hitting the planet surface is
$\thi=15.8$\degr for \Simrmaxkleiner and
$\thi=3.9$\degr for \Simrmaxgroesser,
\neuIII{which bracket the corresponding $\thi=7.1$\degr for the fiducial case. The same applies to}
the \neuII{mass fluxes \neuIII{relative to the respective \MPktnettoHill, which}} are
$\MPktPdir/\MPktnettoHill= 2.4$~percent for \Simrmaxkleiner and
$\MPktPdir/\MPktnettoHill= 0.052$~percent for \Simrmaxgroesser;
\neuIII{the fiducial case had $\MPktPdir/\MPktnettoHill= 0.7$~percent.}

Thus, \neuIII{at fixed \MPktnettoHill,} $\MPktPdir$ depends sensitively on \rmax
but, overall,
the fraction of \MPktnettoHill that shocks on the planetary surface is at most of order of one percent.
The dependence of \MPktHa on $\rmax/\RHill$ will be similar.
We discuss this further in Section~\ref{sec:verschAsp}. %
For reference, we obtain $\LHa=1.0\times10^{-7}$ and $7.3\times10^{-9}~\LSonne$ for \Simrmaxkleiner and \Simrmaxgroesser, respectively, \neuIII{again bracketing the fiducial case with its $\LHa=3.7\times10^{-8}~\LSonne$.}
\subsection{Varying the planet mass}
  \label{sec:varmass}
\begin{figure}
 \centering
 \includegraphics[width=0.47\textwidth]{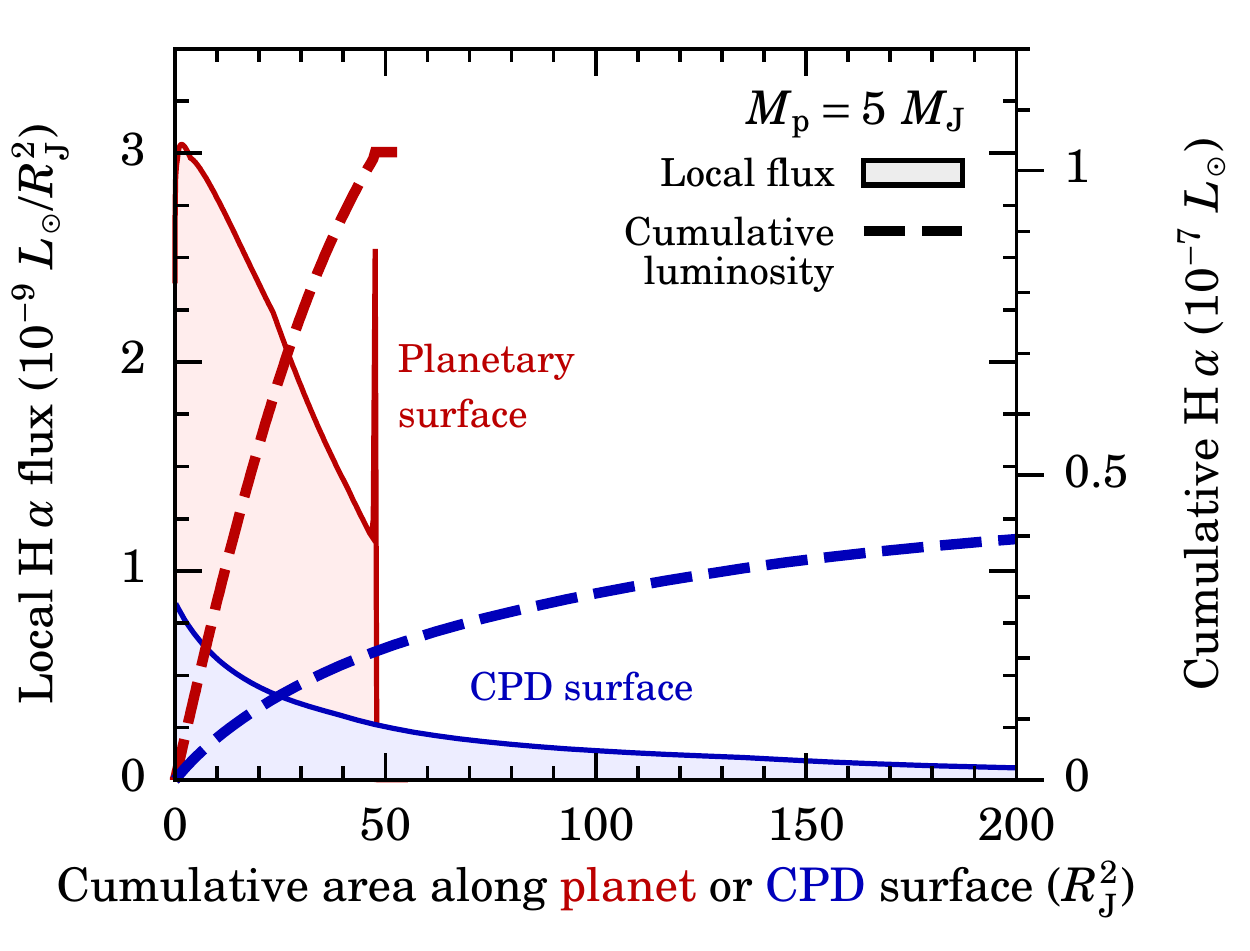}
\caption{%
As in Figure~\ref{fig:LHaPzpSch} but for 5~\MJ. The total is $\LHa=1.4\times10^{-7}~\LSonne$. The planetary surface is twice as bright as the CPD. %
}
\label{fig:LHaPzpSch5MJ}
\end{figure}

Short of doing a full exploration of the whole parameter space, we consider also a higher \MP.
The planetary mass is an important parameter that controls the dynamics of the gas. 
We wish to see whether here too only a fraction of the gas entering the Hill sphere falls directly onto the planet, and whether the planetary surface still dominates the \Ha emission relative to the CPD as it does for $\MP=2~\MJ$ (Figure~\ref{fig:LHaPzpSch}).
A priori, especially the latter
could change for higher masses because
$v_0$ scales to zeroth order with $\vFf\propto r^{-1/2}$,
which will lead to a larger \Ha-emitting area
on the CPD.

\begin{figure*}
 \centering
 \includegraphics[width=0.47\textwidth]{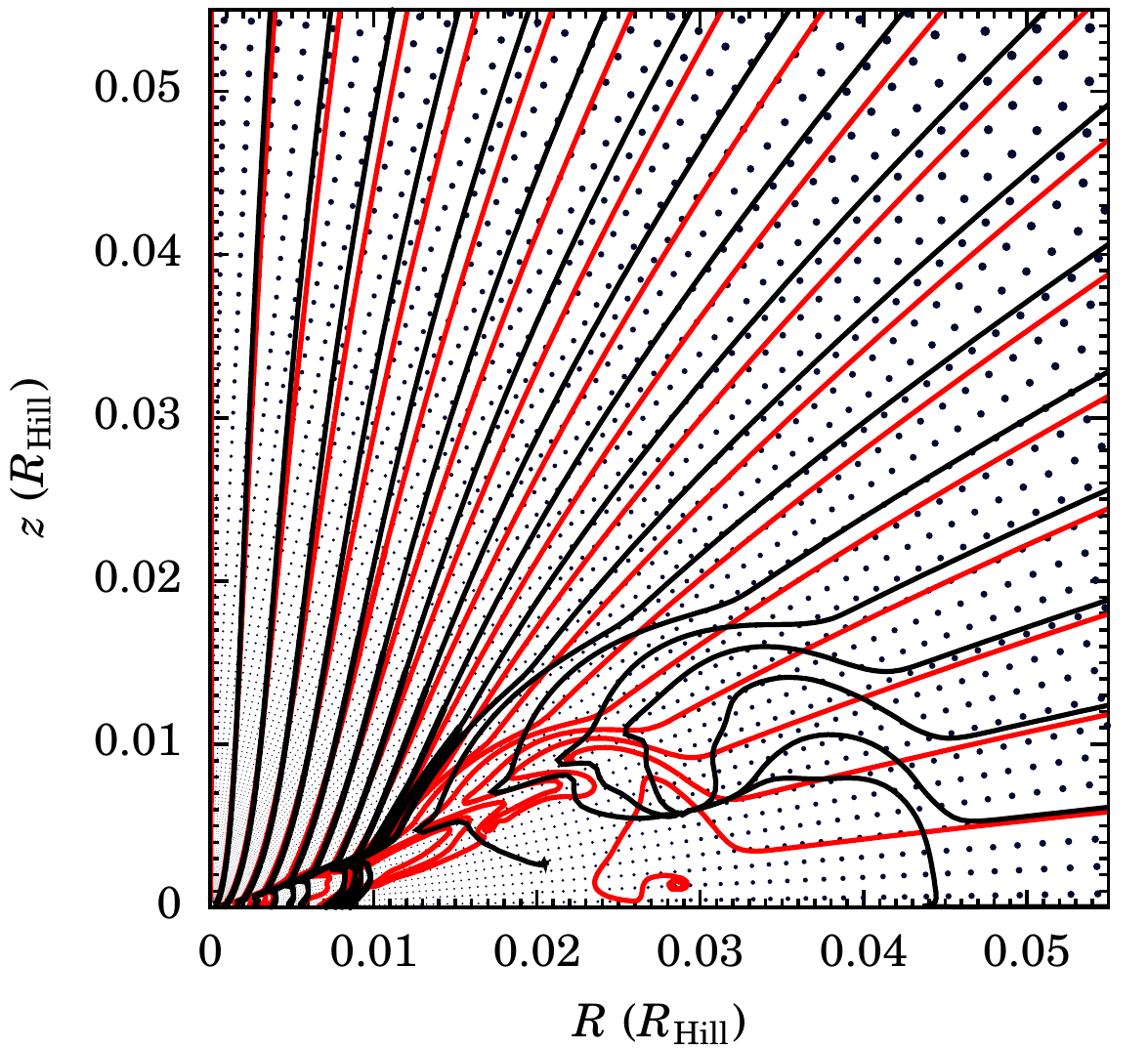}
 \includegraphics[width=0.47\textwidth]{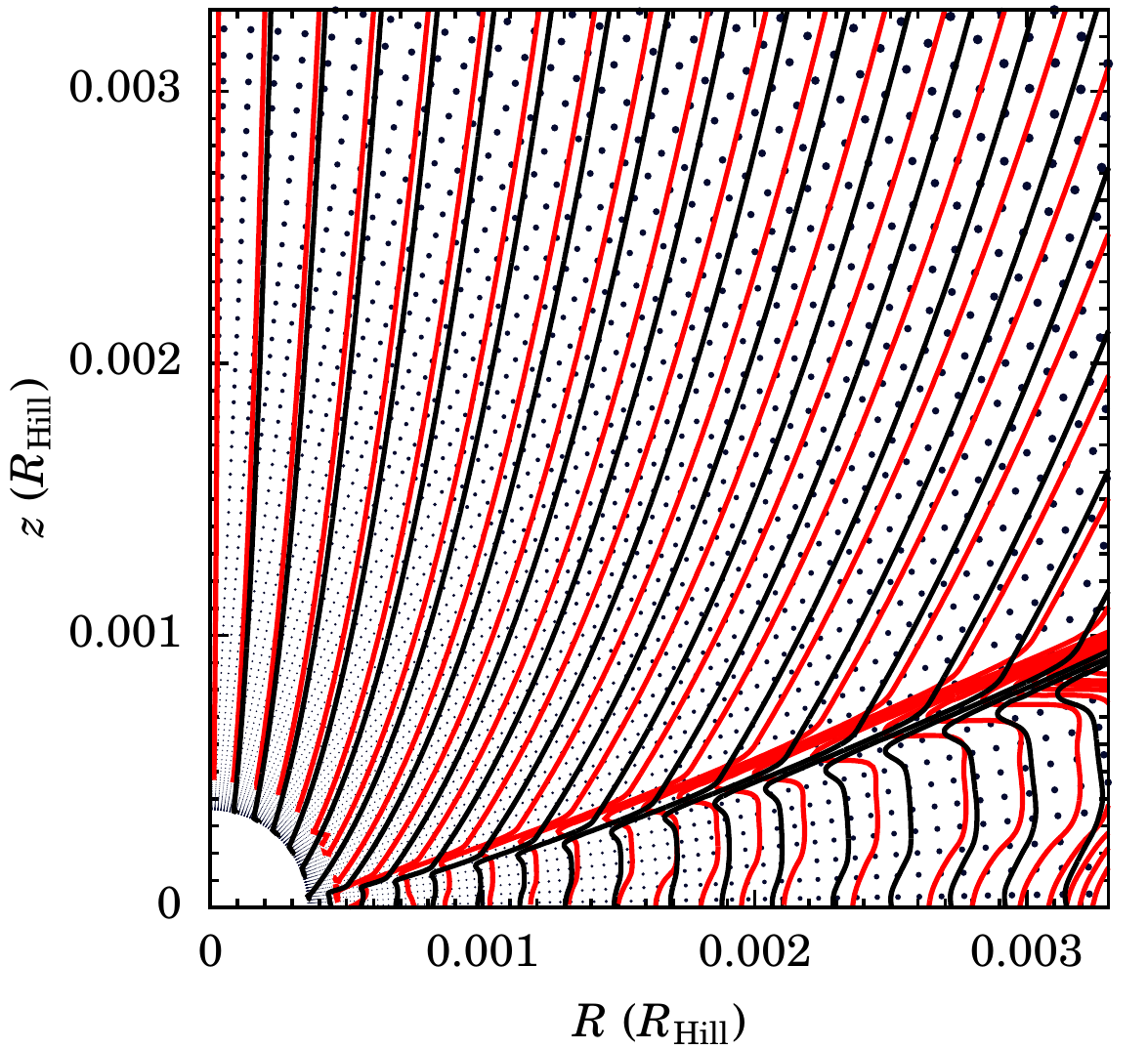}
\caption{%
Comparison of the flow in the $\MP=5$-\MJ simulation (black) to the fiducial 2-\MJ run (red).
Streamlines are for $\thi=0$, 5, \ldots, 90\degr (left) or $\thi=0$, 1, 2\degr, \ldots (right).
The axes, \neuV{different for each panel}, are scaled by the respective Hill radius \neuV{of each simulation}.
\neuV{Dark blue} dots show the cell centers in the 5-\MJ run.
}
\label{fig:flow2vs5MJ}
\end{figure*}

We therefore simulate an accreting planet as in the fiducial run but with $\MP=5~\MJ$ \neuII{and name this \SimMPgroesser}.
\neuII{\citet{bae19} used this value for \PDSb.} We set again \rmax equal to $\RHill(\MP)$, and use $\Nth=51$.
The other parameters are the same, \neuIII{leading to a large $\qth=18$}.
Figure~\ref{fig:flow2vs5MJ} shows that the flow pattern for the 5-\MJ simulation is similar to the one of the 2-\MJ simulation.
Even at this lower resolution in $\theta$, a thin ($\sim1$\degr) fast inward surface flow is seen again. %
Repeating the analysis above, we find that for $5~\MJ$ only
$\MPktPdir/\MPktnettoHill=1.2$~percent of
$\MPktnettoHill=1.4\times10^{-5}~\MPktEJ$
reaches
the planet directly,
and that $\MPktHa/\MPktnettoHill=7$~percent shocks with $v\gtrsim\vHakrit$.
For $\MP=2~\MJ$, we recall that we had smaller fractions of 0.4~and 0.7~percent, respectively.

In Figure~\ref{fig:LHaPzpSch5MJ}, we show the \neuIII{\Ha emission from} the planetary surface and the CPD, as in Figure~\ref{fig:LHaPzpSch}. Again, the inward flow below the CPD surface does not generate an overall important \Ha flux; there is a local spike but its relative contribution is negligible. 
The planetary surface generates
$\LHa=1.2\times10^{-7}~\LSonne$ and the CPD surface generates $\LHa=0.54\times10^{-7}~\LSonne$.
Summing the two terms again in lieu of detailed radiation transport yields in total $\LHa\approx1.7\times10^{-7}~\LSonne$.
Thus, the \neuIII{CPD surface} emits about \neuIII{30}~percent of the total flux, \neuIII{up from 15}~percent in the 2-\MJ case.
\neuIII{This relative increase is because 
$v_0>\vHakrit$ out to a cylindrical radius of $R\approx9~\RJ$ instead of $R\approx3~\RJ$ in the fiducial case (see the coloured regions in both panels of Figure~\ref{fig:vnormSchock}).}

\section{Discussion}
 \label{sec:disc}
 
Our main results are that (a)~only a small fraction of the \neuIII{net mass flux into} the Hill sphere 
falls directly onto the planet, (b)~\neuIII{only a slightly larger fraction}
produces any \Ha (Figures~\ref{fig:rho-Ueberblick} and~\ref{fig:vnormSchock}),
\neuIII{and (c)~\neuIV{the emitted} \Ha comes from both the planetary and the CPD surfaces, and not from the fast flow beneath the CPD surface (Figure~\ref{fig:vnormSchock}).}
\neuII{Our simulations were conducted in 2.5D but}
\citet{tanigawa12} obtained qualitatively the same \neuII{Hill-sphere} flow structure in their larger-scale isothermal 3D simulations. This lends support to our 2.5D approach and suggests that these findings are robust.
The advantage of 2.5D is that it makes it computationally much more accessible not to smooth the gravitational potential while including radiation transport. This allowed us to simulate down to sub-planet scales, crucial for calculating \LHa\neuII{ since the highest-velocity regions strongly dominate the emission}.

In Section~\ref{sec:discPDS}, we 
\neuII{look at}
\PDSb.
In Section~\ref{sec:vglHa}, we compare our results with other predictions of \Ha emission from planets accreting \neuII{other than} by magnetospheric accretion (for the latter, see discussion in \citealt{AMIM21L}).
Finally, in Section~\ref{sec:verschAsp} we comment \neuII{on a few aspects of our models}.

\subsection[Comparison with PDS 70 b]{Comparison with PDS\,70\,b}
 \label{sec:discPDS}

\refstepcounter{numKommG}
\neuII{As a check, we compare} the \LHa estimated from the simulations with the observational data for \PDSb. This planet had motivated our parameter choices (Table~\ref{tab:par}).
Assuming that the \Ha photons are leaving the system isotropically, its measured luminosity is $\LHa=7\times10^{-7}~\LSonne$ \citep{zhou21,sanghi22}.
At such luminosities, absorption within the system is likely unimportant for a very wide range of dust opacities \citep{maea21}, and absorption by the PPD is more likely to be low given that the planet is found in a gap.
\neuII{Therefore, a direct comparison is meaningful.}  

For the different runs,   %
we obtained $\LHa\sim10^{-8}$--$10^{-7}~\LSonne$,  %
which is 10--100~times smaller than \neuII{the observationally derived value}.
\neuII{This}
is in fact %
\neuII{satisfactory}
given that we took nominal model parameters (Table~\ref{tab:par}) from the literature without efforts to match the \LHa.
\neuII{Reducing the incoming angular momentum  %
or radius of the planet, or using $\MP=12$~\MJ as in \citet{dong21},}
would make it easy to \neuII{raise our \LHa closer to the measured value}. 
\neuII{Beyond this, magnetospheric accretion columns, if present, could also be contributing to the flux.}

\subsection[Comparison to other predictions of H alpha]{Comparison to other predictions of \Ha}
 \label{sec:vglHa}

Only few studies so far predict the \Ha emission of forming planets.
\citet{thanathibodee19} applied magnetospheric-accretion radiation-transfer models for stars to the planetary regime, but
\citet{szul20} were the first to present hydrogen-line luminosities based on 3D radiation-hydrodynamics simulations. However, their smoothing of the gravitational potential (Table~\ref{tab:sammenligning/jevnfoering})
makes their results challenging to interpret, as \citet{Aoyama+2020,AMIM21L} discuss.
Therefore, we restrict our comparison here to the work of \citet{takasao21}, who also set $\epsPot=0$.

We follow a similar approach to \citet{takasao21}
to calculate the \Ha emission from the simulation data,
\neuIII{by integrating the local \Ha flux predicted by \citet{aoyama18} as a function of the preshock velocity and density} (Section~\ref{sec:Ha}).
However, \citet{takasao21} find that nearly 95~percent of the \Ha emitted from the planetary surface comes from an approximately 10--$15$\degr-thick surface layer of the CPD where the flow hits the planetary surface \neuIV{(see their schematic Figure~13)}.
This differs significantly from our results, in which the sub-CPD-surface flow is very thin ($\sim1$\degr) and \neuIII{contributes} negligibly to the total luminosity \neuII{(Figures~\ref{fig:LHaPzpSch} and~\ref{fig:LHaPzpSch5MJ})}.

\begin{figure} %
 \centering
 \includegraphics[width=0.47\textwidth]{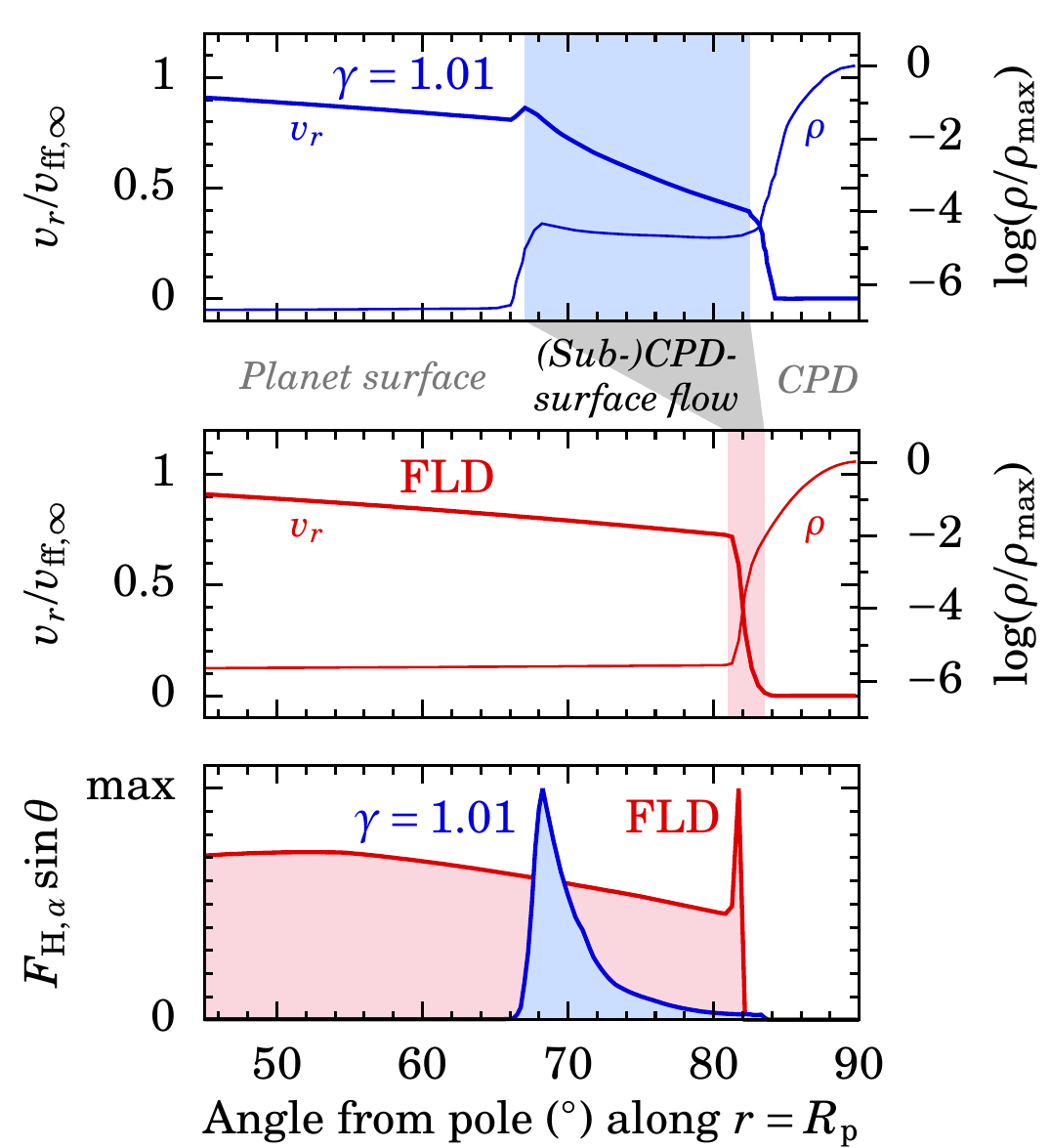}
\caption{%
\neuIII{%
Structures in the $\gamma=1.01$ hydrodynamics
simulation of \citet{takasao21} compared to ours (with FLD).
Top and middle panels:
Radial velocity immediately above the planet surface, normalised to the respective \vFf (Equation~(\ref{eq:vFfinfty})), and density (normalised to the maximum, i.e., midplane density) in their (top panel; blue curves)
and our (middle panel; pink curves) simulations.
Bottom panel:
Local \Ha flux times $\sin\theta$, normalised to the respective maxima.
This way, the relative areas under the curves are proportional to their contribution to the total flux.%
}%
}
\label{fig:vglT21}
\end{figure}

\neuIII{To understand the difference, we compare in Figure~\ref{fig:vglT21} the radial velocity and the density at $r=\RP$ in the work of \citet{takasao21} (top panel) and our work (middle panel).
Three similar zones are present in both work: the free surface of the planet from the pole down to (geometrically, or ``up to'' in angle) $\theta=67.5\degr$ in their case or $\theta=80\degr$ in ours; the fast (sub-)CPD-surface flow, which is quite thick in their case and thin in ours (coloured regions in Figure~\ref{fig:vglT21}ab); and the CPD connecting to the planet surface, at $\theta\gtrsim83\degr$ for both.
In both simulations the radial velocity above the planet is equal to \vFfinfty at the pole (not shown) and decreases with $\theta$, as discussed in Section~\ref{sec:Ha}.
Also, in both cases the density increases quickly in the CPD zone, as expected for (approximately) isothermal structures (see Figure~\ref{fig:T2D}) in hydrostatic equilibrium.

However, there is a large difference in the layer located between the free-falling gas and the CPD.
\citet{takasao21} obtained a zone roughly 10--$15\degr$ thick
in which the radial velocity decreases only slowly by half,
\neuIV{which they called a ``postshock, converged accretion flow''.
However,} we find a zone that is only a few degrees thin and in which}
the
radial velocity %
decreases \neuIII{quickly} as a function of distance below the surface.
\neuIV{In our simulations too is there a visible convergence of the postshock accretion flow, seen as the approximately constant-$\theta$ segments of the sub-CPD-surface streamlines in Figure~\ref{fig:flow2vs5MJ}. However, this converged flow has a smaller $|v_r|$, is thin, and involves relatively little mass. Since the flow is the same while more than doubling the mass ($\MP=2$ vs. 5~\MJ), it most likely does not matter that \citet{takasao21} set an even higher mass $\MP=12~\MJ$.}

This qualitative difference \neuIII{in the flows at the CPD surface must} come
\neuIV{instead from the different thermodynamics that \citet{takasao21} assumed, namely no radiative transfer but an adiabatic equation of state with $\gamma$ close to, but above, unity.}
\neuIII{\neuIV{Even though $\gamma$ is close to the isothermal value of unity,
their adiabatic equation of state leads to a thick and hot post-shock region
surrounding the CPD, as their Figure~6 shows.
Consequently, the radial velocity remains high after the shock in $\theta$, so that the gas hits the planet surface quickly: the gas is subsonic but the Mach number is large, so that the absolute velocity is high.}
On the contrary, with}
radiative transfer, \neuIII{the gas cools quickly and} the density \neuII{increases much more} across the shock.
By mass conservation, the postshock radial velocity is correspondingly smaller,
\neuIII{which decreases} significantly the amount of emission. %

\neuIV{Figure~\ref{fig:vglT21}c shows the local \Ha line emission at $r=\RP=2~\RJ$ as a function of angle from the pole. We plot $\FHa\times\sin(\theta)$, as in Equation~(13) of \citet{takasao21}, and normalise the curves from \citet{takasao21} and our simulation independently to their respective maximum. This way, the areas under the curves are proportional to the contribution of each region to the total flux from a model. This shows very clearly that in the simulation of \citet{takasao21}, only the sub-CPD-surface flow, where it hits the planets, generates appreciable amounts of \Ha, whereas in our case that zone is negligible for the integral (seen also in Figures~\ref{fig:LHaPzpSch} and~\ref{fig:LHaPzpSch5MJ}).}

In our finding that only a small fraction of the large-scale flow falls directly onto the planet, however, we agree qualitatively with \citet{takasao21}.
Excluding the radial flow below the CPD surface,
the accretion rate directly onto the planetary surface in their case\footnote{We use
$\MPktPdir\approx(1-\cos\thmax)4\pi\RP^2\rho_0v_0$ and read off the values from their Figure~8. Similarly, \MPktnettoHill is close to their maximal \MPktnettoHill, which is set by the boundary conditions at their \rmax (see their Figure~2b).%
} is
$\MPktPdir\approx3\times10^{-10}~\MPktEJ$,
which is
0.2~percent of their
net mass influx rate $\MPktnettoHill=2\times10^{-7}~\MPktEJ$. This is smaller than but similar to our fractions for 2~and 5~\MJ (0.4 and 1.2~percent), with the difference likely coming from their
choice of a purely vertical mass flow at their $\rmax = 100~\RP$. 

We conclude that including radiation transfer in hydrodynamical simulations is important for accurate predictions of \Ha emission because of their sensitivity to the velocity structure. The flow in the large-scale, supersonic region can likely be well captured by isothermal simulations, \neuIII{but the post-shock behaviour of the gas depends on the thermodynamics.}
Smoothing-free %
3D simulations
\neuIII{in the high-mass (high-\qth), low-$\RP/\RHill$ regime would be a worthwhile complement to the existing work}
(Table~\ref{tab:sammenligning/jevnfoering}).

\subsection{Further aspects within and beyond our model}
 \label{sec:verschAsp}

We comment on a few aspects within or beyond our model:
\textit{Regions traced by the \Ha.}---%
The results of Section~\ref{sec:varmass} suggest that as the planet mass increases, the contribution of the CPD to the total \Ha becomes increasingly important. However, the two terms remain of the same order of magnitude,
\neuV{and if the CPD is thicker than in Figure~\ref{fig:flow2vs5MJ}, the reduced shock velocity would lead to a smaller contribution.}
Modelling of the line shapes should therefore take both components into account.

\textit{Varying the planetary radius.}---%
We can do this approximately without additional simulations by measuring the \MPkt's according to Equations~(\ref{eq:MPktPdir}) or~(\ref{eq:MPktHa}) at a different $r=\RP'$, and similarly beginning at $\RP'$ the outwards integration of the \Ha emission along the CPD surface.
To first order, the choice of \RP will not affect 
the \neuIV{supersonic} flow. %
In a similar approach, \citet{takasao21} set an open boundary at their \rmin and used the density and velocity there to calculate the \Ha emission that would come from a shock at that position.
Doing this, we find roughly $\LHa\propto1/\RP$,  %
\neuII{which can be used to scale approximately the results of one simulation to other \RP values.}
\textit{Choice of \rmax.}---%
We have varied \rmax by 30~percent, with the case $\rmax=1.3\RHill$ corresponding to $\rmax\approx\HP$. It would be surprising if 3D simulations corresponded effectively to a much larger \rmax, but the effective \rmax could conceivably be smaller. Then, a larger fraction of \MPktnettoHill would reach the planet directly and emit \Ha.
\textit{Models of 1D planet structure.}---%
Global formation models use \MPktnettoHill
to set the ram pressure at the surface of the planet when calculating its radius and luminosity \citep{morda12_I}. %
\neuIII{However, this is an incorrect assumption, since}
only the much smaller rate $\MPktPdir\ll\MPktnettoHill$ will set the pressure on the surface of the planet, which could affect its postformation luminosity \citep[e.g.,][]{morda13,berardo17}.
In our simulations so far the \neuII{(radial)} ram pressure $\Pram(\theta)=\rho(\theta)v_r(\theta)^2$ turns out to be almost constant with polar angle (not shown), so that the reduced ram pressure could be easily included in 1D planet models.
However, an appropriate treatment of the boundary layer with its \fomg-dependent transfer of mass, angular momentum, and energy would be needed
\citep[e.g.,][]{dong21}.

\textit{Other hydrogen lines.}---%
\neuIII{Other hydrogen lines such as \Hb, \Pab, or \Brg have been} observed at a few planetary-mass objects such as \Dlrmb \citep{eriksson20,betti22b,betti22c}. These lines also require a similar minimum shock velocity $\vkrit\approx\vHakrit=30~\kms$ to be emitted \neuII{since their excitation energies are similar} \citep{aoyama18}. Therefore, our analysis 
could have applied to the other lines as well.  %
\textit{Magnetospheric accretion.}---%
If it proceeds as for young stars \citep[e.g.,][]{romanova02,hartmann16},
magnetospheric accretion
is an interesting mechanism 
that could let 
the gas slide ballistically along the magnetic field lines connecting the inner edge of the CPD and the planet surface \citep{lovelace11}.
This would lead to a shock at the planet surface at
almost
free-fall velocity and thus to \Ha emission.
This would be in addition to what the direct infall \MPktPdir generates, contrasting with the stellar case in which \MPktPdir is essentially zero.
In fact, %
magnetospheric accretion would let \neuII{almost} the same amount of \Ha be generated as in the 1D, spherically symmetric classical picture \citep[e.g.,][]{boden00} since ultimately \neuII{most}  %
accreting gas would reach the planet at (nearly) free-fall velocity\footnote{\neuII{One half of the potential energy is dissipated in the CPD if it extends to the surface of the planet (e.g., \citealp{pringle81,hartmann97}).}}.

Whether magnetospheric accretion from the CPD onto the planet actually takes place or not is not yet clear. It requires a few conditions to be met: (i)~the CPD must be an accretion and not a decretion disc; (ii)~the magnetic field of a young planet needs to be able to disrupt the CPD; and (iii)~the gas must be sufficiently ionised to couple to the magnetic field \citep[e.g.,][]{keith14,hasegawa21}. In the picture painted by \citet{batygin18}, in which gas falls towards the pole and a decretion disc, CPD disruption would not be needed and the apex of the magnetic field lines would increase the effective area of the planetary surface intercepting the flow. So far, interesting but only tentative scaling arguments support the main hypothesis of a sufficiently strong magnetic field
\citep{christensen09,katarzy16}, requiring further studies for a robust assessment.
Further motivation might come tentative observational evidence for magnetospheric accretion in the somewhat older, essentially isolated object \Dlrmb \citep{ringqvist23}.

\section{Summary and conclusions}
 \label{sec:summconc}

We studied the gas flow from the Hill radius down to the surface of a forming super-Jupiter planet able to generate hydrogen lines such as \Ha.
We performed axisymmetric, 2.5D %
radiation-hydrodynamical simulations in a vertical frame centered on the planet and following it on its orbit around the star.
\neuII{These simulations connect to global-disc simulations through the net mass inflow into the domain \MPktnettoHill and the angular momentum of the gas, both of which are input parameters here. We argued that the flow structure should depend only little on \MPktnettoHill.} \neuIII{Therefore, this should also apply to the partial accretion rates or the relative contributions to line emission by the planetary surface and the CPD surface.}

\neuII{Two important features compared to previous work are that we (i)}~included radiation transfer, with tabulated dust and gas opacities,
to model correctly thermal effects that can influence the flow \neuII{especially below the CPD-surface shock},
\neuII{and (ii)~did not smooth the}
gravitational potential,
\neuIV{and used} a high spatial resolution close to the planetary surface ($\sim0.01~\RJ$).
Whereas previous work with a \neuII{non-zero} smoothing length (e.g., \citealp{tanigawa12}) was concerned with the accretion of mass and angular momentum onto the CPD, we focus on the planet surface and the innermost regions of the CPD close to it.

We confirmed that most of the mass flux flowing towards the CPD and the planet forms an accretion shock on the surface of the CPD
(Figure~\ref{fig:vnormSchock}). Only a very small fraction of the order of a percent reaches the planet surface directly, and the fraction shocking at sufficiently high velocity ($v_0>\vHakrit= 30~\kms$) to generate hydrogen lines such as \Ha is similarly small (Figure~\ref{fig:varrmax}).
\neuIII{We found that these results are robust to variations in the planetary mass and the angular momentum of the incoming gas.}
The large-scale flow pattern agrees qualitatively with 3D isothermal simulations with a smoothed gravitational potential \citep{tanigawa12,fung19}, lending support to our approach.

\neuII{For all simulations, we estimated the \Ha emission through the non-equilibrium shock models of \citet{aoyama18}.
Our inclusion of radiative transfer
\neuIV{keeps thin the fast flow beneath the CPD shock surface, so that only the free surfaces of the planet and the CPD emit shock tracers appreciably.
This contrasts to the results of} \neuIII{hydrodynamics-only} simulations (Section~\ref{sec:vglHa}),  %
showing the importance of including radiation transfer while not smoothing the gravitational potential.
\refstepcounter{numKommG}
}

\neuII{
In summary, we have studied one aspect determining how many planets can be detected at accretion tracers such as \Ha:
\neuIII{what parts of the flow can generate accretion-line emission.
However, we have not addressed the relation between this \Ha-generating mass flux and the growth rate of the planet.
This is a different question, beyond the scope of our work, and involves studying the timescale for mass transport in the CPD. In one limit, only what falls directly onto the planet will let it grow at a given time, but in the other the CPD would be able to feed the planet appreciably \citep{ab22}. \neuIV{Dedicated simulations are required.}}

It is \neuIII{moreover} a separate issue whether the statistics of known accreting planets matches expectations given our current understanding of planet formation and the empirical demographics of %
directly-imaged planets \citep[e.g.,][]{nielsen19,vigan21}.  %
Both the migration and formation timescales influence this, but also the non-Gaussianity in the residuals in high-contrast images
 (\citealp{marois08a}; see \texttt{applefy} by \citealt{bonse23}).
A careful statistical treatment %
would be welcome \citep{dong23}, as would more detections---%
for which there is hope thanks to instrumentational progress such as VIS-X \citep{haffert21visx}, KPIC \citep{delorme21}, or RISTRETTO \citep{chazelas20}, to name a few.}

\refstepcounter{numKommG}
\refstepcounter{numKommG}
\refstepcounter{numKommG}

\begin{acknowledgments}
{\small
We dedicate this paper to the memory of Willy Kley (Universit\"at T\"ubingen, $\dagger$\,21.12.2021), with whom several of us had the privilege of collaborating and who provided useful advice at an early stage of this project. His kindness and expertise are deeply missed.
We are indebted to Yuhiko Aoyama for his model data and for many answers to many questions, and warmly thank Shinsuke Takasao, Kazuhiro Kanagawa, and Matthew Bate for helpful discussions.
\neuV{G-DM is particularly grateful to Thomas Henning for his vision, support, and encouragement to begin investigating the accretion shock in planet formation.
We thank the referee for a report that helped us significantly clarify the structure and content of this paper.}
This research was supported by the Munich Institute for Astro-, Particle and BioPhysics (MIAPbP) which is funded by the Deutsche Forschungsgemeinschaft (DFG, German Research Foundation) under Germany's Excellence Strategy (EXC-2094 -- 390783311).
G-DM and RK acknowledge the support of the DFG priority program SPP 1992 ``Exploring the Diversity of Extrasolar Planets'' (MA~9185/1, KU~2849/7, and KU~2849/10).
G-DM and CM also acknowledge the support from the Swiss National Science Foundation under grant
200021\_204847
``PlanetsInTime''.
RK acknowledges financial support via the %
Heisenberg Research Grant funded by the DFG under grant %
KU~2849/9.
WB acknowledges funding by the DFG under grant KL~650/31-1.
Parts of this work have been carried out within the framework of the NCCR PlanetS supported by the Swiss National Science Foundation.
This research has made use of NASA's Astrophysics Data System Bibliographic Services.
We acknowledge gratefully the use of Ankit Rohatgi's \texttt{WebPlotDigitizer}.
All figures were produced using \texttt{gnuplot} with the terminal \texttt{epslatex} with the font package \texttt{fouriernc}.
}
\end{acknowledgments}

\appendix

\section{Radial gridding}
 \label{sec:Gitter}

\begin{figure} %
 \centering
 \includegraphics[width=0.47\textwidth]{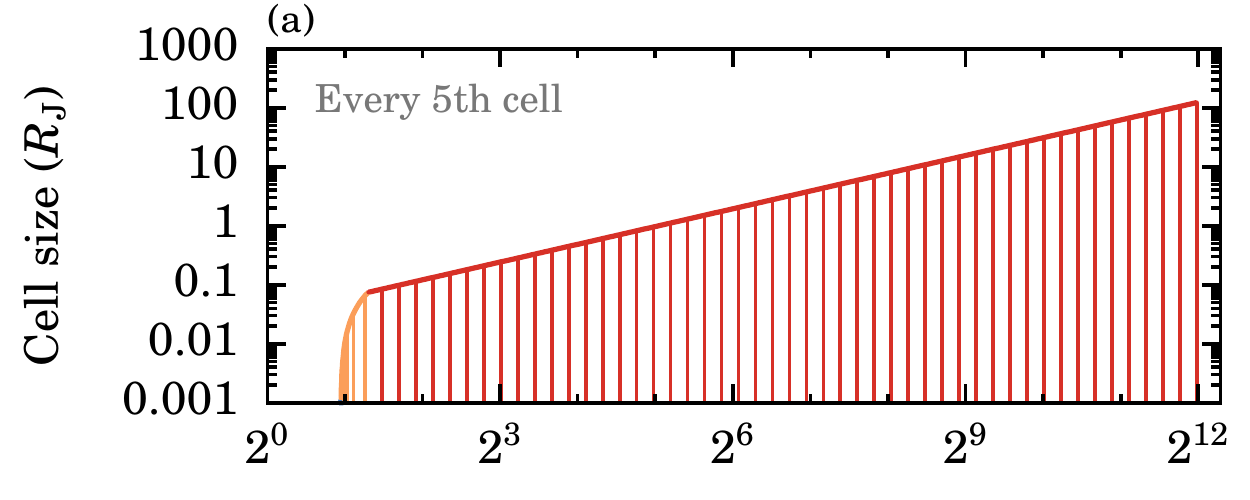}\\
 \includegraphics[width=0.47\textwidth]{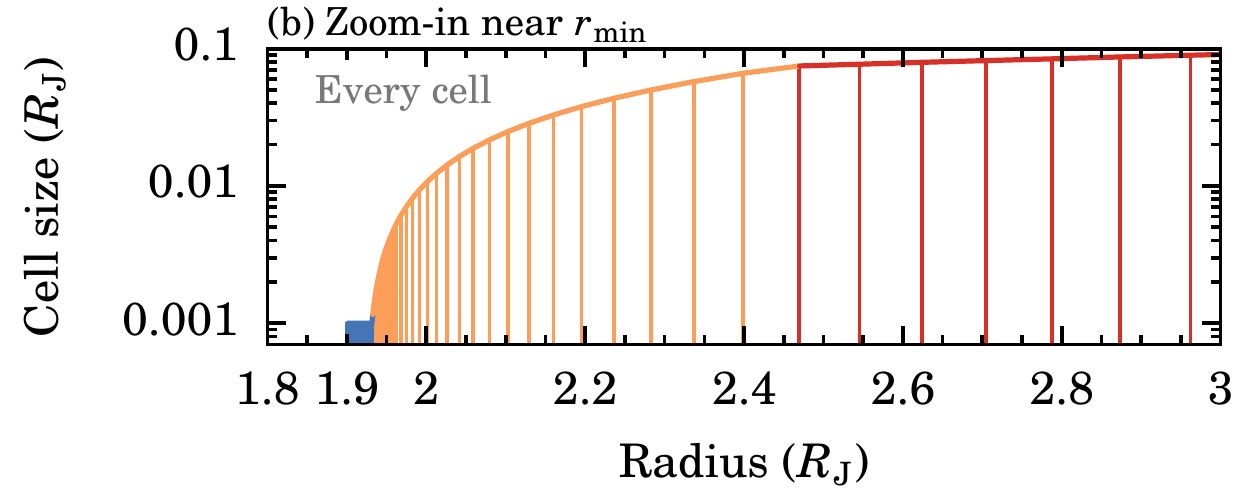}
\caption{%
Cell size of the radial grid, from $\rmin=1.9~\RJ$ to $\rmax=\RH=4100~\RJ$ \neuIII{in the fiducial case},
with its uniform (blue), geometrically stretched (peach), and logarithmic (red) segments.
(a)~Global view. %
The central position of only every 5th cell is shown (vertical lines).
(b)~Zoom-in near \rmin. Every cell centre is shown.
}
\label{fig:Gitter}
\end{figure}

The radial gridding is made of three parts and is shown in Figure~\ref{fig:Gitter}. The inner section at $r\in[\rmin,\rmin+\Lu]$ has 32 uniformly-spaced cells $\Dru=0.001$~\RJ long (hence $\Lu=0.032~\RJ$); the outer section at $r\in[\rmin+\Lu+\Ls,\rmax]$, with $\Ls=0.5~\RJ$, is logarithmically stretched with 76~cells  %
per decade in radius; and the transition section at $r\in[\rmin+\Lu,\rmin+\Lu+\Ls]$ has geometrically stretched cells \citep{mignone07} chosen to have a smooth increase in cell size between \Dru and the first cell size in the logarithmic part; we take 30~cells for the middle section. This gives 307 zones in total.
We have tested that the results do not change appreciably when using a higher resolution for the different parts of the grid. As in our 1D simulations \citepalias{m16Schock,m18Schock}, a lower resolution in the inner, uniform part would lead to artificially high luminosities in the settling zone below the shock, with a rapid increase in time.
However, for some simulations,
we were able to increase the cell size in the inner part to $\Dru=0.002~\RJ$ (adjusting the stretched transition region to have a smooth change in cell size) and still obtain a correct-looking solution.

\section{Dependence of the CPD surface flow on the resolution in the polar direction}
 \label{sec:thres}

\begin{figure*}
 \centering
 \includegraphics[width=0.47\textwidth]{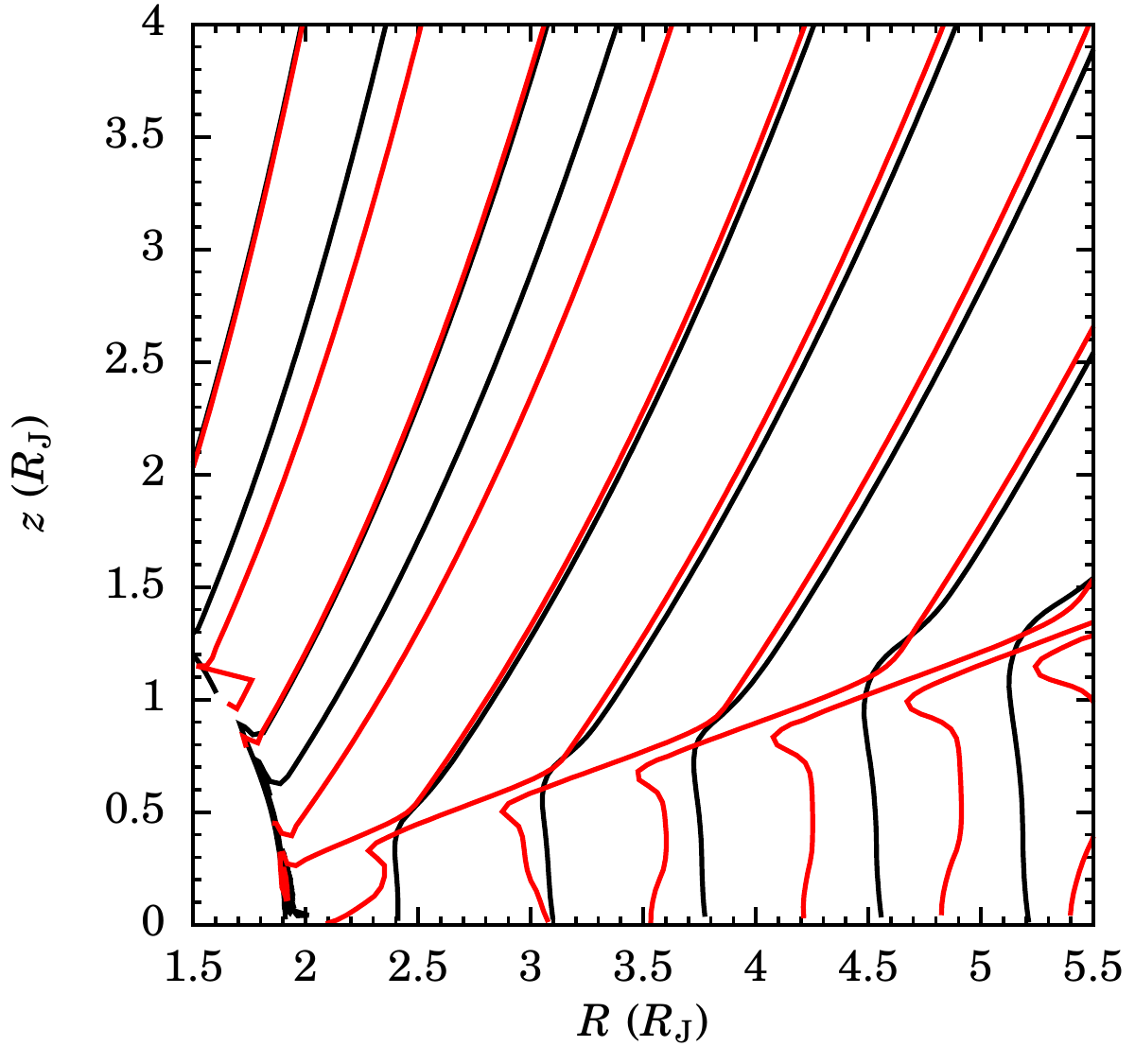}
 \includegraphics[width=0.47\textwidth]{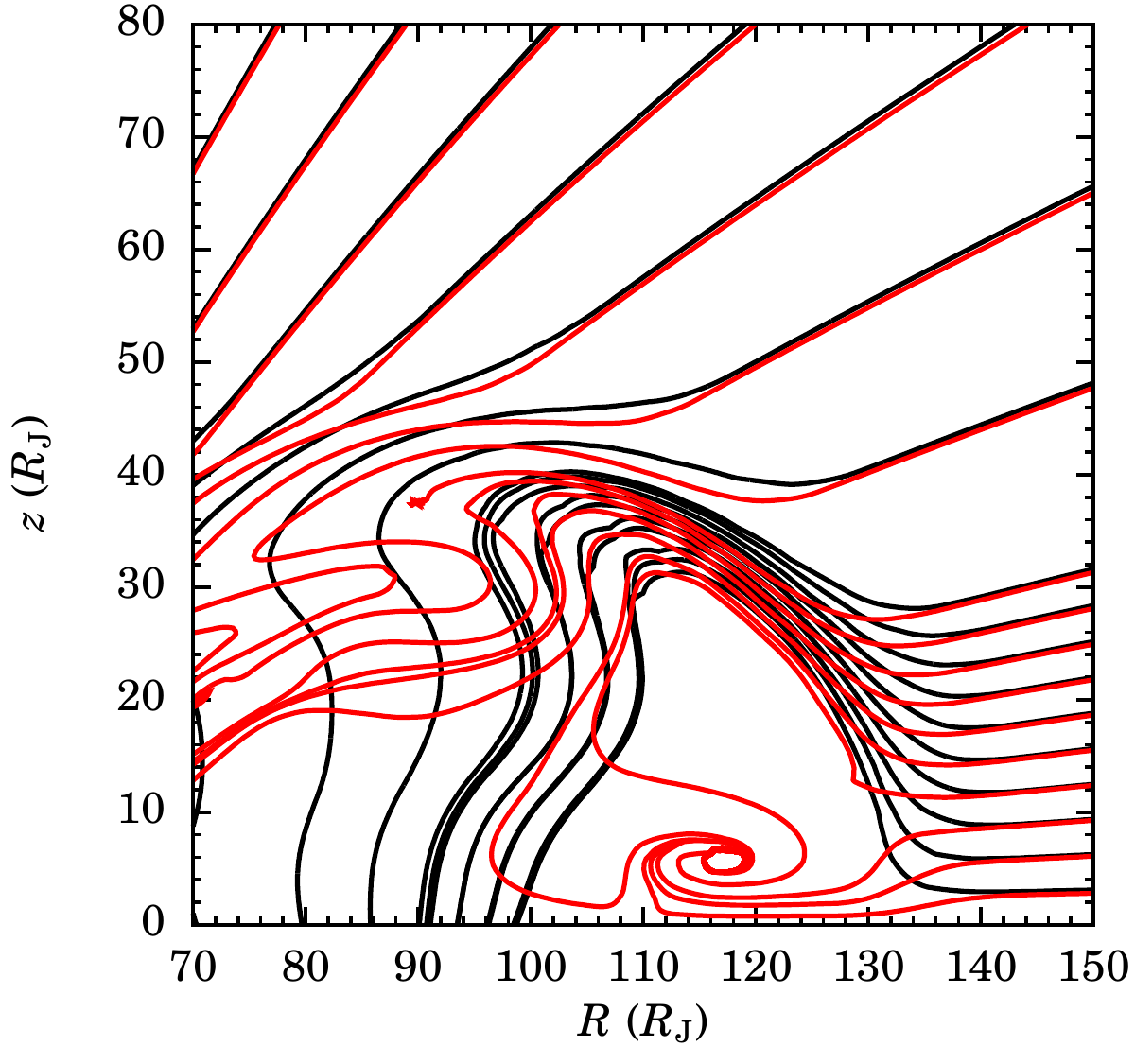}
\caption{%
Streamlines for simulations at two $\theta$ resolutions: $\Delta\theta=0.5$\degr (fiducial run; red) and $\Delta\theta=1.8$\degr (black). The density structure 
(essentially the same for both)
is shown in Figure~\ref{fig:rho-Ueberblick}.
In each panel, the $\thi=\theta(\rmax)$ values of the streamlines are the same for both simulations. Each panel focuses on a different region at a different scale. The flow beneath the CPD shock depends qualitatively on the resolution in the polar direction but this does not affect our results.
}
\label{fig:thres}
\end{figure*}

Beneath the shock on the CPD surface, in which the polar component of the velocity goes from super- to subsonic, there is a thin layer of $\approx1$\degr in which the gas is radially still supersonic.
This layered accretion is described in \citet{tanigawa12} and seen also by \citet{takasao21}. In their simulations, the layer, which is resolved, is however much thicker and of order of 15\degr. We discussed this in Section~\ref{sec:vglHa}. 

With our fiducial resolution of $\Nth=181$, we obtain layered accretion and outward-directed ``backflows'' in the layer right below the shock \citep{tanigawa12,takasao21}, in which $v_\theta$ is subsonic but $v_r$ still supersonic
(see Figure~\ref{fig:thres}).
With $\Nth=51$, the gas settles vertically directly to the midplane instead of performing a ``U-turn'' in a thin layer beforehand. 
However, the fully supersonic part of the flow, especially close to the planet, is independent of the resolution, and the \Ha emission also does not depend on the resolution.

Finally, we see that the behaviour of the gas at the outer edge depends somewhat on the resolution. For $\Nth=181$, the streamlines within about 4\degr of the midplane flow downwards, while the others are lifted up (Figure~\ref{fig:rho-Ueberblick}b). 
At $\Nth=51$, all streamlines are lifted up.
Also, at the outer edge of the CPD, the streamlines are lifted up at the outer edge of the CPD in the $\Nth=181$ but not the $\Nth=51$ simulation.

\section{Variations in the flow pattern}
 \label{sec:sigh-pattern}

Ideally, we would be able to wait for a quasi-steady state to establish in the flow at large and small scales, and measure from this the different properties (\MPktHa, 
etc.). In practice, despite months of wall-clock run-time, in the fiducial simulation a density wave was still travelling out in Phase~I (as mentioned in Section~\ref{sec:timeevol}). It is associated with the growing CPD outer radius and reflects our set-up in which we let the simulation begin without a CPD. This wave changes somewhat the angular distribution of the mass infall close to the freeze radius and thus, in principle, close to the shock radius for Phase~II.

We assess how much variation in the \Ha-generating accretion rate could come from this wave. For this, we measure as a function of time the \neuII{mass flux} in the supersonic region within 45\degr of the pole at a distance of $r=15$, 20, 30, and 50~\RJ from the planet. Since the flow is smooth, these partial accretion rates will correlate directly with \MPktHa, which is not accessible in Phase~I because the freeze radius \rfrz is farther out than the maximal radius for \Ha generation. Furthermore, since even after $t=900~\tFfglob$ the density wave has not yet reached \rmax but rather is still moving out, we look at simulations with different parameters in which the evolution happens more quickly.

We find that in a simulation with identical parameters but a surface density increased by a factor of ten, large-scale oscillations begin around $700~\tFfglob$. The maximum \neuII{mass flux} is $\MPkt=10^{-5}~\MPktEJ$ at $r=50~\RJ$ and goes smoothly as a power law down to $5\times10^{-6}~\MPktEJ$ at $15~\RJ$. The minimum \neuII{mass flux} decreases more steeply from $\MPkt=2\times10^{-7}~\MPktEJ$ at $r=50~\RJ$ to $2\times10^{-8}~\MPktEJ$ at $15~\RJ$. The total influx at \RHill is $\MPktnettoHill=7\times10^{-5}~\MPktEJ$ at all times. Therefore, extrapolating down to a radial distance of $r\approx2~\RJ$, the planet-reaching or \Ha-generating accretion rate is
\neuIII{in the range of $0.05$ to $\sim10^{-5}$ times \MPktnettoHill}.
\neuIII{The minimum} value has a considerable uncertainty \neuII{due to} the extrapolation. These are partial rates and are integrated in angle only down to 45\degr from the pole but the correction down to the CPD height would not be too large.
Assuming that these relative numbers are independent of the surface density and thus also apply to the fiducial run, we would obtain \MPktPdir or \MPktHa values only up to a factor $\approx20$ larger than what we found in the fiducial run (see Figure~\ref{fig:varrmax}b) if we let Phase~II begin from a different moment of Phase~I. At the other extreme, the partial \neuII{mass fluxes} could be orders of magnitude smaller than what we found.

The upshot of this estimate is that \neuIII{there are} transient oscillations
but \neuIII{they will not affect} the basic and crucial point that only a fraction, clearly below 100\,\%, of the gas \neuII{falling} onto the CPD can generate emission lines.

\section{Temperature and luminosity structure}
  \label{sec:T+L-Strukt}

\begin{figure}
 \centering
 \includegraphics[width=0.47\textwidth]{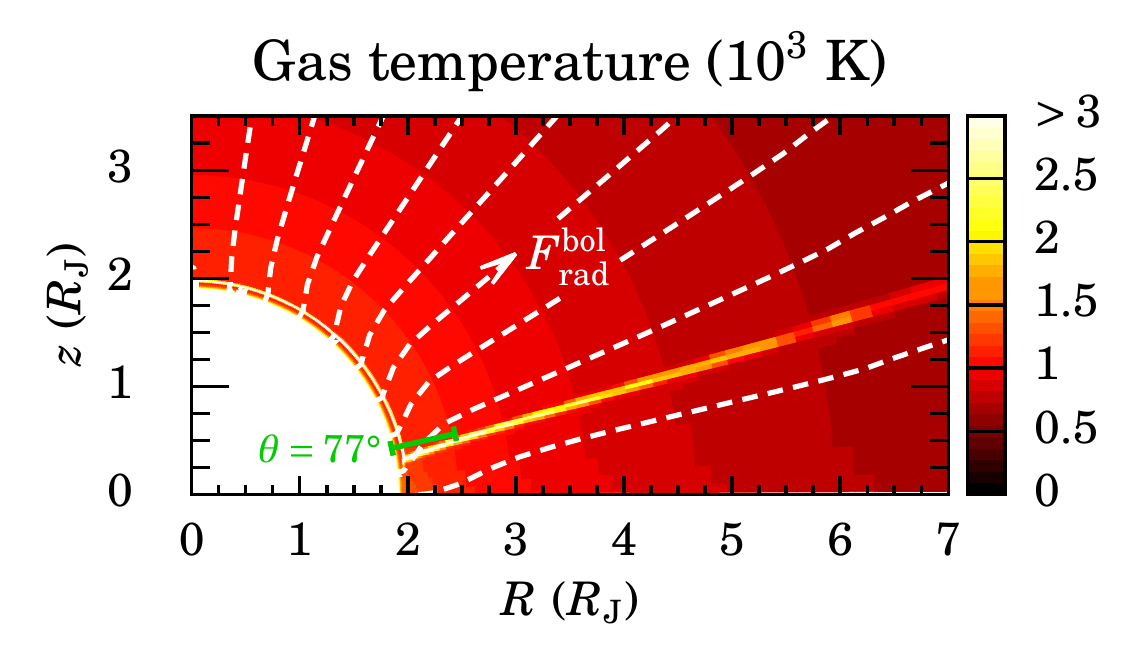}
\caption{%
Temperature close to the planet. The colour scale is capped at 3000~K but the Zel'dovich spikes---the extremely thin bright regions above the planetary surface and the CPD surface---as well as the innermost hydrostatic parts near \rmin reach much higher temperatures.
Streamlines for the bolometric radiation are shown (dashed white lines), whereas
hydrogen lines, including \Ha, originate from both shocks (see regions with a non-grey preshock velocity in Figure~\ref{fig:vnormSchock}).
The radial segment at $\theta=77$\degr (green) is analysed in Figure~\ref{fig:Schnitte}
and compared to a vertical segment at $R=2.5~\RJ$.
}
\label{fig:T2D}
\end{figure}

\begin{figure*} %
 \centering
 \includegraphics[width=0.7\textwidth]{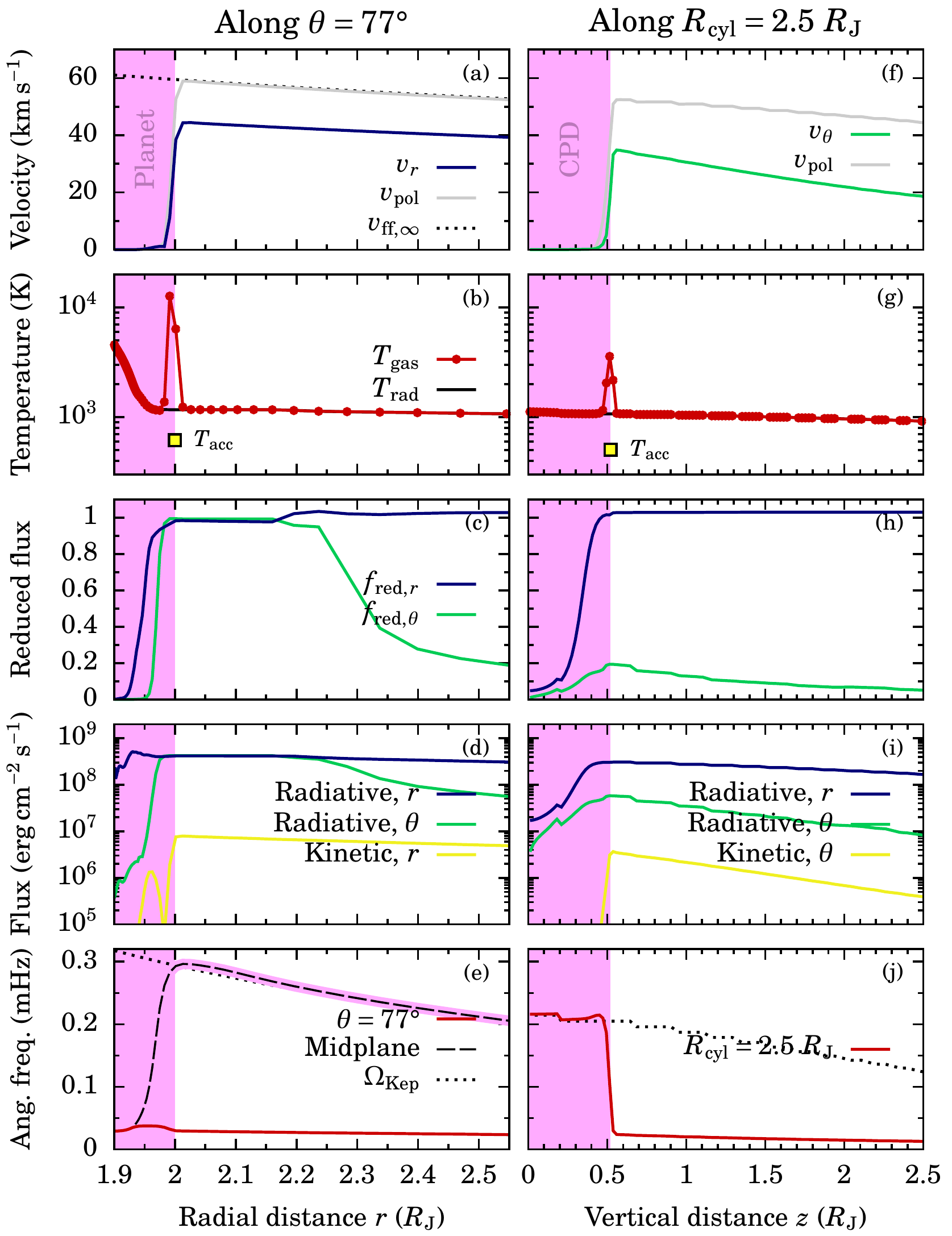}
\caption{%
One-dimensional cuts along a line of constant angle $\theta=77$\degr (left column; corresponding to the green line segment in Figure~\ref{fig:T2D}, just above the CPD surface)
and of constant cylindrical radius $R=2.5~\RJ$ (right column) in the $\Nth=181$ simulation.
In the pink regions, the gas is in hydrostatic equilibrium.
(a/f)~Radial $v_r$, total poloidal \vtr, and free-fall velocities \vFfinfty;
(b/g)~gas and radiation temperatures, showing in~(b) the radial zoning, and with \TAkk;
(c/h)~reduced flux in $r$ and in $\theta$;
(d/i)~bolometric radiative flux \Fradx and mechanical $\Fkinx = 0.5\rho |v_x|^3$ in direction $x=r$ or $x=\theta$, corresponding to \TAkk; and
(e/j)~angular frequency $\Omega=v_\phi/R$ compared to the Keplerian frequency \OKep.
Panel~(e) also shows $\Omega$ %
in the midplane (dashed), which is in (rotation-modified) hydrostatic equilibrium in $R$ and $z$ out to the CPD outer edge (not shown). The apparent steps in the right column are a plotting artefact. %
}
\label{fig:Schnitte}
\end{figure*}

Figure~\ref{fig:T2D} shows the temperature near the planet surface, and Figure~\ref{fig:Schnitte} the velocity, temperature, fluxes, and angular frequency along two cuts.
In Figure~\ref{fig:T2D}, the shocks on the planet and the CPD surfaces are clearly visible as Zel'dovich spikes (\citealp{zeldovich67}; see also discussion in \citetalias{m18Schock}).
Thanks to the small cell sizes (see Figure~\ref{fig:Gitter}), they reach respectively $T>25000$~K and $T>4000$~K, off the colourscale (capped at 3000~K), but this is resolution-dependent. The true physical peak temperature would be of order $10^4$--$10^5$~K \citep{aoyama18}. Fortunately, this need not be resolved to follow the radiation transfer correctly \citepalias{m18Schock}. The pre- and postshock temperatures, which set more directly the thermal structure of the accretion flow and the settling layers below the shock, are equal and near $T=1100$~K. %

Below the CPD surface shock, the temperature is nearly constant (except close to the midplane), which reflects the low opacity. Nevertheless, the polar reduced flux\footnote{The reduced flux, or ``streaming factor'' \citep{kley89a}, measures the extent to which radiation is diffusing ($\fred\rightarrow0$) or freely streaming ($\fred\rightarrow1$).},
\begin{equation}
 \fredth \equiv \frac{|\Fradth|}{c\Erad},
\end{equation}
where \Fradth is the radiation flux in the polar direction, %
is at most $\fredth\approx0.1$, while the radial reduced flux $\fredr$ goes smoothly from $\fredr=0.01$--0.1 near the midplane to $\fredr=1$ below, at, and above the CPD surface shock.
Thus the radiation diffuses in the polar direction while also diffusing radially (below the CPD shock) or flowing freely (above it).

The temperature at the shock on the planet surface is $\TSch\approx1150$~K. This however is set mostly by the luminosity below the shock coming from the compression of the gas. Namely, the free-streaming ``accretion temperature'' for an $\etakin=100$\,\% shock efficiency \citepalias{m16Schock}, given by
\begin{equation}
 \label{eq:TAkk}
  \sigSB \TAkk^4 = \Fkinr = \frac{1}{2} \rho {|v_r|}^3,
\end{equation}
where \sigSB is the Stefan--Boltzmann constant,
is only $\TAkk=615$~K at $\theta=77$\degr\xspace or $\TAkk=715$~K at the pole. In both cases this is much smaller than \TSch.
(This is the limit $\ell=1$ of Equation~(33) in \citetalias{m18Schock}, while here $\ell\gg1$ since the downstream luminosity dominates.)
In the classical assumption of pure radial infall, the direct-infall \MPktPdir would be predicted to lead to an accretion temperature
$\TAkkP=(G\MP\MPktPdir/[4\pi\RP^3\sigSB])^{1/4} = 684$~K, ignoring here a factor $\ffill/\zeta$ \citep{zhu15}, of order unity. As it should, \TAkkP lies between the pole and equator values for \TAkk.
However, the pendant to this (from a global-simulation point of view)
is implicitly to assume that the entire mass flux \MPktnettoHill shocks on the planetary surface, leading to 
$\TAkkklass=2665$~K, which would dominate the interior luminosity.
Neither this radiation temperature \TAkkklass nor the corresponding gas temperature in the free-streaming limit $T=\TAkkklass/4^{1/4}$ have any relevance in describing the system: the gas falls in more slowly and spread over a much larger area than assumed by the formula.

On the surface of the CPD at $R=2.5~\RJ$, the temperature is $T=1070$~K, with the actual $\Tacc=505$~K again much smaller in terms of the radiation fluxes $F\propto T^4$ (the gas and radiation temperatures are equal).
Thus also for the CPD, it is the interior luminosity, not the kinetic energy of the gas, that is responsible for setting the temperature.

In the midplane there is no shock at the planet surface. Instead, the planet and CPD are connected by a boundary layer (e.g., \citealp{hertfelder17,dong21}) in which the angular velocity in the midplane peaks, somewhat above the Keplerian value $\OKep=\sqrt{G\MP/r^3}$, before decreasing smoothly to join the boundary condition at \rmin (Figure~\ref{fig:Schnitte}e). This region will not be studied further here. At least at $R=2.5~\RJ$, the whole vertical extent of the CPD is in Keplerian rotation: in the pink regions in Figure~\ref{fig:Schnitte}j, $\Omega=\OKep$.
The boundary layer leads to a higher temperature close to the midplane but only slightly so.

Away from the CPD (for $r\gtrsim100~\RJ$), the temperature distribution is independent of polar angle, with temperatures below 100~K. In that regime, the dust opacity $\kapStbfpgRoss\approx0.01~\kapE$ dominates by 3--4~dex over the gas opacity even for our choice of a low $\fpg=10^{-4}$.
The radial Rosseland optical depth from \rmax to the shock is $\DtauR\sim3\times10^{-3}$ along the pole, roughly a factor of two higher on a path just above the CPD, and $\DtauR=10^{-3}$ in the midplane down to the outer edge of the CPD. The low overall optical depth reflects the low \fpg and modest mass inflow into the Hill sphere $\MPktnettoHill\sim10^{-5}~\MPktEJ$ (Table~\ref{tab:par}).

In this particular example, most of the bolometric flux reaching the observer is coming from the interior of the planet and from the CPD itself. These fluxes
do not come from the immediate conversion of kinetic energy but rather from the cooling of the hydrostatic regions below the shocks. This in turn depends on the accretion history. In a given simulation, this history is set by the numerical approach (here, the two-phase system we used, which spans several free-fall timescales), and in general by the variation of the accretion rate over formation timescales of order 1~Myr.

In Section~\ref{sec:discPDS}, we compared the \Ha flux that we predict to the observed one for \PDSb.
Assuming, roughly, a linear scaling of the \Ha luminosity with the mass inflow rate,
the latter would need to be 7--100 times larger than in our simulation
in order to match the observed \LHa. From Equation~(\ref{eq:TAkk}),
this would imply $\Tacc\approx1000$--1900~K at the planet's surface near the CPD,
and up to $\Tacc\approx2300$~K at the pole.
In this case, the accretion luminosity from the shock would likely dominate the temperature structure,
and the highest accretion rates would be more challenging to reconcile with the constraints
from \citet{wang21vlti} on \Teff from the $K$-band spectral shape.
Details such as the viewing geometry or complex radiation-transfer effects
could however play an important role.
Next-generation spectroscopic observations would be helpful to develop a robust and self-consistent picture.
\refstepcounter{numKommG}

\bibliography{std}{}

\begin{thebibliography}{}
\providecommand\natexlab[1]{#1}
\providecommand\JournalTitle[1]{#1}

\bibitem[{{Adams} \& {Batygin}(2022)}]{ab22}
{Adams}, F.~C., \& {Batygin}, K. 2022,
  \href{http://dx.doi.org/10.3847/1538-4357/ac7a3e}{\JournalTitle{\apj}, 934,
  111}

\bibitem[{{Aoyama} {et~al.}(2018){Aoyama}, {Ikoma}, \& {Tanigawa}}]{aoyama18}
{Aoyama}, Y., {Ikoma}, M., \& {Tanigawa}, T. 2018,
  \href{http://dx.doi.org/10.3847/1538-4357/aadc11}{\JournalTitle{\apj}, 866,
  84}

\bibitem[{{Aoyama} {et~al.}(2021){Aoyama}, {Marleau}, {Ikoma}, \&
  {Mordasini}}]{AMIM21L}
{Aoyama}, Y., {Marleau}, G.-D., {Ikoma}, M., \& {Mordasini}, C. 2021,
  \href{http://dx.doi.org/10.3847/2041-8213/ac19bd}{\JournalTitle{\apjl}, 917,
  L30}

\bibitem[{{Aoyama} {et~al.}(2020){Aoyama}, {Marleau}, {Mordasini}, \&
  {Ikoma}}]{Aoyama+2020}
{Aoyama}, Y., {Marleau}, G.-D., {Mordasini}, C., \& {Ikoma}, M. 2020,
  \href{http://dx.doi.org/10.48550/arxiv.2011.06608}{\JournalTitle{arXiv
  e-prints}, arXiv:2011.06608}

\bibitem[{{Asensio-Torres} {et~al.}(2021){Asensio-Torres}, {Henning},
  {Cantalloube}, {Pinilla}, {Mesa}, {Garufi}, {Jorquera}, {Gratton}, {Chauvin},
  {Szul{\'a}gyi}, {van Boekel}, {Dong}, {Marleau}, {Benisty}, {Villenave},
  {Bergez-Casalou}, {Desgrange}, {Janson}, {Keppler}, {Langlois}, {M{\'e}nard},
  {Rickman}, {Stolker}, {Feldt}, {Fusco}, {Gluck}, {Pavlov}, \&
  {Ramos}}]{asensiotorres21}
{Asensio-Torres}, R., {Henning}, T., {Cantalloube}, F., {et~al.} 2021,
  \href{http://dx.doi.org/10.1051/0004-6361/202140325}{\JournalTitle{\aap},
  652, A101}

\bibitem[{{Ayliffe} \& {Bate}(2009{\natexlab{a}})}]{ab09a}
{Ayliffe}, B.~A., \& {Bate}, M.~R. 2009{\natexlab{a}},
  \href{http://dx.doi.org/10.1111/j.1365-2966.2008.14184.x}{\JournalTitle{\mnras},
  393, 49}

\bibitem[{{Ayliffe} \& {Bate}(2009{\natexlab{b}})}]{ab09b}
{Ayliffe}, B.~A., \& {Bate}, M.~R. 2009{\natexlab{b}},
  \href{http://dx.doi.org/10.1111/j.1365-2966.2009.15002.x}{\JournalTitle{\mnras},
  397, 657}

\bibitem[{{Ayliffe} \& {Bate}(2012)}]{ayliffe12}
{Ayliffe}, B.~A., \& {Bate}, M.~R. 2012,
  \href{http://dx.doi.org/10.1111/j.1365-2966.2012.21979.x}{\JournalTitle{\mnras},
  427, 2597}

\bibitem[{{Bae} {et~al.}(2019){Bae}, {Zhu}, {Baruteau}, {Benisty}, {Dullemond},
  {Facchini}, {Isella}, {Keppler}, {P{\'e}rez}, \& {Teague}}]{bae19}
{Bae}, J., {Zhu}, Z., {Baruteau}, C., {et~al.} 2019,
  \href{http://dx.doi.org/10.3847/2041-8213/ab46b0}{\JournalTitle{\apjl}, 884,
  L41}

\bibitem[{{Bailey} {et~al.}(2021){Bailey}, {Stone}, \& {Fung}}]{bailey21}
{Bailey}, A., {Stone}, J.~M., \& {Fung}, J. 2021,
  \href{http://dx.doi.org/10.3847/1538-4357/ac033b}{\JournalTitle{\apj}, 915,
  113}

\bibitem[{{Batygin}(2018)}]{batygin18}
{Batygin}, K. 2018,
  \href{http://dx.doi.org/10.3847/1538-3881/aab54e}{\JournalTitle{\aj}, 155,
  178}

\bibitem[{{Berardo} {et~al.}(2017){Berardo}, {Cumming}, \&
  {Marleau}}]{berardo17}
{Berardo}, D., {Cumming}, A., \& {Marleau}, G.-D. 2017,
  \href{http://dx.doi.org/10.3847/1538-4357/834/2/149}{\JournalTitle{\apj},
  834, 149}

\bibitem[{{B{\'e}thune} \& {Rafikov}(2019{\natexlab{a}})}]{b19a}
{B{\'e}thune}, W., \& {Rafikov}, R.~R. 2019{\natexlab{a}},
  \href{http://dx.doi.org/10.1093/mnras/stz1427}{\JournalTitle{\mnras}, 487,
  2319}

\bibitem[{{B{\'e}thune} \& {Rafikov}(2019{\natexlab{b}})}]{b19b}
{B{\'e}thune}, W., \& {Rafikov}, R.~R. 2019{\natexlab{b}},
  \href{http://dx.doi.org/10.1093/mnras/stz1870}{\JournalTitle{\mnras}, 488,
  2365}

\bibitem[{{Betti} {et~al.}(2022{\natexlab{a}}){Betti}, {Follette},
  {Ward-Duong}, {Aoyama}, {Marleau}, {Bary}, {Robinson}, {Janson}, {Balmer},
  {Chauvin}, \& {Palma-Bifani}}]{betti22b}
{Betti}, S.~K., {Follette}, K.~B., {Ward-Duong}, K., {et~al.}
  2022{\natexlab{a}},
  \href{http://dx.doi.org/10.3847/2041-8213/ac85ef}{\JournalTitle{\apjl}, 935,
  L18}

\bibitem[{{Betti} {et~al.}(2022{\natexlab{b}}){Betti}, {Follette},
  {Ward-Duong}, {Aoyama}, {Marleau}, {Bary}, {Robinson}, {Janson}, {Balmer},
  {Chauvin}, \& {Palma-Bifani}}]{betti22c}
{Betti}, S.~K., {Follette}, K.~B., {Ward-Duong}, K., {et~al.}
  2022{\natexlab{b}},
  \href{http://dx.doi.org/10.3847/2041-8213/aca331}{\JournalTitle{\apjl}, 941,
  L20}

\bibitem[{{Bodenheimer} {et~al.}(2000){Bodenheimer}, {Hubickyj}, \&
  {Lissauer}}]{boden00}
{Bodenheimer}, P., {Hubickyj}, O., \& {Lissauer}, J.~J. 2000,
  \href{http://dx.doi.org/10.1006/icar.1999.6246}{\JournalTitle{\icarus}, 143,
  2}

\bibitem[{{Bonse} {et~al.}(2023){Bonse}, {Garvin}, {Gebhard}, {Dannert},
  {Cantalloube}, {Cugno}, {Absil}, {Hayoz}, {Milli}, {Kasper}, \&
  {Quanz}}]{bonse23}
{Bonse}, M.~J., {Garvin}, E.~O., {Gebhard}, T.~D., {et~al.} 2023,
  \href{http://dx.doi.org/10.48550/arXiv.2303.12030}{\JournalTitle{arXiv
  e-prints}, arXiv:2303.12030}

\bibitem[{{Brittain} {et~al.}(2020){Brittain}, {Najita}, {Dong}, \&
  {Zhu}}]{brittain20}
{Brittain}, S.~D., {Najita}, J.~R., {Dong}, R., \& {Zhu}, Z. 2020,
  \href{http://dx.doi.org/10.3847/1538-4357/ab8388}{\JournalTitle{\apj}, 895,
  48}

\bibitem[{{Bryan} {et~al.}(2018){Bryan}, {Benneke}, {Knutson}, {Batygin}, \&
  {Bowler}}]{bryan18}
{Bryan}, M.~L., {Benneke}, B., {Knutson}, H.~A., {Batygin}, K., \& {Bowler},
  B.~P. 2018,
  \href{http://dx.doi.org/10.1038/s41550-017-0325-8}{\JournalTitle{\natas}, 2,
  138}

\bibitem[{{Bryan} {et~al.}(2020){Bryan}, {Ginzburg}, {Chiang}, {Morley},
  {Bowler}, {Xuan}, \& {Knutson}}]{bryan20}
{Bryan}, M.~L., {Ginzburg}, S., {Chiang}, E., {et~al.} 2020,
  \href{http://dx.doi.org/10.3847/1538-4357/abc0ef}{\JournalTitle{\apj}, 905,
  37}

\bibitem[{{Chachan} {et~al.}(2021){Chachan}, {Lee}, \& {Knutson}}]{chachan21}
{Chachan}, Y., {Lee}, E.~J., \& {Knutson}, H.~A. 2021,
  \href{http://dx.doi.org/10.3847/1538-4357/ac0bb6}{\JournalTitle{\apj}, 919,
  63}

\bibitem[{{Chazelas} {et~al.}(2020){Chazelas}, {Lovis}, {Blind}, {K{\"u}hn},
  {Genolet}, {Hughes}, {Turbet}, {Hagelberg}, {Restori}, {Kasper}, \& {Cerpa
  Urra}}]{chazelas20}
{Chazelas}, B., {Lovis}, C., {Blind}, N., {et~al.} 2020,
  \href{http://dx.doi.org/10.1117/12.2576316}{in Society of Photo-Optical
  Instrumentation Engineers (SPIE) Conference Series, Vol. 11448, Society of
  Photo-Optical Instrumentation Engineers (SPIE) Conference Series}, 1144875

\bibitem[{{Chen} \& {Bai}(2022)}]{chenbai22}
{Chen}, Z., \& {Bai}, X. 2022,
  \href{http://dx.doi.org/10.3847/2041-8213/ac4ca9}{\JournalTitle{\apjl}, 925,
  L14}

\bibitem[{{Choksi} {et~al.}(2023){Choksi}, {Chiang}, {Fung}, \&
  {Zhu}}]{choksi23}
{Choksi}, N., {Chiang}, E., {Fung}, J., \& {Zhu}, Z. 2023,
  \href{http://dx.doi.org/10.48550/arXiv.2305.01684}{\JournalTitle{arXiv
  e-prints}, arXiv:2305.01684}

\bibitem[{{Christensen} {et~al.}(2009){Christensen}, {Holzwarth}, \&
  {Reiners}}]{christensen09}
{Christensen}, U.~R., {Holzwarth}, V., \& {Reiners}, A. 2009,
  \href{http://dx.doi.org/10.1038/nature07626}{\JournalTitle{\nat}, 457, 167}

\bibitem[{{Cimerman} {et~al.}(2017){Cimerman}, {Kuiper}, \&
  {Ormel}}]{cimerman17}
{Cimerman}, N.~P., {Kuiper}, R., \& {Ormel}, C.~W. 2017,
  \href{http://dx.doi.org/10.1093/mnras/stx1924}{\JournalTitle{\mnras}, 471,
  4662}

\bibitem[{{Close}(2020)}]{close20}
{Close}, L.~M. 2020,
  \href{http://dx.doi.org/10.3847/1538-3881/abb375}{\JournalTitle{\aj}, 160,
  221}

\bibitem[{{Cugno} {et~al.}(2019){Cugno}, {Quanz}, {Hunziker}, {Stolker},
  {Schmid}, {Avenhaus}, {Baudoz}, {Bohn}, {Bonnefoy}, {Buenzli}, {Chauvin},
  {Cheetham}, {Desidera}, {Dominik}, {Feautrier}, {Feldt}, {Ginski}, {Girard},
  {Gratton}, {Hagelberg}, {Hugot}, {Janson}, {Lagrange}, {Langlois}, {Magnard},
  {Maire}, {Menard}, {Meyer}, {Milli}, {Mordasini}, {Pinte}, {Pragt},
  {Roelfsema}, {Rigal}, {Szul{\'a}gyi}, {van Boekel}, {van der Plas}, {Vigan},
  {Wahhaj}, \& {Zurlo}}]{Cugno+2019}
{Cugno}, G., {Quanz}, S.~P., {Hunziker}, S., {et~al.} 2019,
  \href{http://dx.doi.org/10.1051/0004-6361/201834170}{\JournalTitle{\aap},
  622, A156}

\bibitem[{{Currie} {et~al.}(2022){Currie}, {Lawson}, {Schneider}, {Lyra},
  {Wisniewski}, {Grady}, {Guyon}, {Tamura}, {Kotani}, {Kawahara}, {Brandt},
  {Uyama}, {Muto}, {Dong}, {Kudo}, {Hashimoto}, {Fukagawa}, {Wagner}, {Lozi},
  {Chilcote}, {Tobin}, {Groff}, {Ward-Duong}, {Januszewski}, {Norris},
  {Tuthill}, {van der Marel}, {Sitko}, {Deo}, {Vievard}, {Jovanovic},
  {Martinache}, \& {Skaf}}]{currie22}
{Currie}, T., {Lawson}, K., {Schneider}, G., {et~al.} 2022,
  \href{http://dx.doi.org/10.1038/s41550-022-01634-x}{\JournalTitle{\natas}, 6,
  751}

\bibitem[{{Delorme} {et~al.}(2021){Delorme}, {Jovanovic}, {Echeverri}, {Mawet},
  {Kent Wallace}, {Bartos}, {Cetre}, {Wizinowich}, {Ragland}, {Lilley},
  {Wetherell}, {Doppmann}, {Wang}, {Morris}, {Ruffio}, {Martin}, {Fitzgerald},
  {Ruane}, {Schofield}, {Suominen}, {Calvin}, {Wang}, {Magnone}, {Johnson},
  {Sohn}, {L{\'o}pez}, {Bond}, {Pezzato}, {Sayson}, {Chun}, \&
  {Skemer}}]{delorme21}
{Delorme}, J.-R., {Jovanovic}, N., {Echeverri}, D., {et~al.} 2021,
  \href{http://dx.doi.org/10.1117/1.JATIS.7.3.035006}{\JournalTitle{Journal of
  Astronomical Telescopes, Instruments, and Systems}, 7, 035006}

\bibitem[{{Dong} {et~al.}(2021){Dong}, {Jiang}, \& {Armitage}}]{dong21}
{Dong}, J., {Jiang}, Y.-F., \& {Armitage}, P.~J. 2021,
  \href{http://dx.doi.org/10.3847/1538-4357/ac1941}{\JournalTitle{\apj}, 921,
  54}

\bibitem[{{Dong} {et~al.}(in prep.){Dong}, {Hashimoto}, {Haffert}, {Aoyama},
  {Xie}, \& {Marleau}}]{dong23}
{Dong}, R., {Hashimoto}, J., {Haffert}, S., {et~al.} in prep.,
  \JournalTitle{\apj}

\bibitem[{{Dr{\k{a}}{\.z}kowska} {et~al.}(2019){Dr{\k{a}}{\.z}kowska}, {Li},
  {Birnstiel}, {Stammler}, \& {Li}}]{dr19}
{Dr{\k{a}}{\.z}kowska}, J., {Li}, S., {Birnstiel}, T., {Stammler}, S.~M., \&
  {Li}, H. 2019,
  \href{http://dx.doi.org/10.3847/1538-4357/ab46b7}{\JournalTitle{\apj}, 885,
  91}

\bibitem[{{Emsenhuber} {et~al.}(2021){Emsenhuber}, {Mordasini}, {Burn},
  {Alibert}, {Benz}, \& {Asphaug}}]{emsen21a}
{Emsenhuber}, A., {Mordasini}, C., {Burn}, R., {et~al.} 2021,
  \href{http://dx.doi.org/10.1051/0004-6361/202038553}{\JournalTitle{\aap},
  656, A69}

\bibitem[{{Eriksson} {et~al.}(2020){Eriksson}, {Asensio Torres}, {Janson},
  {Aoyama}, {Marleau}, {Bonnefoy}, \& {Petrus}}]{eriksson20}
{Eriksson}, S.~C., {Asensio Torres}, R., {Janson}, M., {et~al.} 2020,
  \href{http://dx.doi.org/10.1051/0004-6361/202038131}{\JournalTitle{\aap},
  638, L6}

\bibitem[{{Follette} {et~al.}(2023){Follette}, {Close}, {Males}, {Ward-Duong},
  {Balmer}, {Redai}, {Morales}, {Sarosi}, {Dacus}, {De Rosa}, {Garcia Toro},
  {Leonard}, {Macintosh}, {Morzinski}, {Mullen}, {Palmo}, {Saitoti}, {Spiro},
  {Treiber}, {Wagner}, {Wang}, {Wang}, {Watson}, \& {Weinberger}}]{follette23}
{Follette}, K.~B., {Close}, L.~M., {Males}, J.~R., {et~al.} 2023,
  \href{http://dx.doi.org/10.3847/1538-3881/acc183}{\JournalTitle{\aj}, 165,
  225}

\bibitem[{{Fu} {et~al.}(2023){Fu}, {Huang}, \& {Yu}}]{fu23}
{Fu}, Z., {Huang}, S., \& {Yu}, C. 2023,
  \href{http://dx.doi.org/10.3847/1538-4357/acac9c}{\JournalTitle{\apj}, 945,
  165}

\bibitem[{{Fung} {et~al.}(2019){Fung}, {Zhu}, \& {Chiang}}]{fung19}
{Fung}, J., {Zhu}, Z., \& {Chiang}, E. 2019,
  \href{http://dx.doi.org/10.3847/1538-4357/ab53da}{\JournalTitle{\apj}, 887,
  152}

\bibitem[{{Goldreich} \& {Lynden-Bell}(1965)}]{gold65}
{Goldreich}, P., \& {Lynden-Bell}, D. 1965,
  \href{http://dx.doi.org/10.1093/mnras/130.2.125}{\JournalTitle{\mnras}, 130,
  125}

\bibitem[{{Haffert} {et~al.}(2019){Haffert}, {Bohn}, {de Boer}, {Snellen},
  {Brinchmann}, {Girard}, {Keller}, \& {Bacon}}]{Haffert+2019}
{Haffert}, S.~Y., {Bohn}, A.~J., {de Boer}, J., {et~al.} 2019,
  \href{http://dx.doi.org/10.1038/s41550-019-0780-5}{\JournalTitle{\natas}, 3,
  749}

\bibitem[{{Haffert} {et~al.}(2021){Haffert}, {Males}, {Close}, {Long},
  {Schatz}, {van Gorkom}, {Hedglen}, {Lumbres}, {Rodack}, {Guyon}, {Knight},
  {Kautz}, \& {Pearce}}]{haffert21visx}
{Haffert}, S.~Y., {Males}, J.~R., {Close}, L., {et~al.} 2021,
  \href{http://dx.doi.org/10.1117/12.2594875}{in Society of Photo-Optical
  Instrumentation Engineers (SPIE) Conference Series, Vol. 11823, Techniques
  and Instrumentation for Detection of Exoplanets X, ed. S.~B. {Shaklan} \&
  G.~J. {Ruane}}, 1182306

\bibitem[{{Hartmann} {et~al.}(1997){Hartmann}, {Cassen}, \&
  {Kenyon}}]{hartmann97}
{Hartmann}, L., {Cassen}, P., \& {Kenyon}, S.~J. 1997,
  \href{http://dx.doi.org/10.1086/303547}{\JournalTitle{\apj}, 475, 770}

\bibitem[{{Hartmann} {et~al.}(2016){Hartmann}, {Herczeg}, \&
  {Calvet}}]{hartmann16}
{Hartmann}, L., {Herczeg}, G., \& {Calvet}, N. 2016,
  \href{http://dx.doi.org/10.1146/annurev-astro-081915-023347}{\JournalTitle{\araa},
  54, 135}

\bibitem[{{Hasegawa} {et~al.}(2021){Hasegawa}, {Kanagawa}, \&
  {Turner}}]{hasegawa21}
{Hasegawa}, Y., {Kanagawa}, K.~D., \& {Turner}, N.~J. 2021,
  \href{http://dx.doi.org/10.3847/1538-4357/ac257b}{\JournalTitle{\apj}, 923,
  27}

\bibitem[{{Hertfelder} \& {Kley}(2017)}]{hertfelder17}
{Hertfelder}, M., \& {Kley}, W. 2017,
  \href{http://dx.doi.org/10.1051/0004-6361/201730847}{\JournalTitle{\aap},
  605, A24}

\bibitem[{{Hill}(1878)}]{hill78}
{Hill}, G.~W. 1878,
  \href{http://dx.doi.org/10.2307/2369430}{\JournalTitle{\amjm}, 1, 5}

\bibitem[{{Hu{\'e}lamo} {et~al.}(2022){Hu{\'e}lamo}, {Chauvin},
  {Mendigut{\'\i}a}, {Whelan}, {Alcal{\'a}}, {Cugno}, {Schmid}, {de
  Gregorio-Monsalvo}, {Zurlo}, {Barrado}, {Benisty}, {Quanz}, {Bouy},
  {Montesinos}, {Beletsky}, \& {Szulagyi}}]{hu22}
{Hu{\'e}lamo}, N., {Chauvin}, G., {Mendigut{\'\i}a}, I., {et~al.} 2022,
  \href{http://dx.doi.org/10.1051/0004-6361/202243918}{\JournalTitle{\aap},
  668, A138}

\bibitem[{{Isella} \& {Natta}(2005)}]{isella05}
{Isella}, A., \& {Natta}, A. 2005,
  \href{http://dx.doi.org/10.1051/0004-6361:20052773}{\JournalTitle{\aap}, 438,
  899}

\bibitem[{{Kanagawa} {et~al.}(2017){Kanagawa}, {Tanaka}, {Muto}, \&
  {Tanigawa}}]{kanagawa17}
{Kanagawa}, K.~D., {Tanaka}, H., {Muto}, T., \& {Tanigawa}, T. 2017,
  \href{http://dx.doi.org/10.1093/pasj/psx114}{\JournalTitle{\pasj}, 69, 97}

\bibitem[{{Kanagawa} {et~al.}(2018){Kanagawa}, {Tanaka}, \&
  {Szuszkiewicz}}]{kanagawa18a}
{Kanagawa}, K.~D., {Tanaka}, H., \& {Szuszkiewicz}, E. 2018,
  \href{http://dx.doi.org/10.3847/1538-4357/aac8d9}{\JournalTitle{\apj}, 861,
  140}

\bibitem[{{Karlin} {et~al.}(2023){Karlin}, {Pani{\'c}}, \& {van
  Loo}}]{karlin23}
{Karlin}, S.~M., {Pani{\'c}}, O., \& {van Loo}, S. 2023,
  \href{http://dx.doi.org/10.1093/mnras/stad157}{\JournalTitle{\mnras}, 520,
  1258}

\bibitem[{{Katarzy{\'n}ski} {et~al.}(2016){Katarzy{\'n}ski}, {Gawro{\'n}ski},
  \& {Go{\'z}dziewski}}]{katarzy16}
{Katarzy{\'n}ski}, K., {Gawro{\'n}ski}, M., \& {Go{\'z}dziewski}, K. 2016,
  \href{http://dx.doi.org/10.1093/mnras/stw1354}{\JournalTitle{\mnras}, 461,
  929}

\bibitem[{{Keith} \& {Wardle}(2014)}]{keith14}
{Keith}, S.~L., \& {Wardle}, M. 2014,
  \href{http://dx.doi.org/10.1093/mnras/stu245}{\JournalTitle{\mnras}, 440, 89}

\bibitem[{{Keppler} {et~al.}(2018){Keppler}, {Benisty}, {M{\"u}ller},
  {Henning}, {van Boekel}, {Cantalloube}, {Ginski}, {van Holstein}, {Maire},
  {Pohl}, {Samland}, {Avenhaus}, {Baudino}, {Boccaletti}, {de Boer},
  {Bonnefoy}, {Chauvin}, {Desidera}, {Langlois}, {Lazzoni}, {Marleau},
  {Mordasini}, {Pawellek}, {Stolker}, {Vigan}, {Zurlo}, {Birnstiel},
  {Brandner}, {Feldt}, {Flock}, {Girard}, {Gratton}, {Hagelberg}, {Isella},
  {Janson}, {Juhasz}, {Kemmer}, {Kral}, {Lagrange}, {Launhardt}, {Matter},
  {M{\'e}nard}, {Milli}, {Molli{\`e}re}, {Olofsson}, {P{\'e}rez}, {Pinilla},
  {Pinte}, {Quanz}, {Schmidt}, {Udry}, {Wahhaj}, {Williams}, {Buenzli},
  {Cudel}, {Dominik}, {Galicher}, {Kasper}, {Lannier}, {Mesa}, {Mouillet},
  {Peretti}, {Perrot}, {Salter}, {Sissa}, {Wildi}, {Abe}, {Antichi},
  {Augereau}, {Baruffolo}, {Baudoz}, {Bazzon}, {Beuzit}, {Blanchard}, {Brems},
  {Buey}, {De Caprio}, {Carbillet}, {Carle}, {Cascone}, {Cheetham}, {Claudi},
  {Costille}, {Delboulb{\'e}}, {Dohlen}, {Fantinel}, {Feautrier}, {Fusco},
  {Giro}, {Gluck}, {Gry}, {Hubin}, {Hugot}, {Jaquet}, {Le Mignant}, {Llored},
  {Madec}, {Magnard}, {Martinez}, {Maurel}, {Meyer}, {M{\"o}ller-Nilsson},
  {Moulin}, {Mugnier}, {Orign{\'e}}, {Pavlov}, {Perret}, {Petit}, {Pragt},
  {Puget}, {Rabou}, {Ramos}, {Rigal}, {Rochat}, {Roelfsema}, {Rousset}, {Roux},
  {Salasnich}, {Sauvage}, {Sevin}, {Soenke}, {Stadler}, {Suarez}, {Turatto}, \&
  {Weber}}]{keppler18}
{Keppler}, M., {Benisty}, M., {M{\"u}ller}, A., {et~al.} 2018,
  \href{http://dx.doi.org/10.1051/0004-6361/201832957}{\JournalTitle{\aap},
  617, A44}

\bibitem[{{Kley}(1989)}]{kley89a}
{Kley}, W. 1989, \JournalTitle{\aap}, 208, 98

\bibitem[{{Kley}(1998)}]{kley98}
{Kley}, W. 1998, \JournalTitle{\aap}, 338, L37

\bibitem[{{Korycansky} \& {Papaloizou}(1996)}]{korycansky96}
{Korycansky}, D.~G., \& {Papaloizou}, J.~C.~B. 1996,
  \href{http://dx.doi.org/10.1086/192311}{\JournalTitle{\apjs}, 105, 181}

\bibitem[{{Krapp} {et~al.}(2022){Krapp}, {Kratter}, \& {Youdin}}]{krapp21}
{Krapp}, L., {Kratter}, K.~M., \& {Youdin}, A.~N. 2022,
  \href{http://dx.doi.org/10.3847/1538-4357/ac5899}{\JournalTitle{\apj}, 928,
  156}

\bibitem[{{Kuiper} {et~al.}(2010){Kuiper}, {Klahr}, {Dullemond}, {Kley}, \&
  {Henning}}]{kuiper10}
{Kuiper}, R., {Klahr}, H., {Dullemond}, C., {Kley}, W., \& {Henning}, T. 2010,
  \href{http://dx.doi.org/10.1051/0004-6361/200912355}{\JournalTitle{\aap},
  511, A81}

\bibitem[{{Kuiper} {et~al.}(2020){Kuiper}, {Yorke}, \& {Mignone}}]{kuiper20}
{Kuiper}, R., {Yorke}, H.~W., \& {Mignone}, A. 2020,
  \href{http://dx.doi.org/10.3847/1538-4365/ab9a36}{\JournalTitle{\apjs}, 250,
  13}

\bibitem[{{Kurokawa} \& {Tanigawa}(2018)}]{kurokawa18}
{Kurokawa}, H., \& {Tanigawa}, T. 2018,
  \href{http://dx.doi.org/10.1093/mnras/sty1498}{\JournalTitle{\mnras}, 479,
  635}

\bibitem[{{Lambrechts} \& {Lega}(2017)}]{lambrechts17}
{Lambrechts}, M., \& {Lega}, E. 2017,
  \href{http://dx.doi.org/10.1051/0004-6361/201731014}{\JournalTitle{\aap},
  606, A146}

\bibitem[{{Lambrechts} {et~al.}(2019){Lambrechts}, {Lega}, {Nelson}, {Crida},
  \& {Morbidelli}}]{lambrechts19}
{Lambrechts}, M., {Lega}, E., {Nelson}, R.~P., {Crida}, A., \& {Morbidelli}, A.
  2019,
  \href{http://dx.doi.org/10.1051/0004-6361/201834413}{\JournalTitle{\aap},
  630, A82}

\bibitem[{{Lovelace} {et~al.}(2011){Lovelace}, {Covey}, \&
  {Lloyd}}]{lovelace11}
{Lovelace}, R.~V.~E., {Covey}, K.~R., \& {Lloyd}, J.~P. 2011,
  \href{http://dx.doi.org/10.1088/0004-6256/141/2/51}{\JournalTitle{\aj}, 141,
  51}

\bibitem[{{Machida} {et~al.}(2008){Machida}, {Kokubo}, {Inutsuka}, \&
  {Matsumoto}}]{machida08}
{Machida}, M.~N., {Kokubo}, E., {Inutsuka}, S.-i., \& {Matsumoto}, T. 2008,
  \href{http://dx.doi.org/10.1086/590421}{\JournalTitle{\apj}, 685, 1220}

\bibitem[{{Maeda} {et~al.}(2022){Maeda}, {Ohtsuki}, {Tanigawa}, {Machida}, \&
  {Suetsugu}}]{maeda22}
{Maeda}, N., {Ohtsuki}, K., {Tanigawa}, T., {Machida}, M.~N., \& {Suetsugu}, R.
  2022, \href{http://dx.doi.org/10.3847/1538-4357/ac7ddf}{\JournalTitle{\apj},
  935, 56}

\bibitem[{{Mai} {et~al.}(2020){Mai}, {Desch}, {Kuiper}, {Marleau}, \&
  {Dullemond}}]{mai20}
{Mai}, C., {Desch}, S.~J., {Kuiper}, R., {Marleau}, G.-D., \& {Dullemond}, C.
  2020, \href{http://dx.doi.org/10.3847/1538-4357/aba4a8}{\JournalTitle{\apj},
  899, 54}

\bibitem[{{Malygin} {et~al.}(2014){Malygin}, {Kuiper}, {Klahr}, {Dullemond}, \&
  {Henning}}]{malygin14}
{Malygin}, M.~G., {Kuiper}, R., {Klahr}, H., {Dullemond}, C.~P., \& {Henning},
  T. 2014,
  \href{http://dx.doi.org/10.1051/0004-6361/201423768}{\JournalTitle{\aap},
  568, A91}

\bibitem[{{Marleau} {et~al.}(2017){Marleau}, {Klahr}, {Kuiper}, \&
  {Mordasini}}]{m16Schock}
{Marleau}, G.-D., {Klahr}, H., {Kuiper}, R., \& {Mordasini}, C. 2017,
  \href{http://dx.doi.org/10.3847/1538-4357/836/2/221}{\JournalTitle{\apj},
  836, 221}

\bibitem[{{Marleau} {et~al.}(2019){Marleau}, {Mordasini}, \&
  {Kuiper}}]{m18Schock}
{Marleau}, G.-D., {Mordasini}, C., \& {Kuiper}, R. 2019,
  \href{http://dx.doi.org/10.3847/1538-4357/ab245b}{\JournalTitle{\apj}, 881,
  144}

\bibitem[{{Marleau} {et~al.}(2022){Marleau}, {Aoyama}, {Kuiper}, {Follette},
  {Turner}, {Cugno}, {Manara}, {Haffert}, {Kitzmann}, {Ringqvist}, {Wagner},
  {van Boekel}, {Sallum}, {Janson}, {Schmidt}, {Venuti}, {Lovis}, \&
  {Mordasini}}]{maea21}
{Marleau}, G.-D., {Aoyama}, Y., {Kuiper}, R., {et~al.} 2022,
  \href{http://dx.doi.org/10.1051/0004-6361/202037494}{\JournalTitle{\aap},
  657, A38}

\bibitem[{{Marley} {et~al.}(2007){Marley}, {Fortney}, {Hubickyj},
  {Bodenheimer}, \& {Lissauer}}]{marl07}
{Marley}, M.~S., {Fortney}, J.~J., {Hubickyj}, O., {Bodenheimer}, P., \&
  {Lissauer}, J.~J. 2007,
  \href{http://dx.doi.org/10.1086/509759}{\JournalTitle{\apj}, 655, 541}

\bibitem[{{Marois} {et~al.}(2008){Marois}, {Lafreni{\`e}re}, {Macintosh}, \&
  {Doyon}}]{marois08a}
{Marois}, C., {Lafreni{\`e}re}, D., {Macintosh}, B., \& {Doyon}, R. 2008,
  \href{http://dx.doi.org/10.1086/523839}{\JournalTitle{\apj}, 673, 647}

\bibitem[{{Mendoza} {et~al.}(2009){Mendoza}, {Tejeda}, \& {Nagel}}]{mendoza09}
{Mendoza}, S., {Tejeda}, E., \& {Nagel}, E. 2009,
  \href{http://dx.doi.org/10.1111/j.1365-2966.2008.14210.x}{\JournalTitle{\mnras},
  393, 579}

\bibitem[{{Mignone} {et~al.}(2007){Mignone}, {Bodo}, {Massaglia}, {Matsakos},
  {Tesileanu}, {Zanni}, \& {Ferrari}}]{mignone07}
{Mignone}, A., {Bodo}, G., {Massaglia}, S., {et~al.} 2007,
  \href{http://dx.doi.org/10.1086/513316}{\JournalTitle{\apjs}, 170, 228}

\bibitem[{{Mignone} {et~al.}(2012){Mignone}, {Zanni}, {Tzeferacos}, {van
  Straalen}, {Colella}, \& {Bodo}}]{mignone12}
{Mignone}, A., {Zanni}, C., {Tzeferacos}, P., {et~al.} 2012,
  \href{http://dx.doi.org/10.1088/0067-0049/198/1/7}{\JournalTitle{\apjs}, 198,
  7}

\bibitem[{{Moldenhauer} {et~al.}(2021){Moldenhauer}, {Kuiper}, {Kley}, \&
  {Ormel}}]{mold21}
{Moldenhauer}, T.~W., {Kuiper}, R., {Kley}, W., \& {Ormel}, C.~W. 2021,
  \href{http://dx.doi.org/10.1051/0004-6361/202040220}{\JournalTitle{\aap},
  646, L11}

\bibitem[{{Moldenhauer} {et~al.}(2022){Moldenhauer}, {Kuiper}, {Kley}, \&
  {Ormel}}]{mold22}
{Moldenhauer}, T.~W., {Kuiper}, R., {Kley}, W., \& {Ormel}, C.~W. 2022,
  \href{http://dx.doi.org/10.1051/0004-6361/202141955}{\JournalTitle{\aap},
  661, A142}

\bibitem[{{Mordasini}(2013)}]{morda13}
{Mordasini}, C. 2013,
  \href{http://dx.doi.org/10.1051/0004-6361/201321617}{\JournalTitle{\aap},
  558, A113}

\bibitem[{{Mordasini} {et~al.}(2012{\natexlab{a}}){Mordasini}, {Alibert},
  {Georgy}, {Dittkrist}, {Klahr}, \& {Henning}}]{morda12_II}
{Mordasini}, C., {Alibert}, Y., {Georgy}, C., {et~al.} 2012{\natexlab{a}},
  \href{http://dx.doi.org/10.1051/0004-6361/201118464}{\JournalTitle{\aap},
  547, A112}

\bibitem[{{Mordasini} {et~al.}(2012{\natexlab{b}}){Mordasini}, {Alibert},
  {Klahr}, \& {Henning}}]{morda12_I}
{Mordasini}, C., {Alibert}, Y., {Klahr}, H., \& {Henning}, T.
  2012{\natexlab{b}},
  \href{http://dx.doi.org/10.1051/0004-6361/201118457}{\JournalTitle{\aap},
  547, A111}

\bibitem[{{M{\"u}ller} {et~al.}(2018){M{\"u}ller}, {Keppler}, {Henning},
  {Samland}, {Chauvin}, {Beust}, {Maire}, {Molaverdikhani}, {van Boekel},
  {Benisty}, {Boccaletti}, {Bonnefoy}, {Cantalloube}, {Charnay}, {Baudino},
  {Gennaro}, {Long}, {Cheetham}, {Desidera}, {Feldt}, {Fusco}, {Girard},
  {Gratton}, {Hagelberg}, {Janson}, {Lagrange}, {Langlois}, {Lazzoni}, {Ligi},
  {M{\'e}nard}, {Mesa}, {Meyer}, {Molli{\`e}re}, {Mordasini}, {Moulin},
  {Pavlov}, {Pawellek}, {Quanz}, {Ramos}, {Rouan}, {Sissa}, {Stadler}, {Vigan},
  {Wahhaj}, {Weber}, \& {Zurlo}}]{mueller18}
{M{\"u}ller}, A., {Keppler}, M., {Henning}, T., {et~al.} 2018,
  \href{http://dx.doi.org/10.1051/0004-6361/201833584}{\JournalTitle{\aap},
  617, L2}

\bibitem[{Mungan(2009)}]{mungan09}
Mungan, C.~E. 2009,
  \href{http://dx.doi.org/10.1119/1.3246467}{\JournalTitle{The Physics
  Teacher}, 47, 502}

\bibitem[{{Nelson} {et~al.}(2023){Nelson}, {Lega}, \& {Morbidelli}}]{nelson23}
{Nelson}, R.~P., {Lega}, E., \& {Morbidelli}, A. 2023,
  \href{http://dx.doi.org/10.1051/0004-6361/202244885}{\JournalTitle{\aap},
  670, A113}

\bibitem[{{Nielsen} {et~al.}(2019){Nielsen}, {De Rosa}, {Macintosh}, {Wang},
  {Ruffio}, {Chiang}, {Marley}, {Saumon}, {Savransky}, \& {Ammons}}]{nielsen19}
{Nielsen}, E.~L., {De Rosa}, R.~J., {Macintosh}, B., {et~al.} 2019,
  \href{http://dx.doi.org/10.3847/1538-3881/ab16e9}{\JournalTitle{\aj}, 158,
  13}

\bibitem[{{Paxton} {et~al.}(2019){Paxton}, {Smolec}, {Schwab}, {Gautschy},
  {Bildsten}, {Cantiello}, {Dotter}, {Farmer}, {Goldberg}, {Jermyn}, {Kanbur},
  {Marchant}, {Thoul}, {Townsend}, {Wolf}, {Zhang}, \& {Timmes}}]{paxton19}
{Paxton}, B., {Smolec}, R., {Schwab}, J., {et~al.} 2019,
  \href{http://dx.doi.org/10.3847/1538-4365/ab2241}{\JournalTitle{\apjs}, 243,
  10}

\bibitem[{{Pringle}(1981)}]{pringle81}
{Pringle}, J.~E. 1981,
  \href{http://dx.doi.org/10.1146/annurev.aa.19.090181.001033}{\JournalTitle{\araa},
  19, 137}

\bibitem[{{Ringqvist} {et~al.}(2023){Ringqvist}, {Viswanath}, {Aoyama},
  {Janson}, {Marleau}, \& {Brandeker}}]{ringqvist23}
{Ringqvist}, S.~C., {Viswanath}, G., {Aoyama}, Y., {et~al.} 2023,
  \href{http://dx.doi.org/10.1051/0004-6361/202245424}{\JournalTitle{\aap},
  669, L12}

\bibitem[{{Romanova} {et~al.}(2002){Romanova}, {Ustyugova}, {Koldoba}, \&
  {Lovelace}}]{romanova02}
{Romanova}, M.~M., {Ustyugova}, G.~V., {Koldoba}, A.~V., \& {Lovelace},
  R.~V.~E. 2002, \href{http://dx.doi.org/10.1086/342464}{\JournalTitle{\apj},
  578, 420}

\bibitem[{{Sanchis} {et~al.}(2020){Sanchis}, {Picogna}, {Ercolano}, {Testi}, \&
  {Rosotti}}]{Sanchis+2020}
{Sanchis}, E., {Picogna}, G., {Ercolano}, B., {Testi}, L., \& {Rosotti}, G.
  2020, \href{http://dx.doi.org/10.1093/mnras/staa074}{\JournalTitle{\mnras},
  492, 3440}

\bibitem[{{Sanghi} {et~al.}(2022){Sanghi}, {Zhou}, \& {Bowler}}]{sanghi22}
{Sanghi}, A., {Zhou}, Y., \& {Bowler}, B.~P. 2022,
  \href{http://dx.doi.org/10.3847/1538-3881/ac477e}{\JournalTitle{\aj}, 163,
  119}

\bibitem[{{Schulik} {et~al.}(2019){Schulik}, {Johansen}, {Bitsch}, \&
  {Lega}}]{schulik19}
{Schulik}, M., {Johansen}, A., {Bitsch}, B., \& {Lega}, E. 2019,
  \href{http://dx.doi.org/10.1051/0004-6361/201935473}{\JournalTitle{\aap},
  632, A118}

\bibitem[{{Schulik} {et~al.}(2020){Schulik}, {Johansen}, {Bitsch}, {Lega}, \&
  {Lambrechts}}]{schulik20}
{Schulik}, M., {Johansen}, A., {Bitsch}, B., {Lega}, E., \& {Lambrechts}, M.
  2020,
  \href{http://dx.doi.org/10.1051/0004-6361/202037556}{\JournalTitle{\aap},
  642, A187}

\bibitem[{{Semenov} {et~al.}(2003){Semenov}, {Henning}, {Helling}, {Ilgner}, \&
  {Sedlmayr}}]{semenov03}
{Semenov}, D., {Henning}, T., {Helling}, C., {Ilgner}, M., \& {Sedlmayr}, E.
  2003,
  \href{http://dx.doi.org/10.1051/0004-6361:20031279}{\JournalTitle{\aap}, 410,
  611}

\bibitem[{{Shakura} \& {Sunyaev}(1973)}]{shak73}
{Shakura}, N.~I., \& {Sunyaev}, R.~A. 1973, \JournalTitle{\aap}, 500, 33

\bibitem[{{Stolker} {et~al.}(2020){Stolker}, {Marleau}, {Cugno},
  {Molli{\`e}re}, {Quanz}, {Todorov}, \& {K{\"u}hn}}]{Stolker+20b}
{Stolker}, T., {Marleau}, G.~D., {Cugno}, G., {et~al.} 2020,
  \href{http://dx.doi.org/10.1051/0004-6361/202038878}{\JournalTitle{\aap},
  644, A13}

\bibitem[{{Szul{\'a}gyi}(2017)}]{szul17b}
{Szul{\'a}gyi}, J. 2017,
  \href{http://dx.doi.org/10.3847/1538-4357/aa7515}{\JournalTitle{\apj}, 842,
  103}

\bibitem[{{Szul{\'a}gyi} {et~al.}(2022){Szul{\'a}gyi}, {Binkert}, \&
  {Surville}}]{szul22}
{Szul{\'a}gyi}, J., {Binkert}, F., \& {Surville}, C. 2022,
  \href{http://dx.doi.org/10.3847/1538-4357/ac32d1}{\JournalTitle{\apj}, 924,
  1}

\bibitem[{{Szul{\'a}gyi} {et~al.}(2019){Szul{\'a}gyi}, {Dullemond}, {Pohl}, \&
  {Quanz}}]{szul19II}
{Szul{\'a}gyi}, J., {Dullemond}, C.~P., {Pohl}, A., \& {Quanz}, S.~P. 2019,
  \href{http://dx.doi.org/10.1093/mnras/stz1326}{\JournalTitle{\mnras}, 487,
  1248}

\bibitem[{{Szul{\'a}gyi} \& {Ercolano}(2020)}]{szul20}
{Szul{\'a}gyi}, J., \& {Ercolano}, B. 2020,
  \href{http://dx.doi.org/10.3847/1538-4357/abb5a2}{\JournalTitle{\apj}, 902,
  126}

\bibitem[{{Takasao} {et~al.}(2021){Takasao}, {Aoyama}, \& {Ikoma}}]{takasao21}
{Takasao}, S., {Aoyama}, Y., \& {Ikoma}, M. 2021,
  \href{http://dx.doi.org/10.3847/1538-4357/ac0f7e}{\JournalTitle{\apj}, 921,
  10}

\bibitem[{{Tanigawa} {et~al.}(2012){Tanigawa}, {Ohtsuki}, \&
  {Machida}}]{tanigawa12}
{Tanigawa}, T., {Ohtsuki}, K., \& {Machida}, M.~N. 2012,
  \href{http://dx.doi.org/10.1088/0004-637X/747/1/47}{\JournalTitle{\apj}, 747,
  47}

\bibitem[{{Thanathibodee} {et~al.}(2019){Thanathibodee}, {Calvet}, {Bae},
  {Muzerolle}, \& {Hern{\'a}ndez}}]{thanathibodee19}
{Thanathibodee}, T., {Calvet}, N., {Bae}, J., {Muzerolle}, J., \&
  {Hern{\'a}ndez}, R.~F. 2019,
  \href{http://dx.doi.org/10.3847/1538-4357/ab44c1}{\JournalTitle{\apj}, 885,
  94}

\bibitem[{{Thommes} {et~al.}(2008){Thommes}, {Matsumura}, \&
  {Rasio}}]{thommes08}
{Thommes}, E.~W., {Matsumura}, S., \& {Rasio}, F.~A. 2008,
  \href{http://dx.doi.org/10.1126/science.1159723}{\JournalTitle{Science}, 321,
  814}

\bibitem[{{Toci} {et~al.}(2020){Toci}, {Lodato}, {Christiaens}, {Fedele},
  {Pinte}, {Price}, \& {Testi}}]{toci20}
{Toci}, C., {Lodato}, G., {Christiaens}, V., {et~al.} 2020,
  \href{http://dx.doi.org/10.1093/mnras/staa2933}{\JournalTitle{\mnras}, 499,
  2015}

\bibitem[{{Ulrich}(1976)}]{ulrich76}
{Ulrich}, R.~K. 1976,
  \href{http://dx.doi.org/10.1086/154840}{\JournalTitle{\apj}, 210, 377}

\bibitem[{{Uyama} {et~al.}(2021){Uyama}, {Xie}, {Aoyama}, {Beichman},
  {Hashimoto}, {Dong}, {Hasegawa}, {Ikoma}, {Mawet}, {McElwain}, {Ruffio},
  {Wagner}, {Wang}, \& {Zhou}}]{uyama21}
{Uyama}, T., {Xie}, C., {Aoyama}, Y., {et~al.} 2021,
  \href{http://dx.doi.org/10.3847/1538-3881/ac2739}{\JournalTitle{\aj}, 162,
  214}

\bibitem[{{Vigan} {et~al.}(2021){Vigan}, {Fontanive}, {Meyer}, {Biller},
  {Bonavita}, {Feldt}, {Desidera}, {Marleau}, {Emsenhuber}, {Galicher}, {Rice},
  {Forgan}, {Mordasini}, {Gratton}, {Le Coroller}, {Maire}, {Cantalloube},
  {Chauvin}, {Cheetham}, {Hagelberg}, {Lagrange}, {Langlois}, {Bonnefoy},
  {Beuzit}, {Boccaletti}, {D'Orazi}, {Delorme}, {Dominik}, {Henning}, {Janson},
  {Lagadec}, {Lazzoni}, {Ligi}, {Menard}, {Mesa}, {Messina}, {Moutou},
  {M{\"u}ller}, {Perrot}, {Samland}, {Schmid}, {Schmidt}, {Sissa}, {Turatto},
  {Udry}, {Zurlo}, {Abe}, {Antichi}, {Asensio-Torres}, {Baruffolo}, {Baudoz},
  {Baudrand}, {Bazzon}, {Blanchard}, {Bohn}, {Brown Sevilla}, {Carbillet},
  {Carle}, {Cascone}, {Charton}, {Claudi}, {Costille}, {De Caprio},
  {Delboulb{\'e}}, {Dohlen}, {Engler}, {Fantinel}, {Feautrier}, {Fusco},
  {Gigan}, {Girard}, {Giro}, {Gisler}, {Gluck}, {Gry}, {Hubin}, {Hugot},
  {Jaquet}, {Kasper}, {Le Mignant}, {Llored}, {Madec}, {Magnard}, {Martinez},
  {Maurel}, {M{\"o}ller-Nilsson}, {Mouillet}, {Moulin}, {Orign{\'e}}, {Pavlov},
  {Perret}, {Petit}, {Pragt}, {Puget}, {Rabou}, {Ramos}, {Rickman}, {Rigal},
  {Rochat}, {Roelfsema}, {Rousset}, {Roux}, {Salasnich}, {Sauvage}, {Sevin},
  {Soenke}, {Stadler}, {Suarez}, {Wahhaj}, {Weber}, \& {Wildi}}]{vigan21}
{Vigan}, A., {Fontanive}, C., {Meyer}, M., {et~al.} 2021,
  \href{http://dx.doi.org/10.1051/0004-6361/202038107}{\JournalTitle{\aap},
  651, A72}

\bibitem[{{{\v{Z}}erjal} {et~al.}(2023){{\v{Z}}erjal}, {Ireland}, {Crundall},
  {Krumholz}, \& {Rains}}]{rZerjal23}
{{\v{Z}}erjal}, M., {Ireland}, M.~J., {Crundall}, T.~D., {Krumholz}, M.~R., \&
  {Rains}, A.~D. 2023,
  \href{http://dx.doi.org/10.1093/mnras/stac3693}{\JournalTitle{\mnras}, 519,
  3992}

\bibitem[{{Wagner} {et~al.}(2019){Wagner}, {Apai}, \& {Kratter}}]{wagner19}
{Wagner}, K., {Apai}, D., \& {Kratter}, K.~M. 2019,
  \href{http://dx.doi.org/10.3847/1538-4357/ab1904}{\JournalTitle{\apj}, 877,
  46}

\bibitem[{{Wagner} {et~al.}(2018){Wagner}, {Follete}, {Close}, {Apai}, {Gibbs},
  {Keppler}, {M{\"u}ller}, {Henning}, {Kasper}, {Wu}, {Long}, {Males},
  {Morzinski}, \& {McClure}}]{wagner18}
{Wagner}, K., {Follete}, K.~B., {Close}, L.~M., {et~al.} 2018,
  \href{http://dx.doi.org/10.3847/2041-8213/aad695}{\JournalTitle{\apjl}, 863,
  L8}

\bibitem[{{Wang} {et~al.}(2021){Wang}, {Vigan}, {Lacour}, {Nowak}, {Stolker},
  {De Rosa}, {Ginzburg}, {Gao}, {Abuter}, {Amorim}, {Asensio-Torres},
  {Baub{\"o}ck}, {Benisty}, {Berger}, {Beust}, {Beuzit}, {Blunt}, {Boccaletti},
  {Bohn}, {Bonnefoy}, {Bonnet}, {Brandner}, {Cantalloube}, {Caselli},
  {Charnay}, {Chauvin}, {Choquet}, {Christiaens}, {Cl{\'e}net}, {Coud{\'e} Du
  Foresto}, {Cridland}, {de Zeeuw}, {Dembet}, {Dexter}, {Drescher}, {Duvert},
  {Eckart}, {Eisenhauer}, {Facchini}, {Gao}, {Garcia}, {Garcia Lopez},
  {Gardner}, {Gendron}, {Genzel}, {Gillessen}, {Girard}, {Haubois},
  {Hei{\ss}el}, {Henning}, {Hinkley}, {Hippler}, {Horrobin}, {Houll{\'e}},
  {Hubert}, {Jim{\'e}nez-Rosales}, {Jocou}, {Kammerer}, {Keppler}, {Kervella},
  {Meyer}, {Kreidberg}, {Lagrange}, {Lapeyr{\`e}re}, {Le Bouquin}, {L{\'e}na},
  {Lutz}, {Maire}, {M{\'e}nard}, {M{\'e}rand}, {Molli{\`e}re}, {Monnier},
  {Mouillet}, {M{\"u}ller}, {Nasedkin}, {Ott}, {Otten}, {Paladini}, {Paumard},
  {Perraut}, {Perrin}, {Pfuhl}, {Pueyo}, {Rameau}, {Rodet},
  {Rodr{\'\i}guez-Coira}, {Rousset}, {Scheithauer}, {Shangguan}, {Shimizu},
  {Stadler}, {Straub}, {Straubmeier}, {Sturm}, {Tacconi}, {van Dishoeck},
  {Vincent}, {von Fellenberg}, {Ward-Duong}, {Widmann}, {Wieprecht},
  {Wiezorrek}, {Woillez}, \& {Gravity Collaboration}}]{wang21vlti}
{Wang}, J.~J., {Vigan}, A., {Lacour}, S., {et~al.} 2021,
  \href{http://dx.doi.org/10.3847/1538-3881/abdb2d}{\JournalTitle{\aj}, 161,
  148}

\bibitem[{{Xie} {et~al.}(2020){Xie}, {Haffert}, {de Boer}, {Kenworthy},
  {Brinchmann}, {Girard}, {Snellen}, \& {Keller}}]{xie20}
{Xie}, C., {Haffert}, S.~Y., {de Boer}, J., {et~al.} 2020,
  \href{http://dx.doi.org/10.1051/0004-6361/202038242}{\JournalTitle{\aap},
  644, A149}

\bibitem[{{Zel'dovich} \& {Raizer}(1967)}]{zeldovich67}
{Zel'dovich}, Y.~B., \& {Raizer}, Y.~P. 1967, {Physics of Shock Waves and
  High-Temperature Hydrodynamic Phenomena} (Academic Press)

\bibitem[{{Zhou} {et~al.}(2021){Zhou}, {Bowler}, {Wagner}, {Schneider}, {Apai},
  {Kraus}, {Close}, {Herczeg}, \& {Fang}}]{zhou21}
{Zhou}, Y., {Bowler}, B.~P., {Wagner}, K.~R., {et~al.} 2021,
  \href{http://dx.doi.org/10.3847/1538-3881/abeb7a}{\JournalTitle{\aj}, 161,
  244}

\bibitem[{{Zhou} {et~al.}(2022){Zhou}, {Sanghi}, {Bowler}, {Wu}, {Close},
  {Long}, {Ward-Duong}, {Zhu}, {Kraus}, {Follette}, \& {Bae}}]{zhou22}
{Zhou}, Y., {Sanghi}, A., {Bowler}, B.~P., {et~al.} 2022,
  \href{http://dx.doi.org/10.3847/2041-8213/ac7fef}{\JournalTitle{\apjl}, 934,
  L13}

\bibitem[{{Zhu} \& {Dong}(2021)}]{zhudong21}
{Zhu}, W., \& {Dong}, S. 2021,
  \href{http://dx.doi.org/10.1146/annurev-astro-112420-020055}{\JournalTitle{\araa},
  59, 291}

\bibitem[{{Zhu}(2015)}]{zhu15}
{Zhu}, Z. 2015,
  \href{http://dx.doi.org/10.1088/0004-637X/799/1/16}{\JournalTitle{\apj}, 799,
  16}

\bibitem[{{Zurlo} {et~al.}(2020){Zurlo}, {Cugno}, {Montesinos}, {Perez},
  {Canovas}, {Casassus}, {Christiaens}, {Cieza}, \& {Huelamo}}]{Zurlo+2020}
{Zurlo}, A., {Cugno}, G., {Montesinos}, M., {et~al.} 2020,
  \href{http://dx.doi.org/10.1051/0004-6361/201936891}{\JournalTitle{\aap},
  633, A119}

\end{thebibliography}
\bibliographystyle{yahapj.bst}

\end{document}